\documentclass[12pt]{article}
\usepackage[pdftex]{graphicx}
\usepackage{amsfonts,amsmath,amssymb,epsf}
\usepackage{amsmath,braket}

\usepackage{etoolbox}
\patchcmd{\thebibliography}
  {\settowidth}
  {\setlength{\itemsep}{0pt plus 0.1pt}\settowidth}
  {}{}
\apptocmd{\thebibliography}
  {\small}
  {}{}
  
  \usepackage{latexsym}
\usepackage{here}
\usepackage{cite}
\usepackage[pdftex]{hyperref}
\usepackage[dvipsnames]{xcolor}
       \usepackage{graphicx,color,xcolor}
\hypersetup{
    bookmarks=true,         
    unicode=false,          
    pdftoolbar=true,        
    pdfmenubar=true,        
    pdffitwindow=false,     
    pdfstartview={FitH},    
    pdfnewwindow=true,      
    colorlinks=false,       
    linkbordercolor=Blue,          
    citebordercolor=SkyBlue,
}
\usepackage[margin=0.5cm, labelfont=bf]{caption}
\newcommand{\ep}{\epsilon}

\numberwithin{equation}{section}
\begin{document}

\begin{titlepage}
\thispagestyle{empty}

\begin{flushright}
\end{flushright}

\bigskip

\begin{center}
\noindent{{\Large \textbf{Island for Gravitationally Prepared State and \\Pseudo Entanglement Wedge}}}\\
\vspace{1cm}

Masamichi Miyaji
\vspace{0.4cm}\\

{\it Berkeley Center for Theoretical Physics, \\
Department of Physics, University of California, Berkeley, \\
CA 94720, USA }\\

\end{center}

\begin{abstract}We consider spacetime initiated by a finite-sized initial boundary as a generalization of the Hartle-Hawking no-boundary state. We study entanglement entropy of matter state prepared by such spacetime. We find that the entanglement entropy for large subregion is given either by the initial state entanglement or the entanglement island, preventing the entropy to grow arbitrarily large. Consequently, the entanglement entropy is always bounded from above by the boundary area of the island, leading to an entropy bound in terms of the island. The island $I$ is located in the analytically continued spacetime, either at the bra or the ket part of the spacetime in Schwinger-Keldysh formalism. The entanglement entropy is given by an average of $complex$ pseudo generalized entropy for each entanglement island. We find a necessary condition of the initial state to be consistent with the strong sub-additivity, which requires that any probe degrees of freedom are thermally entangled with the rest of the system. We then find a large parameter region where the spacetime with finite-sized initial boundary, which does not have the factorization puzzle at leading order, dominates over the Hartle-Hawking no-boundary state or the bra-ket wormhole. Due to the absence of a moment of time reflection symmetry, the island in our setup is a generalization of the entanglement wedge, called pseudo entanglement wedge. In pseudo entanglement wedge reconstruction, we consider reconstructing the bulk matter transition matrix on $A\cup I$, from a fine-grained state on $A$. The bulk transition matrix is given by a thermofield double state with a projection by the initial state. We also provide an AdS/BCFT model by considering EOW branes with corners. We also find the exponential hardness of such reconstruction task using a generalization of Python's lunch conjecture to pseudo generalized entropy.\end{abstract}

\end{titlepage}

\newpage

\tableofcontents


\section{Introduction and Summary}

The resolution of the black hole information paradox \cite{Penington:2019npb, Almheiri:2019psf, Almheiri:2019hni, Penington:2019kki, Almheiri:2019qdq} 
has revealed that semiclassical gravity with Euclidean wormholes can explain Page curve in the black hole evaporation, capturing the finiteness of Hilbert space dimension of gravitational theories. The key machineries behind this surprising finding are the generalized versions of Ryu-Takayanagi formula \cite{Ryu:2006bv, Ryu:2006ef, Hubeny:2007xt, Takayanagi:2011zk, Fujita:2011fp, Lewkowycz:2013nqa, Faulkner:2013ana, Engelhardt:2014gca, Jafferis:2015del}, the brane world models \cite{Randall:1999vf, Karch:2000ct}, the entanglement wedge reconstruction \cite{Almheiri:2014lwa, Pastawski:2015qua, Dong:2016eik, Harlow:2016vwg, Cotler:2017erl, Hayden:2018khn, Chen:2019gbt} and the replica wormholes \cite{Penington:2019kki, Almheiri:2019qdq}. The application of the entanglement wedge reconstruction allows us a natural interpretation of Hayden-Preskill protocol, which states we can recover information thrown into black hole after the Page time from a finite number of quanta of Hawking radiation \cite{Hayden:2007cs}, in terms of a penetration of the entanglement wedge into the black hole interior. The entanglement island has been considered in \cite{Almheiri:2019yqk, Rozali:2019day, Almheiri:2019psy, Bousso:2019ykv, Marolf:2020xie, Balasubramanian:2020hfs, Gautason:2020tmk, Anegawa:2020ezn, Hashimoto:2020cas, Hartman:2020swn, Bousso:2020kmy, Chen:2020uac,  Chandrasekaran:2020qtn, Li:2020ceg, Bak:2020enw,  Chen:2020tes, Dong:2020uxp, Chen:2020hmv, Hernandez:2020nem, Marolf:2020rpm, Matsuo:2020ypv, Numasawa:2020sty, Colin-Ellerin:2020mva, Geng:2020fxl, Caceres:2020jcn, Akal:2020twv, Bousso:2021sji, Kawabata:2021hac, Miyata:2021ncm, Kawabata:2021vyo}, which has been applied to the absence of global symmetry in gravity \cite{Harlow:2020bee, Chen:2020ojn,Hsin:2020mfa, Yonekura:2020ino}. Closely related discussions on wormholes are in \cite{Saad:2018bqo,Saad:2019pqd, Cotler:2020lxj, Cotler:2021cqa}, and parallel approach toward interior of black hokes using ensembles is given in \cite{Nomura:2019dlz, Langhoff:2020jqa, Nomura:2020ewg}. 

In this paper, we consider entanglement entropy in cosmological spacetimes and study the lessons from the finiteness of Hilbert space dimension in gravity, using the entanglement island. The entanglement islands in various cosmological spacetimes have been studied in \cite{Penington:2019kki, Hartman:2020khs, Balasubramanian:2020coy, Balasubramanian:2020xqf, Dong:2020uxp, Chen:2020tes, Geng:2021wcq, Aalsma:2021bit, Balasubramanian:2021wgd, Kames-King:2021etp, Langhoff:2021uct}. From the holographic principle and the extension of Bekenstein bound to cosmological spacetimes, the semiclassical entropy of a region $A$ of a time slice is upper bounded by its boundary area \cite{Bekenstein:1980jp, Fischler:1998st, Bousso:1999xy, Bousso:1999cb, Bousso:2000nf, Bousso:2000md, Bousso:2015mna}

\begin{equation}
S_A\leq\frac{\text{Area}[\partial A]}{4G_N},\label{areabound2}
\end{equation}\\
when the light sheet from $\partial A$ with non positive expansion can terminate at a point, with no singularities in the future or past domain of dependence. This formula indicates that the number of independent, effective matter states is bounded from above by the boundary area of $A$. For Bekenstein-Hawking entropy in asymptotically de Sitter spacetimes, see \cite{Gibbons:1977mu, Shiromizu:1993mt, Hayward:1993tt, Maeda:1997fh}. Since the finiteness of Hilbert space dimension of a black hole can be captured by the island formula, it is natural to expect that the island formula with appropriate generalization would allow us to derive an analog of (\ref{areabound2}). Indeed, by introducing {\it{bra-ket wormhole}} \cite{Penington:2019kki, Anous:2020lka, Chen:2020tes, Dong:2020uxp} and applying the island formula, \cite{Chen:2020tes} showed that an analog of (\ref{areabound2}) holds for spacetimes with a bra-ket wormhole. For closely related constructions were considered in \cite{Akal:2020wfl} in terms of wedge holography, and from continuous tensor network viewpoint in \cite{Caputa:2017urj, Caputa:2017yrh, Boruch:2020wax, Boruch:2021hqs}. Importantly, the island formula with a bra-ket wormhole is free from the strong sub-additivity paradox, which arises when we consider decoupled CFT with a small central charge and apply the island formula. 

\begin{figure}[t]
 \begin{center}
 \includegraphics[width=12.0cm,clip]{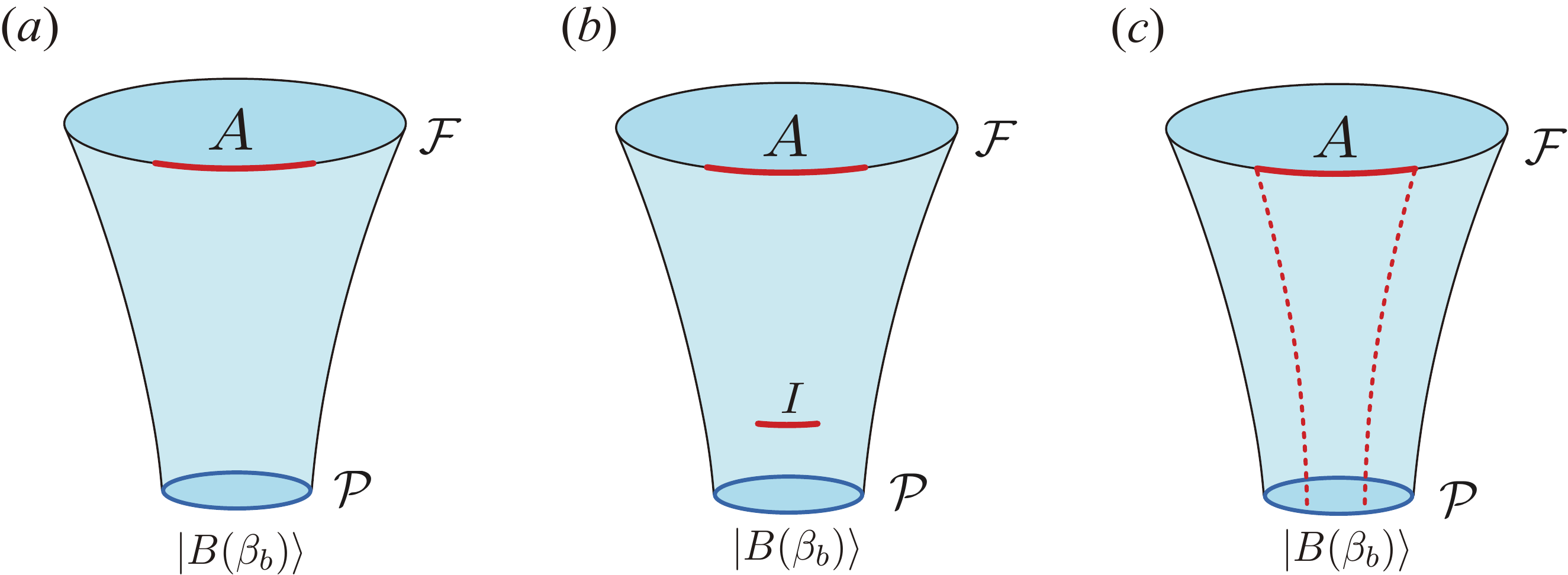}
 \end{center}
 \caption{The spacetime we consider is initiated by an End-of-the-world brane. We consider the matter state prepared by such spacetime, and study its entanglement entropy. The entanglement entropy has three phases. ($a$) Naive thermal entropy phase, computed via effective matter theory. ($b$) The island phase, the entropy is an average of pseudo entropy. ($c$) The boundary phase, the entropy is given by the boundary entropy of the initial state.}
 \label{fig:total}
\end{figure}

In this paper, we will consider a different class of initial states and initial conditions for gravity. The Hartle-Hawking no-boundary state, which was considered in \cite{Chen:2020tes}, is a particular choice of initial conditions of gravity, setting the initial size of its timeslice as zero. In Einstein gravity, the canonical conjugate of the induced metric on a time slice are extrinsic curvatures, and fixing those extrinsic curvatures is equally legitimate initial condition as fixing the initial induced metric \cite{Bousso:1998na}. Therefore, instead of Hartle-Hawking no-boundary proposal, the spacetime we consider has a finite-sized initial-boundary $\mathcal{P}$ with boundary conditions on extrinsic curvatures and matter. Such boundary conditions were considered as off-shell configurations in \cite{DiTucci:2019dji, DiTucci:2019xcr, DiTucci:2019bui} to obtain the Hartle-Hawking no boundary state as the dominant contribution in the path integral using Picard-Lefschetz theory. Note that here we consider such finite initial-boundary spacetime as a on-shell saddle point instead of off-shell configuration, as a natural candidate spacetime which emerges when a phase transition takes place at a transition from UV theory to the semiclassical gravity regime.

On this finite-sized initial-boundary $\mathcal{P}$, there is a freedom for the matter boundary condition. We will consider the pure state as our initial state that behaves as a thermal state from local probes. One of the most convenient, simplest choices is the {\it{regularized boundary state}}

\begin{equation}
|B(\beta)\rangle:=\text{e}^{-\beta H_{\text{eff}}/4}|B\rangle,\label{regularizedbstate}\end{equation}\\
here $|B\rangle$ is the unnormalized boundary state. These states look thermal from local probes with the inverse temperature $\beta$. Indeed, for small subregions, the entanglement entropy behaves thermally with inverse temperature $\beta$, while the entanglement entropy of the entire region is zero. The regularized boundary states are used as the models of the initial state in global quench induced by relevant deformation \cite{Calabrese:2005in}. Note that a regularized boundary state is close to the CFT vacuum state at UV while close to the CFT boundary state at IR. Using such regularized boundary states, we can model a transition from fundamental UV complete theory of gravity to semiclassical gravity description. 

It is important to note the difference between the initial boundary state $|B(\beta)\rangle$ and the Hartle-Hawking state. The initial boundary state $|B(\beta)\rangle$ is a locally thermal state by itself, which is distinct from the thermal state obtained via taking Rindler or Milne patch of a Minkowski CFT vacuum state as in Hartle-Hawking state. 

With these initial conditions on $\mathcal{P}$, we consider a late time slice $\mathcal{F}$, on which we turn off the gravitational interaction afterward, leaving the effective matter CFT with no gravity. The length of final timeslice $\mathcal{F}$ is fixed, and in JT gravity the value of dilaton is also fixed. From quantum mechanical perspective, we are projecting gravity degrees of freedom onto a fixed length, fixed dilaton state at $\mathcal{F}$. Since the fixed length, fixed dilaton state is, at semiclassical level, pure, the projected state is also pure. 

We will then study the entanglement entropy of the matter CFT on this fixed geometry. Since the initial matter state is locally thermal, the state on $\mathcal{F}$ is also locally thermal, and consequently, the entanglement entropy obeys the volume law for small subregions. However, the entropy bound (\ref{areabound}) suggests that this volume law should be modified for the sufficiently large subregions. We will indeed show that either the finiteness of the Hilbert space dimension or the initial state entropy modify large subregion entanglement entropy. We will explain the entanglement entropy of large subregion $A$ is given either by the area of the boundary of entanglement island which lives in the past spacetime or by the initial state entanglement entropy

\begin{equation}
S_A\approx \text{Min}\Big[S^{\text{Thermal}}_A(\beta_M),~S^{\text{Island}}_A(I),~S^{\text{Initial Boundary}}_A\Big],\label{SBisland}
\end{equation}\\
assuming we have an island which is extremal. This is one of the main results of this paper; see Fig.\ref{fig:total}. Here $I$ is the entanglement island and the entropy is given by 

\begin{equation}
S^{\text{Island}}_A(I)=\underset{I}{\text{Min}}~\text{Re}\Big[\underset{I}{\text{Ext}}\Big[\frac{\text{Area}[\partial I]}{4G_N}+S^P(A\cup I)\Big]\Big],
\end{equation}\\
the boundary area of the island. Here we $S^P$ is pseudo entropy, which will be discussed in section \ref{pEntanglement}. The initial boundary entropy is given by $S^{\text{Initial Boundary}}_A\approx 2S_B$ where $S_B$ is the boundary entropy of the initial boundary state $|B\rangle$. Both the area term and the boundary entropy are constant in $2$ dimensions, while they both obey the {\it{area law}} in higher dimensions. The direct consequence of (\ref{SBisland}) is the island version of the entropy bound

\begin{equation}
S_A\leq \text{Re}\Big[\frac{\text{Area}[\partial I]}{4G_N}+S^P(A\cup I)\Big],\label{areabound}
\end{equation}\\ 
which is more general than (\ref{SBisland}), because it can be applied to when we do not have an island that is extremal. This bound tells us, the entropy of gravitationally prepared state is always bounded from above by the boundary area of island, which is located at positive Euclidean time in analytically continued Euclidean spacetime.
\begin{figure}[t]
 \begin{center}
 \includegraphics[width=5.0cm,clip]{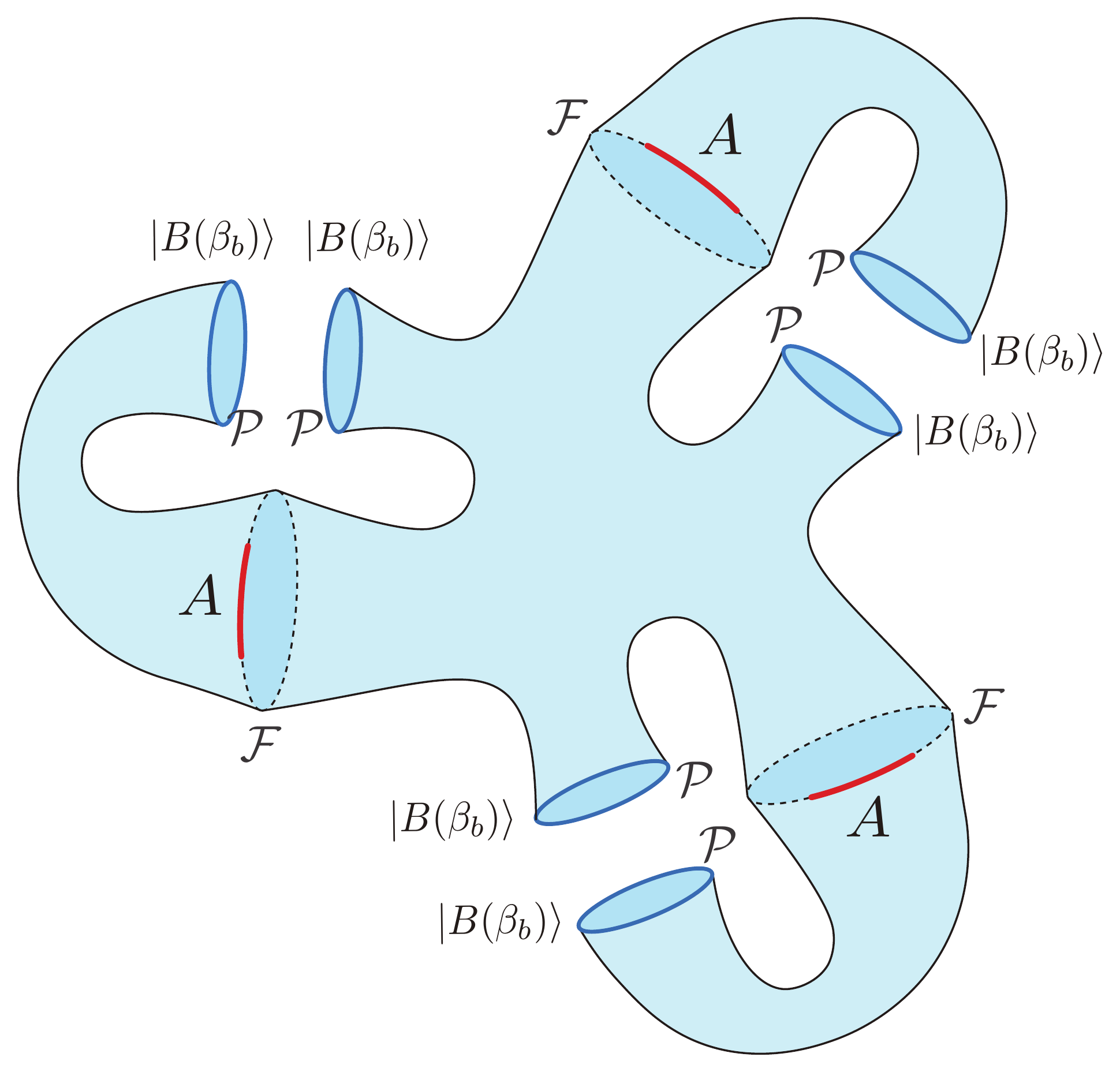}
 \end{center}
 \caption{The replica geometry in Schwinger-Keldysh formalism, with a replica wormhole in the ket part of the bulk. This geometry computes the pseudo entanglement entropy $S^{\text{ket}}_A$. The average of two pseudo entanglement entropy gives the desired entanglement entropy $S^{\text{Island}}_A(I)=(S^{\text{ket}}_A+S^{\text{bra}}_A)/2$, assuming there are no other saddles.}
 \label{fig:geometry}
\end{figure}

The island formula we will use has several distinguishing features compared to the usual island formula in evaporating black holes coupled with bath. Let us consider the relevant gravitational replica trick and the corresponding entanglement island. In order to compute entanglement entropy through replica trick, we consider Schwinger-Keldysh formalism. In the replica geometry, the replica wormholes may connect the bra and ket part of the bulk. The leading geometries which can be analytically continued from integer $n>1$ to non-integer $n$ are the ones whose ket parts are connected with each other via replica wormhole and the ones with connected bra parts, see Fig.\ref{fig:geometry}. These geometries are related via complex conjugation; therefore, the magnitudes are identical so that they maintain the time reflection symmetry as a whole. The resulting entanglement island is described in ($b$) of Fig.\ref{fig:total}. Up to now, there is no known way to understand what the entanglement wedge reconstruction means in such situations, despite its importance.

A main claim we will advocate in this paper is the {\it{pseudo entanglement wedge reconstruction}} does fill this gap and offers a natural interpretation of entanglement wedge reconstruction in such situations. The pseudo entanglement wedge reconstruction is a natural generalization of entanglement wedge reconstruction for the {\it{bulk transition matrix}} which is defined on a subregion of a bulk timeslice, whose boundary is the extremal surface for the {\it{pseudo entropy}} \cite{Nakata:2021ubr, Mollabashi:2020yie, Mollabashi:2021xsd, Nishioka:2021cxe}. The pseudo entropy is a natural generalization of von Neumann entropy of transition matrix, whose value can be complex. In holographic CFT, the pseudo entropy is computed by the area of the extremal surface, which coincides with the usual Ryu-Takayanagi/HRT surface when the bulk dual the Schwinger-Keldysh contour has the moment of time reflection symmetry. On the other hand, when the bulk does not have the moment of time reflection symmetry, the pseudo entropy and corresponding extremal surface give us new information beyond RT/HRT formulae.
\begin{figure}[t]
 \begin{center}
 \includegraphics[width=11.0cm,clip]{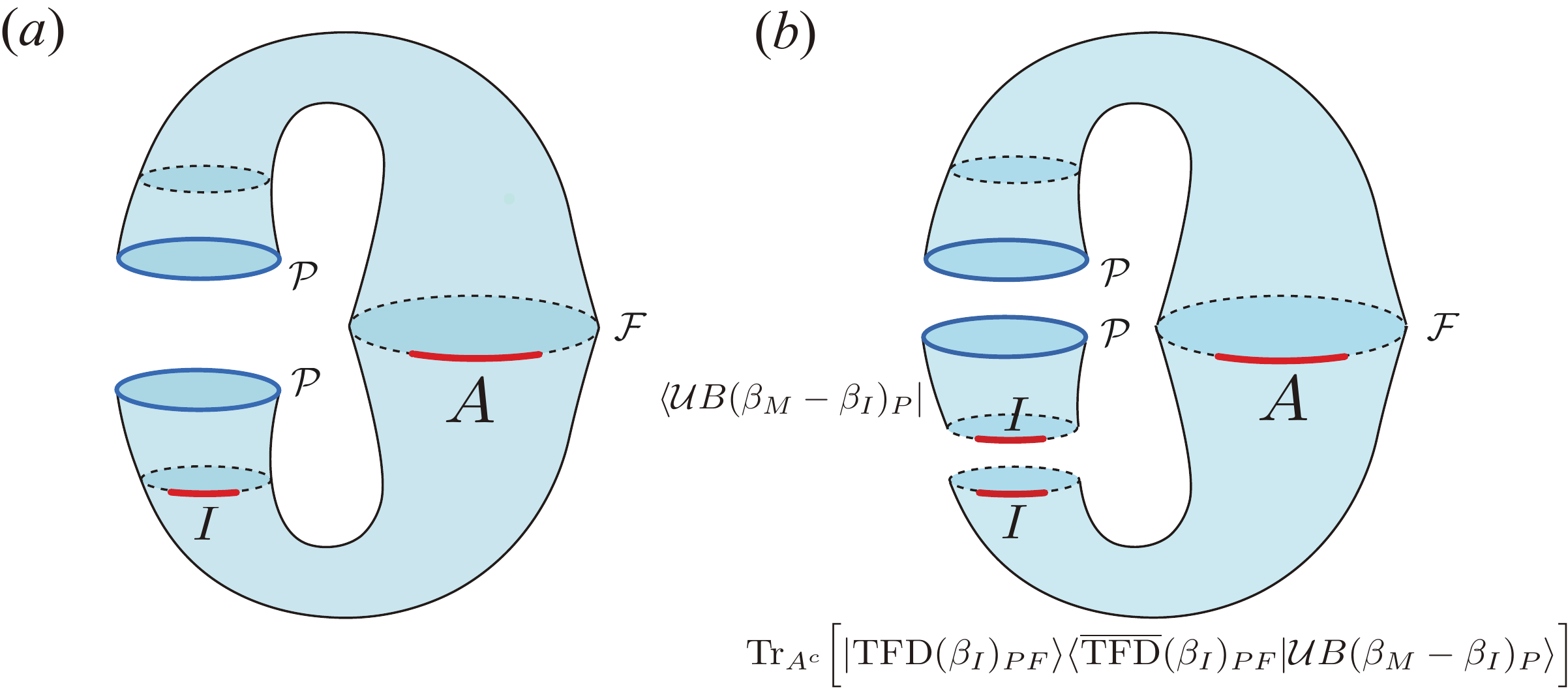}
 \end{center}
 \caption{($a$) The geometry for the bulk transition matrix $\rho_{\text{ket}}^{\text{Bulk}}(A\cup I)$. Its complex conjugate gives the bra counter part. The entanglement island is present only at the ket part, so the time reflection symmetry is absent. ($b$) The bulk transition matrix on $A\cup I$ can be constructed in terms of thermofield double state and by projection onto the initial state.}
 \label{fig:EW}
 \end{figure}

In this paper, we will discuss some basic aspects of the pseudo entanglement wedge reconstruction. We will apply this proposal to understand the cosmological entanglement island with an initial state, whose semiclassical spacetime lacks a moment of time reflection symmetry due to asymmetric islands. We leave the more general analysis to \cite{OPEW}. The reconstruction aims to obtain the bulk transition matrix on $A\cup I$ from the fine-grained state on $A$. Since there are two semiclassical geometries, we have corresponding transition matrices $\rho_{\text{bra}}^{\text{Bulk}}(A\cup I)$ and $\rho_{\text{ket}}^{\text{Bulk}}(A\cup I)=\rho_{\text{bra}}^{\text{Bulk}}(A\cup I)^{\dagger}$. The transition matrix $\rho_{\text{ket}}^{\text{Bulk}}(A\cup I)$ is given by

\begin{equation}
\rho^{\text{Bulk}}_{\text{ket}}(A\cup I:a_I):=\text{Tr}_{I^c, A^c}\Big[|\text{TFD}(\beta_I)_{PF}\rangle\langle \overline{\text{TFD}}(\beta_I)_{PF}|\mathcal{U}B(\beta_M-\beta_I)_P\rangle\langle \mathcal{U}B(\beta_M-\beta_I)_P|\Big],
\end{equation}\\
which is a bulk matter state projected by the initial state, see Fig.\ref{fig:EW}. Here we defined CPT operator $\mathcal{U}$, which originates from the the reverse time evolution to express the transition matrix by thermofield double state with projection. Such thermofield double state can be interpreted as a {\it{closed universe with a final boundary condition}}. 

We can define the corresponding modular Hamiltonian for this transition matrix and generalize the JLMS formula \cite{Jafferis:2015del} to such bulk transition matrix. Such generalization leads to an equality of bulk effective pseudo relative entropy on $A\cup I$ and fine-grained pseudo relative entropy on $A$.

When the entanglement entropy is dominated by the boundary entropy, whether we can apply the usual entanglement wedge reconstruction depends on the detail of the physics of the initial state. When the whole setup is described by AdS/BCFT model in section \ref{hayward1} and the pseudo entanglement wedge contains a subset of the initial time slice, it is possible to reconstruct the bulk state from the boundary fine grained state. See section \ref{hayward1} for more detail. Such case is an example of bra-ket wormholes, and EOW brane may be interpreted as a closed universe.
 
An important sanity check of our proposed island formula is the strong sub-additivity. It is not always guaranteed that the island formula gives a consistent result for the entanglement entropy. Any inconsistency would imply that appropriate modification to the island formula is necessary, or the model itself is pathological. We will derive a condition on the initial state from the strong sub-additivity, that any small fraction of degrees of freedom must be thermally entangled with the rest of the degrees of freedom.
  
The rest of the paper is organized as follows. In section \ref{Egeometry}, we will study states gravitationally prepared by JT gravity on Euclidean AdS with an initial state, which generalizes Hartle-Hawking no-boundary proposal. We will explain how to obtain such geometry from AdS wormhole from $\mathbb{Z}_2$ orbifold, and then introduce a AdS/BCFT model which is dual to the whole set up. We then study when the final time slice remembers the information of the initial state, by comparing with the Hartle-Hawking no boundary state and the bra-ket wormhole, finding a parameter region where the spacetime with initial EOW brane dominates. In section \ref{pEntanglement}, we compute the entanglement entropy of this gravitationally prepared state and obtain the island formula. In particular, we show that the entropy bound (\ref{areabound}) holds due to the emergence of the island. We then derive a condition on the initial state required by the strong sub-additivity of entanglement entropy, which states that any small subset of degrees of freedom is thermally entangled with the rest. We then propose the pseudo entanglement wedge reconstruction, which allows us to reconstruct the transition matrix on the island near the past initial boundary. We will discuss the difficulty of the reconstruction task by generalizing the Python's lunch proposal \cite{Brown:2019rox} to pseudo entropy and give evidence of its exponential difficulty. In section \ref{discussion}, we conclude with discussions on Lorentzian generalization and causality.

\section{Euclidean AdS${}_2$\label{Egeometry}}

We consider the simplest example; Euclidean JT gravity \cite{Maldacena:2016upp}. We set $\Lambda=-1$. The total Euclidean action is

\begin{equation}
S=S_{\text{Top}}+S_{\text{JT}}+S_{\mathcal{P}}+S_{\text{Matter}}.\end{equation}\\
Here $S_{\text{Top}}$ is the topological term

\begin{eqnarray}\nonumber
S_{\text{Top}}&=&-\frac{\phi_0}{16\pi G_N}\int_M d^2x\sqrt{g_M}R-\frac{\phi_0}{8\pi G_N}\int_{\mathcal{F}\cup\mathcal{P}}dx \sqrt{h} K
\\&=&
-\frac{\phi_0}{16\pi G_N}\Big(4\pi(2-2g-b)\Big)=-(2-2g-b)S_0,\end{eqnarray}\\
which suppresses higher topologies when the initial condition is fixed. The second term is the JT gravity action

\begin{equation}
S_{\text{JT}}=-\frac{1}{16\pi G_N}\int_M d^2x\sqrt{g_M}\phi(R+2)-\frac{1}{8\pi G_N}\int_{\mathcal{F}}dx \sqrt{h_{\mathcal{F}}}\phi (K_{\mathcal{F}}-1),\end{equation}\\ 
which fixes the length of $\mathcal{F}$ and the dilaton on $\mathcal{F}$. The third term $S_{\mathcal{P}}$ fixes the boundary condition at the initial timeslice $\mathcal{P}$, and depends on the initial boundary condition chosen. The throughout analysis on the JT boundary conditions and actions was done in \cite{Goel:2020yxl}. The boundary action for externally fixed $\sqrt{h}$ and $\phi$ on $\mathcal{P}$ is

\begin{equation}
S^{\sqrt{h},~\phi}=-\frac{1}{8\pi G_N}\int_{\mathcal{P}}dx \sqrt{h_{\mathcal{P}}}\phi K_{\mathcal{P}}.\end{equation}\\
While the boundary action for externally fixed $\sqrt{h}K$ and $\partial_n\phi$ is

\begin{equation}
S^{\sqrt{h}K,~\partial_n\phi}=\frac{1}{8\pi G_N}\int_{\mathcal{P}}dx \sqrt{h_{\mathcal{P}}}\partial_n\phi.\end{equation}\\
However, we will not use $S^{\sqrt{h},~\phi}$ nor $S^{\sqrt{h}K,~\partial_n\phi}$ in the following. We will instead allow the metric and the dilaton to fluctuate on $\mathcal{P}$. We will use the boundary action 

\begin{equation}
S_{\mathcal{P}}=-\frac{1}{8\pi G_N}\int_{\mathcal{P}}dx \sqrt{h_{\mathcal{P}}}\Big(\phi (K_{\mathcal{P}}-T)-T'\Big).\label{2daction}\end{equation}\\
The equation of motion from this action is

\begin{equation}
\partial_{n}\phi-T\phi-T'=0,~K-T=0.\end{equation}\\
Therefore, $K$ and $\partial_{n}\phi-T\phi$ are fixed on-shell. Then the on-shell past boundary action is 

\begin{equation}
S_{\mathcal{P}}=\frac{1}{8\pi G_N}\int_{\mathcal{P}}dx \sqrt{h_{\mathcal{P}}}T'.\end{equation}\\
Note that here we use outward normal vectors to define the extrinsic curvatures $K$ and normal derivative $\partial_n$. We also use $\tilde{K}=-K$ on $\mathcal{P}$, which is defined by the "future-directed" normal vector $\tilde{n}$. We will now study three solutions corresponding to distinct initial conditions in detail below.

Note that the tensions $T$ and $T'$ is related to parameters of the system dual to 2$d$ gravity, while not to the boundary entropy of the boundary states $|B(\beta)\rangle$\footnote{The boundary entropy of $|B(\beta)\rangle$ is related to the tension of the EOW brane in 3$d$ bulk, when the 2$d$ BCFT has a gravity dual.}. The tensions $T$ and $T'$, and the action (\ref{2daction}) can be derived via dimensional reduction from higher dimensional EOW brane model \cite{Antonini:2021xar}, see also \cite{Takayanagi:2011zk, Fujita:2011fp, Cooper:2018cmb, Antonini:2019qkt}. When the two dimensional model is derived by such dimensional reduction, $T'$ is related to $T$ as $T'=\phi_0T$. This implies that in order to have finite dilaton $\phi$, we need to impose $K=T=\mathcal{O}(\phi_0^{-1})$.


\subsection{Background AdS${}_2$ Geometry\label{background}}

We will analyze pure AdS JT gravity and consider the back reactions to the dilaton and the geometry from the conformal matter and the initial pure state. We will evaluate the gravity+matter partition function, which gives the transition amplitudes from the given initial state to the final state. 

\subsubsection{AdS Wormhole\label{wormhole}}

We assume the spatial direction is homogeneous. We begin with background pure AdS wormhole metric

\begin{equation}
ds^2=\frac{dz^2+dx^2}{\text{sin}^2z},~~\phi=\frac{\phi_r}{-\text{tan}z}~~\text{for} ~~z\leq -\epsilon.
\end{equation}\\
We glue the spacetime at $\mathcal{F}$ defined by $z=-a_c$ to a flat spacetime,

\begin{equation}
ds^2=\frac{-dt_M^2+dx_M^2}{\epsilon^2}, ~~\phi=\frac{\phi_b}{\ep}=\frac{\phi_r}{a_c}.
\end{equation}\\
The relation between the two coordinates is

\begin{equation}
\frac{dx_M}{\ep}=\frac{dx}{a_c}.
\end{equation}\\
Note that $a_c$ parametrizes a family of solutions. DIfferent $a_c$ gives distinct circumference of $x$. Let $L$ represents (large but finite) spatial length $0<x_M<L$. Then $L^{\mathcal{P}}:=\frac{a_c}{\ep}L$ is the circumference of $x$ coordinate. We will use the coordinates $x_M:=\frac{\epsilon}{a_c}x$ and $z_M:=\frac{\epsilon}{a_c}z$ occasionally to write down quantities independent from the solutions.

We consider effective matter state $|C\rangle$ on a timeslice in the flat space, at $t_M=0$. For this purpose, we look for the semiclassical saddle of gravity which gives a dominant contribution to $\langle C|C\rangle$, with fixed initial conditions. We restrict to the simplest topology with $g=0$ and $b=2$, in which case the topological action is $S_{\text{Top}}=0$. Note that higher topology is suppressed by the topological action and breaks translation symmetry. However, it is possible that they dominate over the simplest topology depending on the initial condition. We will study such comparison between topologies in section \ref{comparetop}. One difference from the bra-ket wormhole here is the presence of an explicit initial boundary, which makes it impossible to introduce a bra-ket wormhole without breaking the translation symmetry.

Let us now turn on the effective matter and study its back reaction to the dilaton. We assume translational symmetry in the $x$ direction, and we take the $z$ coordinate of the initial surface $\mathcal{P}$ as $z=-a$. On $\mathcal{P}$, the induced spatial metric and the extrinsic curvatures are

\begin{equation}
\sqrt{h_{x_Mx_M}}=\frac{1}{\text{sin}~a}\left(\frac{a_c}{\epsilon}\right),~~\widetilde{K}=\text{cos}~a.\end{equation}\\
The induced metric $h_{x_Mx_M}$ can take any positive value, while the extrinsic curvature is restricted to $-1<\widetilde{K}<1$. To study the matter CFT, it is convenient to Weyl transform the original metric to the flat space,

\begin{equation}
ds^2_g=\frac{dz^2+dx^2}{\text{sin}^2z}\rightarrow ds^2_{g'}=\frac{dz^2+dx^2}{a_c^2}=\frac{dz_M^2+dx_M^2}{\epsilon^2}.\end{equation}\\
The Weyl factor is

\begin{equation}
ds^2_g=\text{e}^{2\tau}ds^2_{g'} ~~\text{with}~~ e^{\tau}=\frac{a_c}{-\text{sin}z}.\end{equation}\\
This Weyl transformation allows us to consider the problems of matter theory in terms of flat space CFT. Using this flat space CFT, we assume the initial matter state is the regularized boundary state

\begin{equation}
|B(\beta_b)\rangle:=\text{e}^{-\beta_b H_{\text{eff}}/4}|B\rangle.\end{equation}\\
The regularization parameter $\beta_b$ and the Euclidean time evolution $\text{e}^{-\beta_b H_{\text{eff}}/4}$ are defined in terms of this Weyl transformed, flat $(z_M,~x_M)$ coordinate system. The dependence of $\beta_b$ on the Weyl transform can be more generic, however we will not pursue further, as we will eventually take $\beta_b=0$.

Let us evaluate the full action and the backreacted dilaton. In general, the anomaly effective action for Euclidean signature is given by \cite{Herzog:2015ioa},

\begin{eqnarray}\nonumber
W_{\text{Anomaly}}^{E,~g'\rightarrow g}&=&W_{\text{Anomaly}}^E(g)-W_{\text{Anomaly}}^E(g')
\\&=&
-\frac{c}{24\pi}\int_M d^2x\sqrt{g}(R_g\tau-(\partial\tau)_g^2)-\frac{c}{12\pi}\int_{\partial M} dx\sqrt{h_g}K_{h_g}\tau,\label{generalanomaly}\end{eqnarray}\\
 for $g=\text{e}^{2\tau}g'$. Therefore, the anomaly stress tensor $T^{\text{matter}}_{\mu\nu}:=\frac{2}{\sqrt{g_M}}\frac{\delta S^E_{\text{Matter}}}{\delta g^{\mu\nu}}$ is given by

\begin{equation}
T^{\text{Anomaly},g'\rightarrow g}_{\mu\nu}=\frac{c}{12\pi}\Big(\nabla_{\mu}\nabla_{\nu}\tau-g_{\mu\nu}\nabla^2\tau+\nabla_{\mu}\tau\nabla_{\nu}\tau-\frac{1}{2}g_{\mu\nu}(\nabla\tau)^2\Big).
\end{equation}\\
For the Weyl transformation in our case, the anomaly stress tensor is then,

\begin{equation}
T^{\text{Anomaly},g'\rightarrow g}_{\mu\nu}=\widetilde{T}^{\text{Anomaly},g'\rightarrow g}_{\mu\nu}-\frac{c}{24\pi}g_{\mu\nu}.\label{anomalywormhole}
\end{equation}\\
Here we defined

\begin{equation}\widetilde{T}^{\text{Anomaly},g'\rightarrow g}_{z_Mz_M}=-\widetilde{T}^{\text{Anomaly},g'\rightarrow g}_{x_Mx_M}=\frac{c}{24\pi}\Big(\frac{a_{\text{bra},c}}{\epsilon}\Big)^2.
\end{equation}\\
The second term in (\ref{anomalywormhole}) term in the anomaly stress tensor can be canceled by adding the term $\frac{c}{24\pi}\int_M \sqrt{g}$ to the matter Euclidean action. This results in the shift $\frac{\phi_0}{4G_N}\rightarrow\frac{\phi_0}{4G_N}-\frac{c}{12}=:\frac{\phi'_0}{4G_N}$ and an introduction of a new boundary term $S_{\text{New}}=-\frac{c}{24\pi}\int\sqrt{h}K$. In the following, we add $-S_{\text{New}}$ to the gravity action initially to cancel this $S_{\text{New}}$, and will denote $\phi'_0$ as $\phi_0$. 

The flat part of the stress tensor $T^{\text{Matter},g'}_{\mu\nu}$ can be computed by CFT stress tensor on a long Euclidean strip. Its width in $(z_M,~x_M)$ coordinate is

\begin{equation}
\frac{\beta_M}{2}:=a_{\text{bra}}\frac{\epsilon}{a_{\text{bra},~c}}+a_{\text{ket}}\frac{\epsilon}{a_{\text{ket},~c}}+\frac{\beta_b}{2}.
\end{equation}\\
$\beta_M$ is the effective temperature of the matter state produced at the $t_M=0$ in $(z_M,~x_M)$ coordinate. It is convenient to define the bulk inverse temperature when $a_c=a_{\text{bra},~c}=a_{\text{ket},~c}$ as

\begin{equation}
\beta_{\text{Bulk}}:=\frac{a_c}{\epsilon}\beta_M=2a_{\text{bra}}+2a_{\text{ket}}+\beta_b\frac{a_c}{\epsilon}.
\end{equation}\\
The stress tensor is approximately given by, assuming $\beta_M/L\ll 1$, 

\begin{equation}
T^{\text{Matter},g'}_{z_Mz_M}=-T^{\text{Matter},g'}_{x_Mx_M}=-\frac{c}{6\pi}\Big(\frac{\pi}{\beta_M}\Big)^2.
\end{equation}\\
In holographic BCFT the condition $\beta_M/L\ll 1$ can be written in more detail as $\frac{S_L^{\text{Thermal}}(\beta_M)}{4}=\frac{\pi Lc}{12\beta_M}>\frac{\pi c\beta_M}{12L}+2S_B$.

Total matter stress tensor $\widetilde{T}^{\text{Anomaly},g'\rightarrow g}_{\mu\nu}+T^{\text{Matter},g'}_{\mu\nu}$ is traceless and gives a backreaction to the dilaton. Solving the dilaton equation of motion in the ket part, we obtain,
 
\begin{equation}
\phi=\frac{\phi_b}{-\text{tan}z}\frac{a_{\text{ket},~c}}{\epsilon}+\frac{4\pi^2G_Nc}{3}\Big(\frac{1}{\beta_{\text{Bulk}}^2}-\frac{1}{4\pi^2}\Big)\Big(1-\frac{z}{\text{tan}z}\Big).\end{equation}\\
We now seek for the weight for gravity saddle by evaluating the overlap $\langle C|C\rangle$. The Euclidean anomaly effective action for the Weyl transformation $g'\rightarrow g$ is 

\begin{equation}
W_{\text{Anomaly}}^{E,~g'\rightarrow g}+\frac{c}{24\pi}\int_M \sqrt{g}=L\frac{c}{24\pi}\Big(\frac{a_{\text{ket},~c}}{\epsilon}a_{\text{ket}}+\frac{a_{\text{bra},~c}}{\epsilon}a_{\text{bra}}\Big).\end{equation}\\
Here we included the term $\frac{c}{24\pi}\int_M \sqrt{g}$, which was introduced to make the matter stress tensor traceless. 

In order to evaluate and compare matter actions in different background geometries, we will use inner products in radially quantized CFT on a plane and Weyl transform them onto various background geometries. Consider CFT on flat plane $ds^2_{\text{Radial}}=dwd\bar{w}$ in radial coordinate $z$. Conformal transformation to a cylinder with circumference $l$ is $z=\frac{l}{2\pi}\text{log}w$, and the metric is $ds^2_{\text{Radial}}=\frac{2\pi w}{l}\frac{2\pi \bar{w}}{l}dzd\bar{z}$. The Weyl transformation of  $ds^2_{\text{Radial}}$ to $ds^2_{\text{Flat cylinder}}=dzd\bar{z}$ gives rise to the following anomaly effective action

\begin{equation}
W_{\text{Anomaly}}^{E,~\text{Radial}\rightarrow\text{Flat Cylinder}}=\frac{\pi c}{6}\frac{b}{l},
\end{equation}\\
where $b$ is the length of the cylinder in $z$ coordinate. For the present setup, the anomaly effective action is given by

\begin{equation}
W_{\text{Anomaly}}^{E,~\text{Radial}\rightarrow\text{Flat Cylinder}}=\frac{\pi c}{24}\frac{\beta_M}{L}.\end{equation}\\
Suppose that the BCFT is holographic. When $\frac{S_L^{\text{Thermal}}(\beta_M)}{4}=\frac{\pi Lc}{12\beta_M}>\frac{\pi c\beta_M}{12L}+S_B^a+S_B^b$ and the boundary conditions are identical $a=b$, the CFT transition amplitude between boundary states on the Weyl transformed flat cylinder is approximated by the vacuum contribution,

\begin{equation}
\langle B_a(\beta_M)|B_a(\beta_M)\rangle_{\text{Flat Cylinder}}\approx e^{L\frac{\pi c}{12\beta_M}-\frac{\pi c\beta_M}{12L}-2W_{\text{Anomaly}}^{E,~\text{Radial}\rightarrow\text{Flat Cylinder}}-S_B^a-S_B^b}=e^{L\frac{\pi c}{12\beta_M}-\frac{\pi c\beta_M}{6L}-S_B^a-S_B^b}.\label{boundarystateoverlap}
\end{equation}\\
Here the inner product is computed on the flat cylinder, normalized in the radial coordinate. $S^a_B$ and $S_B^b$ are the boundary entropy labeled by $a$ and $b$, and are defined via the overlap $e^{S_B}=\langle 0|B\rangle_{\text{Radial}}$. Since boundary states are unnormalized state, we have a freedom of its normalization. Such normalization corresponds to the probability of the number $N_{\text{EOW}}$ of EOW branes in the gravitational path integral, action that is proportional to $N_{\text{EOW}}$. Here we have chosen the normalization $|B\rangle_{\text{radial}}e^{-S_B}$ for convenience. With this choice, the path integral simply behaves as 

\begin{equation}
\langle C|C\rangle\propto e^{-(N_{\text{EOW}}-2+2g)\frac{\phi_0}{4G_N}}.\label{probability}
\end{equation}\\
When $\frac{S_L^{\text{Thermal}}(\beta_M)}{4}<\frac{\pi c\beta_M}{12L}+S_B^a+S_B^b$ or the boundary conditions are distinct, the amplitude is  

\begin{equation}
\langle B_a(\beta_M)|B_b(\beta_M)\rangle_{\text{Flat Cylinder}}\approx e^{-2W_{\text{Anomaly}}^{E,~\text{Radial}\rightarrow\text{Flat Cylinder}}}=e^{-\frac{\pi c\beta_M}{12L}}.
\end{equation}\\
See \cite{Miyaji:2021ktr} for more detail. 

Collecting the results, the overlap is

\begin{equation}
\text{log}\langle C|C\rangle= -L\frac{c}{24\pi}\Big(a_{\text{bra}}\frac{a_{\text{bra},c}}{\epsilon}+a_{\text{ket}}\frac{a_{\text{ket},c}}{\epsilon}\Big)+
\text{log}\langle B(\beta_M)|B(\beta_M)\rangle_{\text{Flat Cylinder}}-S_{\text{JT}_{\text{bra}}}-S_{\text{JT}_{\text{ket}}}-S_{\mathcal{P}_{\text{bra}}}-S_{\mathcal{P}_{\text{ket}}}.\label{60}\end{equation}\\
Here we have

\begin{equation}
S_{\text{JT}_{\text{ket}}}=L\frac{\phi_b}{16\pi G_N}\Big(\frac{a_{\text{ket},c}}{\epsilon}\Big)^2,~S_{\mathcal{P}_{\text{ket}}}=L\frac{T'}{8\pi G_N}\frac{1}{\text{sin}a}\Big(\frac{a_{\text{ket},c}}{\epsilon}\Big).\end{equation}\\
The tension $T$ determines $a$ uniquely, since $\text{cos}a=-K=-T$. For simplicity, we will assume $\beta_b=0$ in the following. Assuming $S^{\text{Thermal}}(\beta_M)/4>\frac{\pi c\beta_M}{12L}+2S_B$, $a=a_{\text{bra}}=a_{\text{ket}}$ and $\beta_M/L\ll1$, the total action has a maximum if 

\begin{equation}
-\frac{a}{12\pi}+\frac{\pi}{48a}-\frac{T'}{4\pi G_Nc~\text{sin}a}>0.\label{constraintwormhole}\end{equation}\\
The effective temperature of the matter state at $\mathcal{F}$ is

\begin{equation}
\beta_M=\frac{a\phi_b}{\pi G_Nc\Big(-\frac{a}{12\pi}+\frac{\pi}{48a}-\frac{T'}{4\pi G_Nc~\text{sin}a}\Big)},
\end{equation}\\
and the maximum value is

\begin{equation}
\text{log}\langle C|C\rangle= L\frac{2(\text{Arccos}T)^2 \phi_b}{\pi G_N\beta_M^2}-2S_B.\label{overlapwormhole}\end{equation}\\
With the above solution for $a_c/\epsilon$, the dilaton is

\begin{equation}
\phi=\frac{4\pi^2G_Nc}{3}\Big(\frac{1}{16a^2}-\frac{1}{4\pi^2}\Big)\Big(1-\frac{z+a}{\text{tan}z}\Big)-\frac{T'}{-\text{sin}a~\text{tan}z},\label{dilatonwormhole}\end{equation}\\
which indeed satisfies $\partial_{n}\phi-T\phi-T'=0$ on $\mathcal{P}$. We note that when we take the limit $T\rightarrow 0$ and $T'\rightarrow 0$, the geometry is exactly the half of the bra-ket wormhole \cite{Chen:2020tes}.


\subsubsection{EOW Brane from $\mathbb{Z}_2$ Orbifold of Wormhole Geometry}

\begin{figure}[t]
 \begin{center}
 \includegraphics[width=7.5cm,clip]{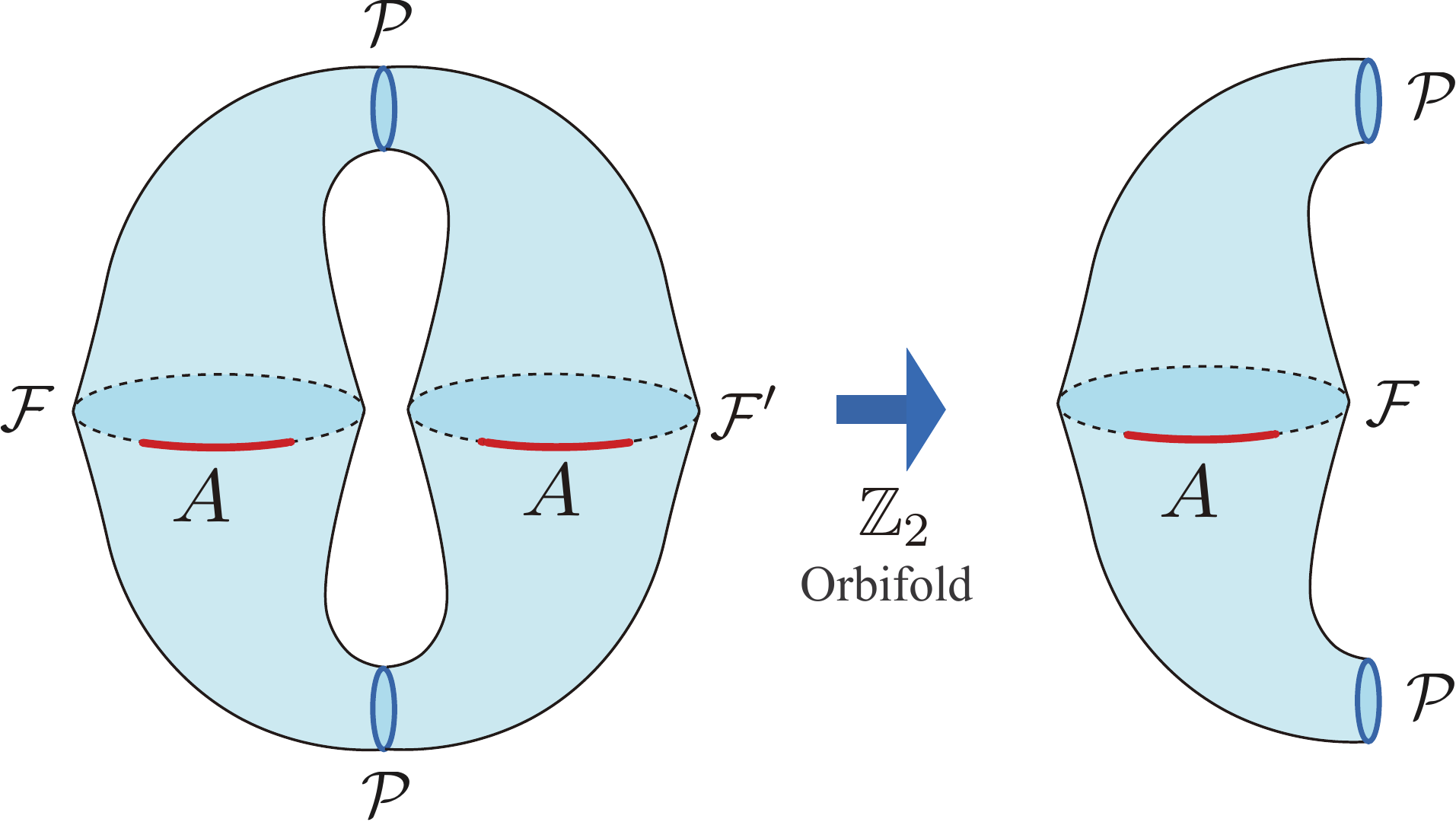}
 \end{center}
 \caption{Double wormholes connecting bra-and-bra and ket-and-ket. Taking $\mathbb{Z}_2$ orbifold of this two wormhole geometry gives the Schwinger-Keldysh geometry with an initial boundary state.}
 \label{fig:orbifold}
\end{figure}

In this section, we will explain how to obtain the geometry with the initial boundary state and EOW brane, from full wormhole geometry. Namely, we will take $\mathbb{Z}_2$ orbifold of two non-gravitating spacetimes connected via two wormholes with interfaces at the middle, see Fig \ref{fig:orbifold}. 

Suppose we have a flat timeslice $\mathcal{F}$ and its duplicate $\mathcal{F}'$, and a gravitationally prepared state on  $\mathcal{F}\mathcal{F}'$ is prepared by a wormhole connecting $\mathcal{F}$ and $\mathcal{F}'$ with an interface at the middle. We assume that the interface is infinitely thin and the extrinsic curvature jumps on this interface giving Israel junction conditions \cite{Israel:1966rt}, and combined geometry has a $\mathbb{Z}_2$ symmetry for the reflection at the interface. We further assume that this interface is a conformal interface for the bulk CFT. 

Let us assume the central charge of the bulk CFT is $c/2$. By folding and taking $\mathbb{Z}_2$ orbifold, we have a conformal boundary state of $\text{CFT}_{c/2}\otimes\text{CFT}_{c/2}/\mathbb{Z}_2$ at the interface \cite{Oshikawa:1996dj}. The geometry  we obtain which has two EOW branes at $\mathcal{P}$, with matter theory $\widetilde{\text{CFT}}_c:=\text{CFT}_{c/2}\otimes\text{CFT}_{c/2}/\mathbb{Z}_2$ and the boundary state $|B\rangle$ of $\widetilde{\text{CFT}}_c$ at $\mathcal{P}$. Note that boundary states of  $\text{CFT}_{c/2}\otimes\text{CFT}_{c/2}$ and $\widetilde{\text{CFT}}_c$ are in general incompatible \cite{Oshikawa:1996dj}. 

The action of this double-wormhole geometry is related to the overlap of the EOW brane geometry as,

\begin{equation}
\text{log}Z_{\text{Double Wormhole}}=2~\text{log}\langle C|C\rangle_{\text{AdS Wormhole}}.\end{equation}


\subsection{AdS/BCFT Model with Corner: Hayward Term\label{hayward1}}

In this section, we consider AdS/BCFT model, which models the Euclidean spacetime with the initial boundary which we have been considering.

\begin{figure}[t]
 \begin{center}
 \includegraphics[width=12.0cm,clip]{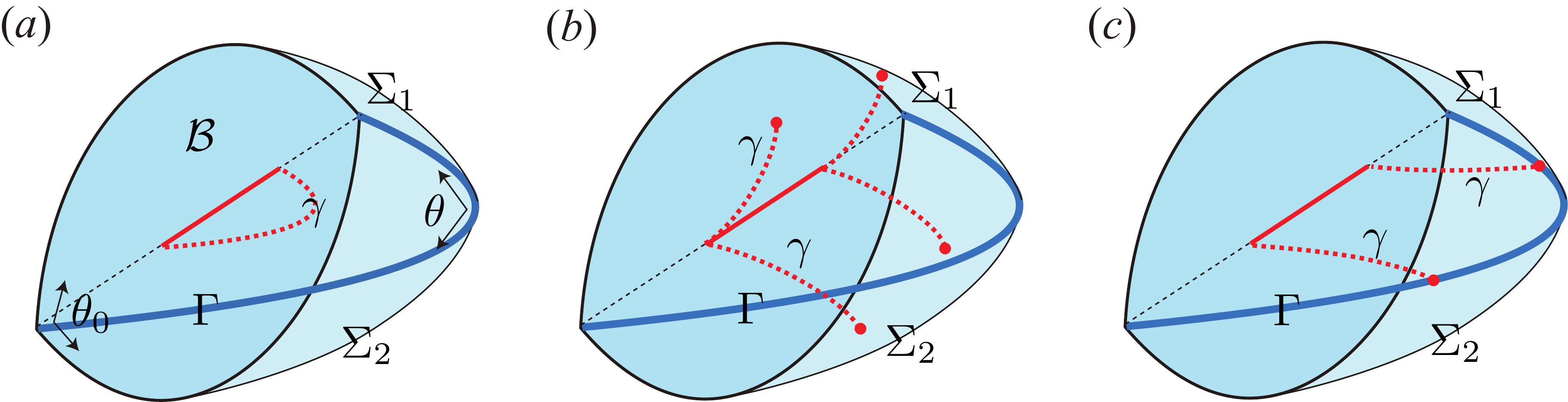}
 \end{center}
 \caption{Three phases of pseudo RT surface $\gamma$ for a boundary subregion. ($a$) $\gamma$ does not intersect with any of $\Sigma_1$, $\Sigma_2$ and $\Gamma$. ($b$) $\gamma$ ends on EOW branes $\Sigma_1$ and $\Sigma_2$. When the set up is time reflection symmetric, there are multiple pseudo RT surfaces which are related via time reflection. ($c$) $\gamma$ ends on the corner $\Gamma$.}
 \label{fig:hayward}
\end{figure}

Let us consider Euclidean three dimensional asymptotically AdS space with two EOW branes $\Sigma_1$ and $\Sigma_2$, which are connected to each other at one dimensional curve $\gamma$, with corner angle $\theta$ which is measured from interior of $M$ \cite{Takayanagi:2019tvn}, see figure \ref{fig:hayward} ($a$). The action of this model is given by

\begin{equation}
S=-\frac{1}{16\pi G_N}\int_M d^3x\sqrt{g_M}(R-2\Lambda)+S_{\Sigma_1}+S_{\Sigma_2}+S_{\Gamma }^{\text{Hayward}},\end{equation}\\ 
where the EOW brane action is

\begin{equation}
S_{\Sigma_i}=-\frac{1}{8\pi G_N}\int_{\Sigma_i}dx \sqrt{h_{\Sigma_i}} (K_{\Sigma_i}-T_i),~(i=1,~2),\end{equation}\\ 
and the Hayward term \cite{Hayward:1993my}

\begin{equation}
S_{\gamma}^{\text{Hayward}}=\frac{1}{8\pi G_N}\int_{\Gamma}dx \sqrt{g_{\Gamma}} (\theta-\theta_0).\end{equation}\\ 
Varying the metric on $M$, $\Sigma_i$ and $\Gamma$ yields the following equations of motion

\begin{equation}
\theta=\theta_0,~K_{\Sigma_i:ab}=(K_{\Sigma_i}-T_i)h_{ab},
\end{equation}\\
as well as usual Einstein equation on AdS. In particular we have $K_{\Sigma_i}=2T_i$. 

The AdS boundary $\mathcal{B}$ is now bounded by $\mathcal{B}\cap\Sigma_1$, $\mathcal{B}\cap\Sigma_2$. $\mathcal{B}\cap\Sigma_1$ and $\mathcal{B}\cap\Sigma_2$ can intersect at $\mathcal{B}\cap\Gamma$, with angle $\theta_0$. Therefore, this gravity model is dual to BCFT, with possible corners at its boundary.

We can consider boundary subregions and their corresponding pseudo entanglement entropy by using the extremal surface. Note that, in general, this setup is not time reflection symmetric, and subregions define transition matrices instead of states. The pseudo RT surface can end on either $\Sigma_1$, $\Sigma_2$ and $\Gamma$.

When this setup has a moment of time reflection symmetry, $\Sigma_1$ and $\Sigma_2$ are related via time reflection, and $\Gamma$ is a time reflection symmetric curve. Importantly, the pseudo-RT surface $\gamma$ for boundary subregion on the time reflection symmetric timeslice may have three phases. We will see how these phases correspond directly to those of gravitationally prepared state in section \ref{pEntanglement}.

In the first phase, the pseudo-RT surface does not end on any of $\Sigma_1$, $\Sigma_2$, and $\Gamma$. This first phase in figure \ref{fig:hayward} ($a$) corresponds to the {\it{thermal phase}} in gravitationally prepared state. In the second phase, figure \ref{fig:hayward} ($b$), the pseudo-RT surface ends on EOW branes, and there are multiple pseudo-RT surfaces with the same length, and they are pairs related by time reversal. This second phase corresponds to the {\it{island phase}}. In the third phase, figure \ref{fig:hayward} ($c$), the pseudo RT surface ends on $\Gamma$, which corresponds to the {\it{boundary phase}}.


\subsection{Comparison with Other Topologies and Factorization\label{comparetop}}

For a given initial boundary condition, there are several topologies and geometries which can contribute to the overlap potentially. We first emphasize that the initial EOW branes are dynamical excitations of the bulk theory, and they are not the boundaries of the bulk where we can insert external sources \footnote{In the previous version of this paper, it was argued that only when $\tilde{K}>0$, spacetime with initial time slice dominates over the bra-ket wormhole and the Hartle-Hawking no-boundary state, by asserting there are no topologically simple geometries with $\tilde{K}>0$ other than the one which is connected to $\mathcal{F}$. This statement on allowed geometries is false, because there are smooth complex geometries whose boundaries have $\tilde{K}>0$ which should be included. Moreover, the EOW branes are bulk excitations rather than spacetime boundaries, therefore the number of EOW branes can vary in the gravitational path integral. In the current version, we reanalyzed the dominant saddles with varying number of EOW branes and fixed future time slice $\mathcal{F}$, finding parameter region where spacetime with EOW brane indeed dominates. We thank Douglas Stanford and Victor Gorbenko for pointing out the possibility of complex geometries connecting $\mathcal{P}$s.}. The only boundary conditions of the gravitational path integral that are fixed are those of $\mathcal{F}$, and we allow arbitrary number of bulk EOW branes $\mathcal{P}$ as bulk excitations. Note that geometries completely disconnected from $\mathcal{F}$ contribute only to the overall constant in the path integral, therefore we will not take these disconnected contributions into account when comparing between various saddles, see Fig \ref{fig:comparison} ($e$). 

At the leading order in topological expansion, there are several types of geometries as discussed below, see Fig \ref{fig:comparison}. We will study when the spacetime with initial EOW brane is the dominant geometry, compared to other geometries including no-boundary state, bra-ket wormhole and off-diagonal geometry.


\subsubsection*{No-Boundary State and Bra-ket Wormhole}

\begin{figure}[t]
 \begin{center}
 \includegraphics[width=13.0cm,clip]{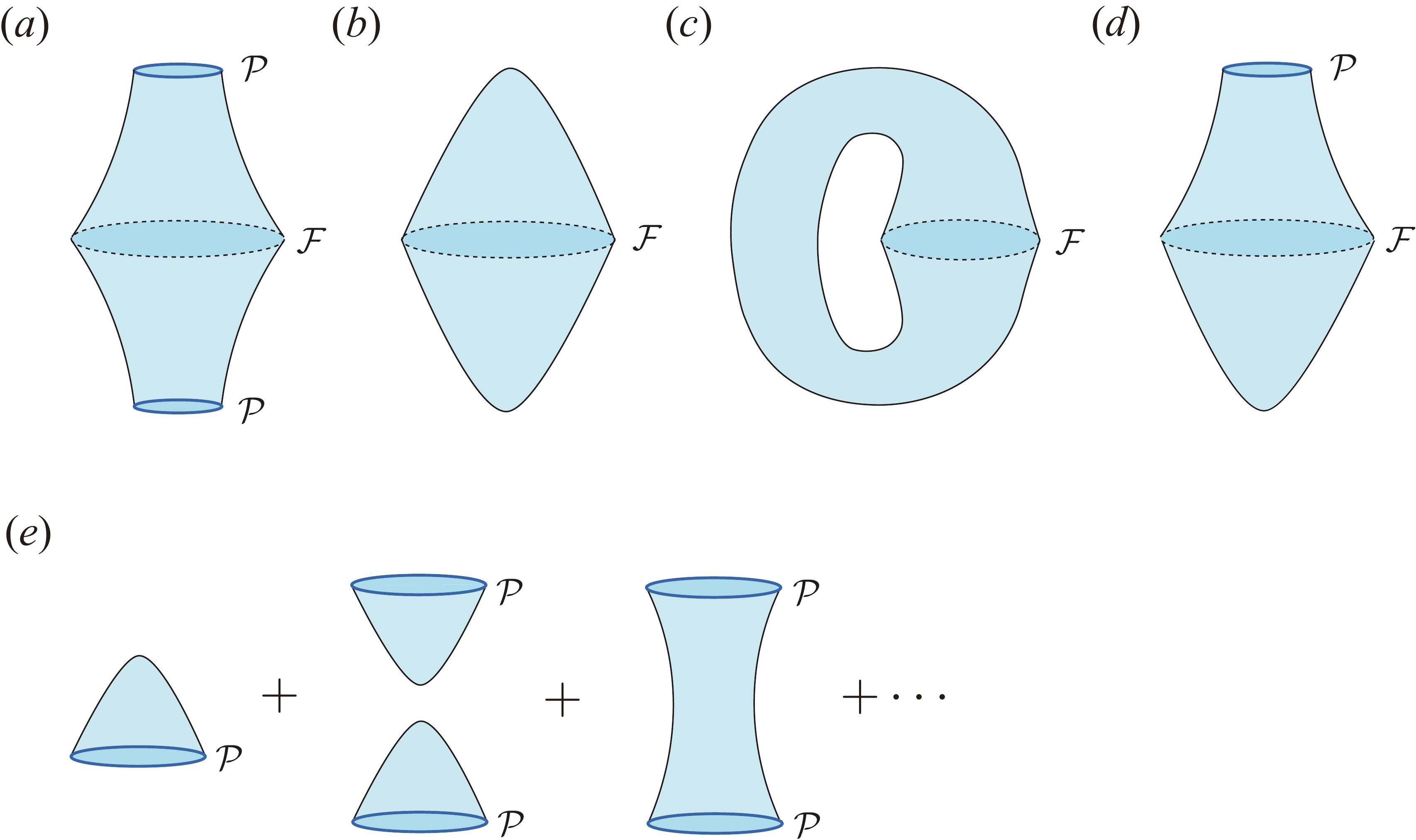}
 \end{center}
 \caption{Geometries that appear in the computation of the overlap $\langle C|C\rangle$. ($a$): Spacetime initiated by the EOW brane. ($b$): Hartle-Hawking no-boundary state. ($c$): Bra-ket wormhole. ($d$): Off-diagonal overlap of spacetime initiated by the EOW brane and no-boundary state. ($d$): Geometries completely disconnected from $\mathcal{F}$ which only contribute to the overall constant of $\langle C|C\rangle$.}
 \label{fig:comparison}
\end{figure}

In the Hartle-Hawking no-boundary state and in the bra-ket wormhole, the EOW branes are absent, and the overlaps were obtained in \cite{Chen:2020tes} which we will explain briefly. In the Hartle-Hawking no-boundary state, the geometry is given by the global AdS, see Fig \ref{fig:comparison} ($b$). The anomaly contribution to the matter partition function comes from the Weyl transformation from a flat disk $ds_{g'}^2=dr^2+r^2d\theta^2$ to the global AdS $ds_g^2=4(dr^2+r^2d\theta^2)/(1-r^2)^2$, whose anomaly effective action is $W_{\text{Anomaly}}^{E,~g'\rightarrow g}+\frac{c}{24\pi}\int_M \sqrt{g}=0$. As the result, we have

\begin{equation}
\text{log}\langle C|C\rangle_{\text{No-Boundary}}=2\frac{\phi_0}{4G_N}+\frac{\pi\phi_b}{2G_NL}.\label{hhoverlap}\end{equation}\\
In the bra-ket wormhole, the geometry is given by the AdS wormhole, see Fig \ref{fig:comparison} ($c$) . Assuming $\beta_M/L\ll1$, similar computation as section \ref{wormhole} yields

\begin{equation}
\text{log}\langle C|C\rangle_{\text{Bra-Ket}}=L\frac{\pi c^2G_N}{32\phi_b}.\label{braketoverlap}\end{equation}\\
When $L\gtrsim\frac{16}{\pi}\frac{\phi_b\phi_0}{(G_Nc)^2}$, the bra-ket wormhole dominates over the no-boundary state.

\subsubsection*{Off-diagonal Overlap of EOW Brane and No-Boundary State}

The another possible geometry has a EOW brane in the bra part and the Hartle-Hawking no-boundary state in the ket part, see Fig \ref{fig:comparison} ($d$). 

Let us first consider the case when $1>\tilde{K}=-T>-1$. The overlap is given by

\begin{equation}
\text{log}\langle C|C\rangle_{\text{Off-diagonal}}=\Big(\frac{\phi_0}{4G_N}+\frac{\pi\phi_b}{4G_NL}\Big)+\Big(-L\frac{c}{24\pi}a\frac{a_{\text{bra},c}}{\epsilon}-\frac{\pi c}{6L}a\frac{\epsilon}{a_{\text{bra},c}}-S_{\text{JT}_{\text{bra}}}-S_{\mathcal{P}_{\text{bra}}}\Big).\end{equation}\\
Here we have

\begin{equation}
S_{\text{JT}_{\text{bra}}}=L\frac{\phi_b}{16\pi G_N}\Big(\frac{a_{\text{bra},c}}{\epsilon}\Big)^2,~S_{\mathcal{P}_{\text{bra}}}=L\frac{T'}{8\pi G_N}\frac{1}{\text{sin}a}\Big(\frac{a_{\text{bra},c}}{\epsilon}\Big).\end{equation}\\
The tension $T$ is $\text{cos}a=-K=-T$. Assuming $a\epsilon/ (a_{\text{bra},c}L)\ll1$, the condition for that the overlap has a maximum is

\begin{equation}
-\frac{a}{24\pi}-\frac{T'}{8\pi G_Nc~\text{sin}a}>0.\end{equation}\\
If this were not the case, the overlap is maximized at $a_c/\epsilon=0$. Defining the off-diagonal effective temperature $\beta^{\text{off-diagonal}}_M:=4a\epsilon/a_{\text{bra},c}$, the solution is given by

\begin{equation}
\beta^{\text{off-diagonal}}_M=\frac{a\phi_b}{\pi G_Nc\Big(-\frac{a}{12\pi}-\frac{T'}{4\pi G_Nc~\text{sin}a}\Big)}.
\end{equation}\\
The maximum overlap is

\begin{equation}
\text{log}\langle C|C\rangle_{\text{Off-diagonal}}=\Big(\frac{\phi_0}{4G_N}+\frac{\pi\phi_b}{4G_NL}\Big)+L\frac{(\text{Arccos}T)^2\phi_b}{\pi G_N(\beta_M^{\text{off-diagonal}})^2}.\label{wormholeoffdiagonal}\end{equation}\\
This overlap is always suppressed compared to the diagonal overlap of EOW branes (\ref{totalactionglobal}) or the Hartle-Hawking no-boundary state (\ref{hhoverlap}) as expected, from the condition $\frac{S_L^{\text{Thermal}}(\beta_M)}{4}=\frac{\pi Lc}{12\beta_M}>\frac{\pi c\beta_M}{12L}+2S_B$.

\subsubsection*{Spacetime Initiated by EOW Brane}

\begin{figure}[t]
 \begin{center}
 \includegraphics[width=6.0cm,clip]{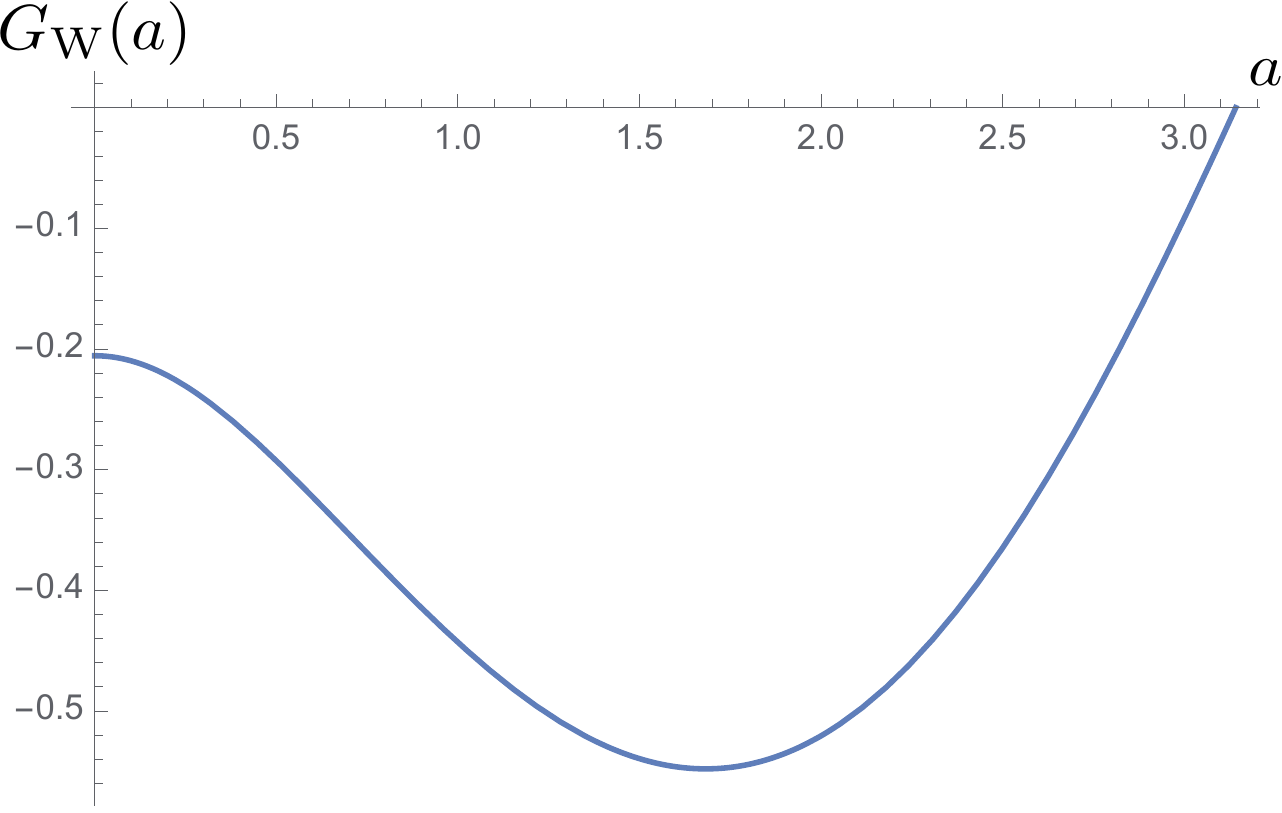}
 \end{center}
 \caption{Plot of the function $G_{W}(a)$. $\frac{T'}{4G_Nc}<G_{W}(a)$ is sufficient (but not necessary) to ensure the spacetime initiated by EOW brane dominates over the bra-ket wormhole.}
 \label{fig:plot}
\end{figure}

We revisit spacetimes initiated by EOW brane, see Fig \ref{fig:comparison} ($a$). We assume the matter BCFT is in thermal phase, so that the condition $\frac{S_L^{\text{Thermal}}(\beta_M)}{4}=\frac{\pi Lc}{12\beta_M}>\frac{\pi c\beta_M}{12L}+2S_B$ is satisfied. We further assume that $S_B>\phi_0/(4G_N)$, so that the initial state entanglement violates the entropy bound (\ref{areabound2}). When $1>\tilde{K}=-T>-1$, the background geometry is given by the AdS wormhole. A sufficient condition for the overlap (\ref{overlapwormhole}) is dominant compared to the bra-ket wormhole, can be derived using $\frac{S_L^{\text{Thermal}}(\beta_M)}{4}=\frac{\pi Lc}{12\beta_M}>\frac{\pi c\beta_M}{12L}+2S_B$. This sufficient condition is given by, combined with (\ref{constraintwormhole}), as

\begin{equation}
\frac{T'}{4G_Nc}<G_{W}(a),\label{sufcondition}\end{equation}\\
where $G_{W}(a)=\pi\text{sin}a(-\frac{1}{8}\sqrt{1+\frac{\pi^2}{36a^2}}-\frac{a}{12\pi})$. This sufficient condition requires $T'<0$ in particular, see Fig (\ref{fig:plot}). Finally, we note that the result of this section relies on the probability distribution for the EOW brane which was assumed in (\ref{boundarystateoverlap}) and (\ref{probability}). 

\subsubsection*{Factorization}

Since the spacetime initiate by EOW brane does not have a bra-ket wormhole, the corresponding gravitationally prepared state does not have the factorization paradox at the leading order. The factorization puzzle in the bra-ket wormhole becomes manifest when we project the gravitationally prepared state $|C\rangle$ onto an energy eigenstate $|E\rangle$ with the same energy density \cite{Chen:2020tes}. Insertion of such projection does not change the stress tensor therefore the geometry, while the gravitational computation of the overlap $\langle C|E\rangle\langle E|C\rangle$ and the usual product of $\langle C|E\rangle$ and $\langle E|C\rangle$ do not straightforwardly match. Moreover, when we consider dilaton profile for the overlap $\langle E|C\rangle$, it is given by (\ref{dilatonglobal}) therefore $\phi\rightarrow +\infty$ at IR, indicating emergent AdS boundary. 

On the other hand, we do not have this factorization puzzle in the spacetime with EOW brane, since the spacetime is explicitly factorized. This, however, does not mean the bulk gravity is not an ensemble average, as there are off-shell configurations which again causes the factorization puzzle. Our interpretation is the EOW brane plays the role of leading bulk excitation which restores the factorization. It is tempting to interpret along the way of half-wormholes\cite{Saad:2021rcu}, eigenbranes \cite{Blommaert:2019wfy} and non-local interaction in spacetime branes\cite{Blommaert:2021fob}.
\section{Entanglement Entropy and Pseudo Entanglement Wedge\label{pEntanglement}}

\subsection{Pseudo Entanglement Island}

We now consider the entanglement entropy of subregions of length $\Delta l$ on $\mathcal{F}$ using replica trick in Schwinger-Keldysh formalism \cite{Dong:2016hjy, Colin-Ellerin:2020mva, Goto:2020wnk, Colin-Ellerin:2021jev}. The gravity is turned off at $\mathcal{F}$ and afterward, and the gravity quantum state is projected onto a fixed length, fixed dilaton pure state. For simplicity, we assume the total system size $L$ is large and only consider the case $t_M=0$. 

Since the initial matter state is locally thermal, the entanglement entropy on the future slice $\mathcal{F}$ is also locally thermal, with inverse temperature $\beta_M$, assuming $\frac{S_L^{\text{Thermal}}(\beta_M)}{4}=\frac{\pi cL}{12\beta_M}>\frac{\pi c\beta_M}{12L}+2S_B$. Therefore, when $\Delta l$ is sufficiently small, the entanglement entropy of $A$ can be approximated by the finite temperature effective matter theory with the inverse temperature $\beta_M$,

\begin{equation}
S^{\text{Thermal}}_A=\frac{c}{3}\text{log}\Big[\frac{\beta_M}{\pi \ep}\text{sinh}\Big(\frac{\pi \Delta l}{\beta_M}\Big)\Big].\label{24s1}
\end{equation}\\
For subsystems with $\Delta l\ll \beta_M$, (\ref{24s1}) gives the usual vacuum entanglement entropy. For subregions with $\Delta l\gg \beta_M$, this expression gives

\begin{equation}
S^{\text{Thermal}}_A\approx \frac{c}{3}\frac{\pi \Delta l}{\beta_M}+\frac{c}{3}\text{log}\Big(\frac{\beta_M}{2\pi\epsilon}\Big),\label{thermalphase}
\end{equation}\\
which is indeed the thermal entropy at inverse temperature $\beta_M$. Since (\ref{thermalphase}) obeys the volume law, we would naively expect that the entanglement entropy increases unboundedly as the subsystem becomes larger until the system size reaches $L/2$. If the system satisfies 

\begin{equation}
2\frac{\phi_0}{4G_N}\ll\frac{1}{2}S^{\text{Thermal}}_L=\frac{c}{6}\frac{\pi L}{\beta_M},
\end{equation}\\
such an unbounded entropy would violate the entropy bound (\ref{areabound}). This problem is similar to the black hole information paradox, assuming turning off the gravity at $\mathcal{F}$ has small effects on the entanglement structure of the matter theory. However, for large enough subsystems, namely when $\beta_M\ll \Delta l$, we will show that there are two other phases that can dominate, saving the entropy bound (\ref{areabound}), which we call {\it{initial boundary phase}} and {\it{island phase}}.

In the {\it{initial boundary phase}}, the boundary contribution dominates the entanglement entropy. When the distance between twist operators is comparable to the distance from the boundary, (\ref{thermalphase}) is no longer a good approximation, and we need to take the effects of the initial boundary into account. In terms of Cardy's doubling trick in BCFT, the original twist operator 2 point function is now a 4 point function of two chiral twist operators and their mirror images. For sufficiently large subregions, the 4 point function in holographic BCFT is dominated by the Wick contractions between the chiral twist operator and its mirror image. Then the entanglement entropy is given by \cite{Calabrese:2004eu, Miyaji:2014mca},

\begin{equation}
S_A^{\text{Boundary}}\approx 2S_B+\frac{c}{3}\text{log}\Big(\frac{\beta_M}{\pi \ep}\Big),\label{boundaryphase}
\end{equation}\\
see also appendix \ref{Abcftpseudo}. $S_B$ is the boundary entropy \cite{Affleck:1991tk}, given by the overlap between the vacuum and the boundary state $S_B=\text{log}~\langle 0|B\rangle_{\text{matter}}$. The entanglement entropy $S_A^{\text{Boundary}}$ does not depend on $\Delta l$. From the assumption $\frac{S_L^{\text{Thermal}}(\beta_M)}{4}=\frac{\pi cL}{12\beta_M}>\frac{\pi c\beta_M}{12L}+2S_B$, the effective matter entanglement entropy of any subregion $A$ with $3L/4>\Delta l>L/4$ is given by (\ref{boundaryphase}).

We now consider another phase from non-perturbative quantum gravity, the {\it{island phase}}, which arises from the replica wormholes. The island phase competes with the thermal phase as well as the initial boundary phase and emerges when the boundary phase and the thermal phase would still violate the entropy bound (\ref{areabound}),

\begin{equation}
2\frac{\phi_0}{4G_N}\ll \text{Min}\Big[2S_B,~S^{\text{Thermal}}_A\Big].\end{equation}\\
We will see that such violation would not occur due to the emergence of an entanglement island located near the past initial surface at a positive Euclidean time. As in the usual derivation of the island formula, the replica wormholes induce emergent twist operators in the bulk spacetime. These emergent twist operators define a subregion $I$ in the bulk spacetime, where $\partial I$ corresponds to the location of these twist operators. These emergent twist operators modify the entanglement entropy $S_A$ largely, and $S_A$ is given by the {\it{generalized pseudo entropy}} of $A\cup I$,

\begin{equation}
S^{\text{Island}}_A=\underset{I}{\text{Min}}~\text{Re}\Big[\underset{I}{\text{Ext}}\Big[S^{\text{eff},~P}_{A\cup I}+\frac{\text{Area}[\partial I]}{4G_N}\Big]\Big]\label{islandformula}.
\end{equation}\\
Note that this expression reduces to the original matter entropy when $I$ is empty. The subregion $I$ realizing the minimal, extremal generalized entropy is called entanglement island. In JT gravity, the area term in (\ref{islandformula}) is replaced by the sum of the values of dilaton $\phi_0+\phi$ at $\partial I$. 

\begin{figure}[t]
 \begin{center}
 \includegraphics[width=15cm,clip]{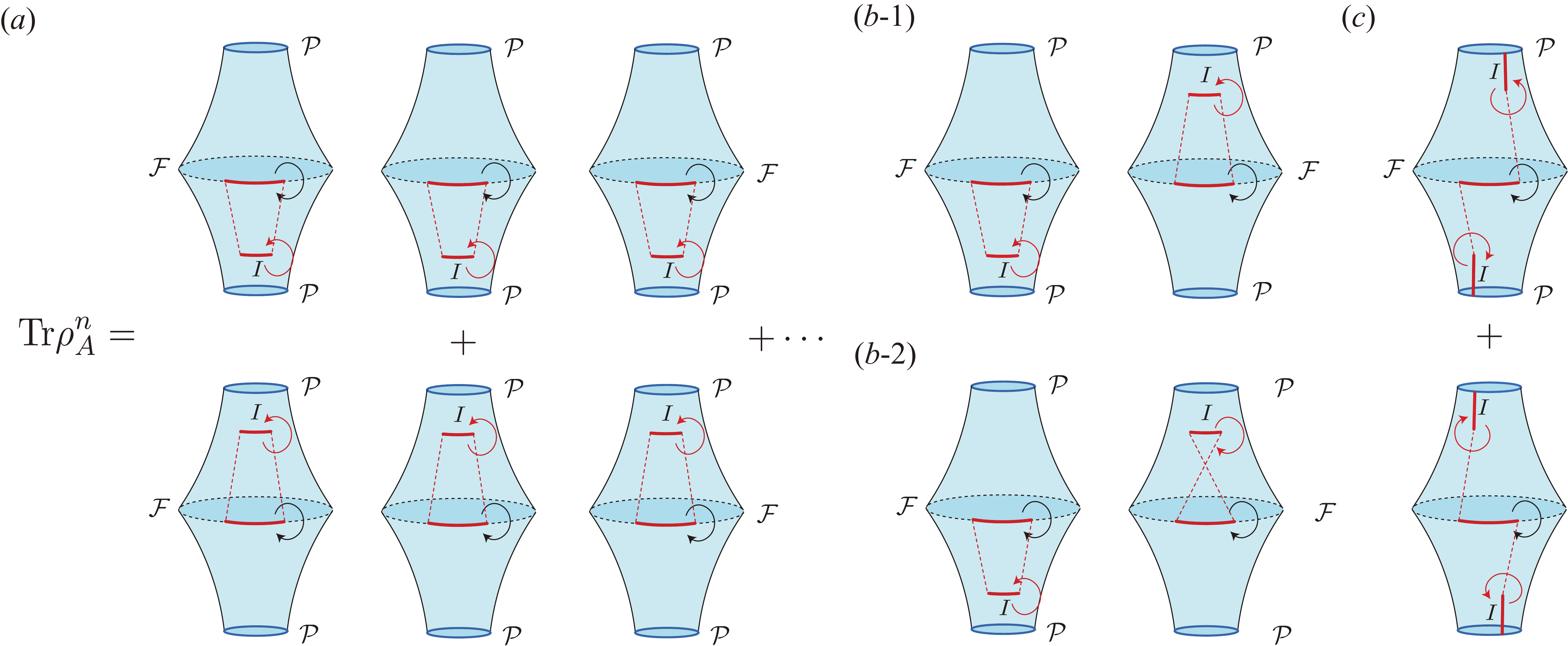}
 \end{center}
 \caption{($a$) The leading contribution to $\text{Tr}\rho_A^n$ from the replica wormholes. This contribution consists of two geometries, one has island in the ket part and the other in the bra part. ($b$) The replica wormholes that connect the bra and the ket part of the geometry. ($c$) The initial time slice is now replicated, and the emergent twist operators appear at both the bra and the ket part of the geometry. The entropy from this configuration is $S_A= 2S_B+S_A^{\text{Island}}$, therefore this saddle is always subdominant, as long as $S_B>0$.}
 \label{fig:replicas}
\end{figure}

In order to find the entanglement island, we must look for the replica wormhole geometry, which gives the dominant contribution to the replica partition function $\text{Tr}\rho_A^n$ for a sufficiently large subregion $A$. When the dominant background geometry is given by the one with all $\mathcal{F}$ and $\mathcal{P}$ are connected, the replica wormholes that are replica symmetric and can be analytically continued to noninteger $n$, are the one that connects bra part of the bulk, and the one that connects ket part of the bulk, see Fig.\ref{fig:replicas} (a).
These two geometries are related by complex conjugation; therefore the resulting entanglement entropy is guaranteed to be real. We can dress these two geometries with handles and wormholes, which give subleading contributions to the entanglement entropy.

The other topologies that can contribute to $\text{Tr}\rho_A^n$ are the replica wormholes that connect the bra and the ket part of the bulk geometries belonging to different replicas, see Fig.\ref{fig:replicas} ($b$). We can analytically continue from even $n$, without leaving conical defects at $\partial I$. It is interesting to understand whether and how these geometry contribute to the entanglement entropy. 

The another topology is the one with the emergent twist operators appearing both at the bra and the ket part of the geometry, see Fig.\ref{fig:replicas} ($c$). This topology enters when the initial surfaces in different replicas are connected via the replica wormhole. The entanglement entropy from this geometry is given by $S_A= 2S_B+S_A^{\text{Island}}$. This implies this saddle is always subdominant compared to the island phase and the boundary phase, as long as $S_B>0$. In the bra-ket wormhole case, we have $S_B=0$ , therefore the saddles in Fig.\ref{fig:replicas} ($a$) and ($c$) are degenerate.

Let us investigate the entanglement entropy with such replica wormholes or entanglement islands. We consider the subsystem of $|C\rangle$ on $\mathcal{F}$ with boundaries at $x_M=\pm\frac{\Delta l}{2}$. We assume the candidate entanglement island is at $z=-a_{I}$, and with boundaries at $x_M=\pm\frac{\Delta x_M}{2}$. The entanglement entropy is given by the average of the ket and the bra contributions. The ket part of the candidate island entropy is

\begin{eqnarray}\nonumber S_A^{\text{Island, ket},P}(a_I)&=&2\frac{\phi_0+\phi(a_I)}{4G_N}+\frac{c}{3}\text{log}\Big[\Big(\frac{\beta_M}{\pi \ep}\Big)^2\Big(\frac{a_c}{F(a_I)}\Big)
\text{sin}\Big(\pi\frac{ a_I\frac{\ep}{a_c}+i(\Delta l-\Delta x)/2}{\beta_M}\Big)\nonumber\\&&\times\text{sin}\Big(\pi\frac{a_I\frac{\ep}{a_c}+i(-\Delta l+\Delta x)/2}{\beta_M}\Big)\Big],\label{generalentropy}\end{eqnarray}\\
Here the function $F(a)$ is either $\text{sin}a$ for AdS wormhole, $\text{sinh}a$ for global AdS. This ket entropy is locally minimized against the variation of $\Delta x$ at $\Delta l=\Delta x$, so we set $\Delta l=\Delta x$. 

In the following, we will explicitly show $S_A^{\text{Island, ket},P}(a_I)$ has extremal, where both $a_I$ and $S_A^{\text{Island, ket},P}(a_I)$ can be complex valued. Since (\ref{generalentropy}) is a real function of $a_I$, $S_A^{\text{Island, ket},P}(a_I)$ is also extremal at $a_I^{*}$ with $S_A^{\text{Island, ket},P}(a_I)^{*}=S_A^{\text{Island, ket},P}(a_I^{*})$. Therefore, combining with the fact that the bra part is the complex conjugate of the ket, the island entanglement entropy is given by

\begin{equation}
S_A^{\text{Island}}=\underset{a_I}{\text{Min}}~\text{Re}\Big[\underset{a_I}{\text{Ext}}~S_A^{\text{Island, ket},P}(a_I)\Big]\Big].
\end{equation}\\
The results can be summarized as

\begin{equation}
S_A=\text{Min}\Big[S^{\text{Thermal}}_A(\beta_M),~S_A^{\text{Island}},~S_A^{\text{Initial Boundary}}\Big],\label{finalresult}
\end{equation}\\
assuming we do not have other saddles, such as Fig.\ref{fig:replicas} ($b$). Note that neither the boundary entropy nor the island phase depend on $\Delta l$ at leading order. Therefore whether the island phase or the boundary phase dominates, is determined by the comparison between the boundary entropy $S_B$ and the area term $\phi_0/(4G_N)$. In the following, we will explicitly work on specific initial boundary conditions and study the island phase entanglement entropy. 


As an explicit example, we consider the case $\tilde{K}=0$. The background geometry is that of an AdS wormhole, and the dilaton is given by (\ref{dilatonwormhole}). The sufficient condition for EOW dominance (\ref{sufcondition}) is $-\frac{\pi(1+\sqrt{10})}{24}>\frac{T'}{4G_Nc}$. In this case, $S_A^{\text{Island}}(a_I)$ is extremal at

\begin{equation}
a_I^{\pm}=\frac{\pi}{2}\pm i\text{Arccosh}\Big(\frac{-3T'}{2G_Nc}\Big),
\end{equation}\\
and the island pseudo entropy is 

\begin{eqnarray}\nonumber
S_A^{\text{Island, ket},P}(a_I^{\pm})&=&
2\frac{\phi_0}{4G_N}+\frac{c}{3}\text{log}
\Big[\frac{2\beta_M}{\pi \ep}\frac{\text{sin}^2\Big(\frac{\pi}{4}\pm \frac{i}{2}\text{Arccosh}\Big(\frac{-3T'}{2G_Nc}\Big)\Big)}{\Big(\frac{-3T'}{2G_Nc}\Big)}
\Big]
\nonumber\\&\mp&\frac{ci}{3}\sqrt{\Big(\frac{-3T'}{2G_Nc}\Big)^2-1} .\end{eqnarray}\\
Since the actions for the background geometry are the same, and the real part of $S_A^{\text{Island}}(a_I^{\pm})$ are identical, we conclude that these four pseudo islands from bra/ket contribute equally to the entanglement entropy. Therefore, we conclude that the island phase entropy is given by

\begin{eqnarray}\nonumber
S_A^{\text{Island}}&=&\frac{1}{2}\Big[S_A^{\text{Island, ket},P}(a_I^{+})+S_A^{\text{Island, ket},P}(a_I^{-})\Big]
\nonumber\\&=&
2\frac{\phi_0}{4G_N}+\frac{c}{3}\text{log}
\Big[\frac{\beta_M}{\pi \ep}
\Big],\end{eqnarray}\\
which has the identical structure as the boundary phase entropy (\ref{boundaryphase}) when we replace $S_B$ by $\phi_0/(4G_N)$.

We can also analyze the saddle for pseudo entropy using linear perturbation around $a=\pi/2$. For given perturbation $a=\pi/2+\delta a$, the saddle point $a_I$ of the island is given by

\begin{equation}
a_I=\Big(\frac{\pi}{2}-\delta a\Big)\pm i\frac{\frac{-3T'}{G_Nc}-\pi}{\pi\sqrt{(\frac{-3T'}{2G_Nc})^2-1}}\delta a\pm i\Big(1-\frac{2}{\pi}\delta a\Big)\text{Arccosh}\Big(\frac{-3T'}{2G_Nc}\Big).\end{equation}\\
Note that we have $\delta\text{Re}[a_I]=-\delta a$. as the result we have $\text{Re}[a_I]<a$ only when $\delta a>0$, in other word $\tilde{K}<0$. In section \ref{ProjectedEW}, we will give an entanglement wedge interpretation for the case when $\text{Re}[a_I]<a$, leaving the interpretation for the case $\text{Re}[a_I]>a$ as an open question.

\subsection{Constraint on the Initial State \label{thermality}}
When the entanglement entropy is given by the entanglement island, it is not obvious that such entropy is consistent with the laws of quantum mechanics. Indeed, the bra-ket wormhole was introduced as a remedy for the violation of the strong sub-additivity \cite{Chen:2020tes}. In this section, we will give a sufficient condition of the initial state for a class of tests of the strong sub-additivity of the entropy, which in our case given by an average of generalized pseudo entropy from islands. We emphasize that there is no well-defined notion of strong sub-additivity for pseudo entropy, and what we will employ is that of usual entanglement entropy. Note that for a pure transition matrix $X=|\psi\rangle\langle\phi|/\langle\phi|\psi\rangle$, the usual purity $S^P_A(X)=S^P_{\tilde{A}}(X)$ holds.

Let us consider the situation where the strong sub-additivity could be violated, as pointed out in \cite{Chen:2020tes}. Consider a CFT which is a tensor product of two CFTs, namely $\text{CFT}_{c+c_p}=\text{CFT}_{c}\otimes\text{CFT}_{c_p}$. $c$ and $c_p$ are the central charges. We assume $c\gg c_p$ and the entanglement entropy is given by the island phase for CFT${}_c$ and CFT${}_{c+c_p}$ for large enough regions, while we also assume the island phase in CFT${}_{c_p}$ is absent, so that CFT${}_{c_p}$ is either in thermal phase or in the boundary phase. We consider a large enough subregion $A$ with length $\mathcal{O}(L)$. We define the complement interval as $\tilde{A}$. We ask the condition required by the strong sub-additivity of the following form

\begin{equation}
S(A_{c_p}+A_c)+S(A_{c_p}+\tilde{A}_{c_p})\geq S(A_{c}+A_{c_p}+\tilde{A}_{c_p})+S(A_{c_p}).\label{23s2}\end{equation}\\
We will write the pseudo island for $A_cA_{c_p}$ as $I$, and for $A_c$ as $I'$. Using the fact that the emergent twist operators in the bulk couple to both CFTs, we have

\begin{equation}
S(A_c+A_{c_p}+\widetilde{A}_{c_p})=S(\widetilde{A}_c)=\text{Re}\Big[\frac{\text{Area}[\partial I'']}{4G_N}+S^{\text{eff},P}_{\widetilde{A}_c\widetilde{I}''_c\widetilde{I}''_{c_p}}\Big],\end{equation}\\
Here $I''$ is the pseudo island for $\widetilde{A}_c$. Assuming $c_p$ is sufficiently small, we can approximate $I'$ and $I''$ by $I$, obtaining

\begin{equation}
S(A_c+A_{c_p}+\widetilde{A}_{c_p})\approx \text{Re}\Big[\frac{\text{Area}[\partial I]}{4G_N}+S^{\text{eff},P}_{\widetilde{A}_c\widetilde{I}_c\widetilde{I}_{c_p}}\Big],~S(A_c)\approx \text{Re}\Big[\frac{\text{Area}[\partial I]}{4G_N}+S^{\text{eff},P}_{A_cI_cI_{c_p}}\Big].\end{equation}\\ The errors in these approximate equalities are sub-leading due to the extremization in the island formula. Note that the size of the island $I$ is approximately the same as $A$ in the gravitationally prepared state, which is not necessarily the case for the islands in evaporating black hole. Then the condition for the strong sub-additivity is

\begin{equation}
S^{\text{eff}}_{A_{c_p}\tilde{A}_{c_p}}\gtrsim S^{\text{eff}}_{A_{c_p}}+\text{Re}\Big[S^{\text{eff},P}_{\widetilde{A}_c\widetilde{I}_c\widetilde{I}_{c_p}}-S^{\text{eff,P}}_{A_cA_{c_p}I_cI_{c_p}}\Big]\label{condition},\end{equation}\\
which is automatically satisfied if these entropy are ordinary entropy of a state, from the strong sub-additivity of entanglement entropy. This is the main reason why the application of this argument to evaporating black hole coupled to bath does not yield any new conditions. However, since we are considering pseudo entropy and therefore there is no analog of the strong sub-additivity, this inequality yield a new condition on the initial state. Let us study the implication of (\ref{condition}) explicitly. From the Weyl transformation, using (\ref{effpseudo}), we have

\begin{equation}
S^{\text{eff}}_{A_{c_p}}=S^{\text{eff},P}_{I_{c_p}}+\frac{c_p}{3}\text{log}\frac{F(a_I)}{a_c},~S^{\text{eff}}_{\widetilde{A}_{c_p}}=S^{\text{eff},P}_{\widetilde{I}_{c_p}}+\frac{c_p}{3}\text{log}\frac{F(a_I)}{a_c},\end{equation}\\
where $F(a)=\text{sin}a$ for AdS wormhole and $F(a)=\text{sinh}a$ for global AdS, and $L_A$ is the length of $A$. Since the twist operators of $A_{c_p}$ and $I_{c_p}$ in $S^{\text{eff,P}}_{A_cA_{c_p}I_cI_{c_p}}$ contract with each other, we have

\begin{equation}
S^{\text{eff}}_{A_{c_p}\tilde{A}_{c_p}}\gtrsim S^{\text{eff}}_{A_{c_p}}+ S^{\text{eff}}_{\widetilde{A}_{c_p}}-\text{Re}\Big[\frac{c_p}{3}\text{log}\Big[\Big(\frac{\beta_M}{\pi\epsilon}\Big)^2\text{sin}^2\Big(\frac{\pi a_I}{4a}\Big)\Big]\Big]\label{condition2}.\end{equation}\\
This condition leads, ignoring the last $L_A$ independent term and using the sub-additivity $S^{\text{eff}}_{A_{c_p}\tilde{A}_{c_p}}\leq S^{\text{eff}}_{A_{c_p}}+ S^{\text{eff}}_{\widetilde{A}_{c_p}}$ of usual entropy,

\begin{equation}
S^{\text{eff}}_{A_{c_p}\tilde{A}_{c_p}}\approx S^{\text{eff}}_{A_{c_p}}+ S^{\text{eff}}_{\widetilde{A}_{c_p}}.\label{saturation}\end{equation}\\
We conclude that the state $\text{Tr}_{\text{CFT}_c}\Big[|B(\beta_M)\rangle\langle B(\beta_M)|_{\text{CFT}_c\otimes\text{CFT}_{c_p}}\Big]$ must be a thermal mixed state with inverse temperature $\beta_M$, in order to be consistent with the strong sub-additivity of the usual entanglement entropy. Note that $\text{Tr}_{\text{CFT}_c}\Big[|B(\beta_M)\rangle\langle B(\beta_M)|_{\text{CFT}_c\otimes\text{CFT}_{c_p}}\Big]$ cannot be a maximally mixed state of regularized boundary states. Indeed, though the boundary states in $\text{CFT}_c\otimes\text{CFT}_{c_p}$ are zero-eigenvectors of $L^{c+c_p}_n-\widetilde{L}^{c+c_p}_{-n}=L^{c}_n-\widetilde{L}^{c}_{-n}+L^{c_p}_n-\widetilde{L}^{c_p}_{-n}$, they are not necessarily zero-eigenvectors of $L^{c_p}_n-\widetilde{L}^{c_p}_{-n}$. This freedom relaxes the boundary conformal symmetry which equates the chiral and anti-chiral conformal dimensions, so that the reduced density matrix $\text{Tr}_{\text{CFT}_c}\Big[|B(\beta_M)\rangle\langle B(\beta_M)|_{\text{CFT}_c\otimes\text{CFT}_{c_p}}\Big]$ can contain nonzero eigenvectors of $L^{c_p}_n-\widetilde{L}^{c_p}_{-n}$. Note that in non-diagonal CFT, the density of regularized boundary states can be estimated as a square root of full density of states, as demonstrated in holographic BCFT \cite{Miyaji:2021ktr}.

It is now clear why the usual Hartle-Hawking state without bra-ket wormhole is not consistent with an island, violating the strong sub-additivity (\ref{23s2}). Since the matter state prepared by the Hartle-Hawking no boundary prescription is the CFT vacuum state, when CFT takes a form as $\text{CFT}_{c+c_p}=\text{CFT}_{c}\otimes\text{CFT}_{c_p}$, the matter state is an unentangled state $|0\rangle_{c+c_p}=|0\rangle_c\otimes|0\rangle_{c_p}$. Therefore, the state lacks sufficient entanglement necessary for the island formula to satisfy (\ref{23s2}) or (\ref{saturation}). In appendix \ref{Alowentropy}, we explicitly show low entropy mixed state of boundary states violates (\ref{saturation}). 

It is tempting to interpret the condition (\ref{saturation}) as a consequence of ensemble average in semiclassical gravity, which is closely related to the absence of global symmetry in gravity \cite{Harlow:2020bee, Chen:2020ojn,Hsin:2020mfa, Yonekura:2020ino}. 

\subsection{Comparison with Other Islands}

Let us describe the differences between the islands we will discuss, islands in black hole evaporation context, cosmological islands \cite{Hartman:2020khs}, and islands in the bra-ket wormhole \cite{Chen:2020tes}. 

The cosmological island set up in \cite{Hartman:2020khs} describes open universe entangled with a reservoir. When there is sufficiently large entanglement between the subregion in the cosmological spacetime and the reservoir, an island will be formed in the cosmological spacetime, and the entropy is described by the boundary area of the island. This description is in parallel with the standard island story in black hole evaporation.

For the islands in bra-ket wormhole \cite{Chen:2020tes}, we can interpret them in two ways. The first way is to understand the bra-ket wormhole as a closed universe entangled with the universe which ends at $\mathcal{F}$ \cite{Dong:2020uxp, Chen:2020tes}. In this interpretation, the bra-ket wormhole appears because we need to trace out the closed universe to obtain the reduced density matrix on $\mathcal{F}$. Interestingly, the matter entanglement entropy of whole $\mathcal{F}$ is zero at leading order, so the matter state on $\mathcal{F}$ is always pure. This implies that we should not take the closed universe degrees of freedom as independent degrees of freedom. This interpretation is in parallel with the standard island story in black hole evaporation, except for the continuation to the non-gravitational spacetime is done on a space-like timeslice.
\begin{figure}[t]
 \begin{center}
 \includegraphics[width=12.0cm,clip]{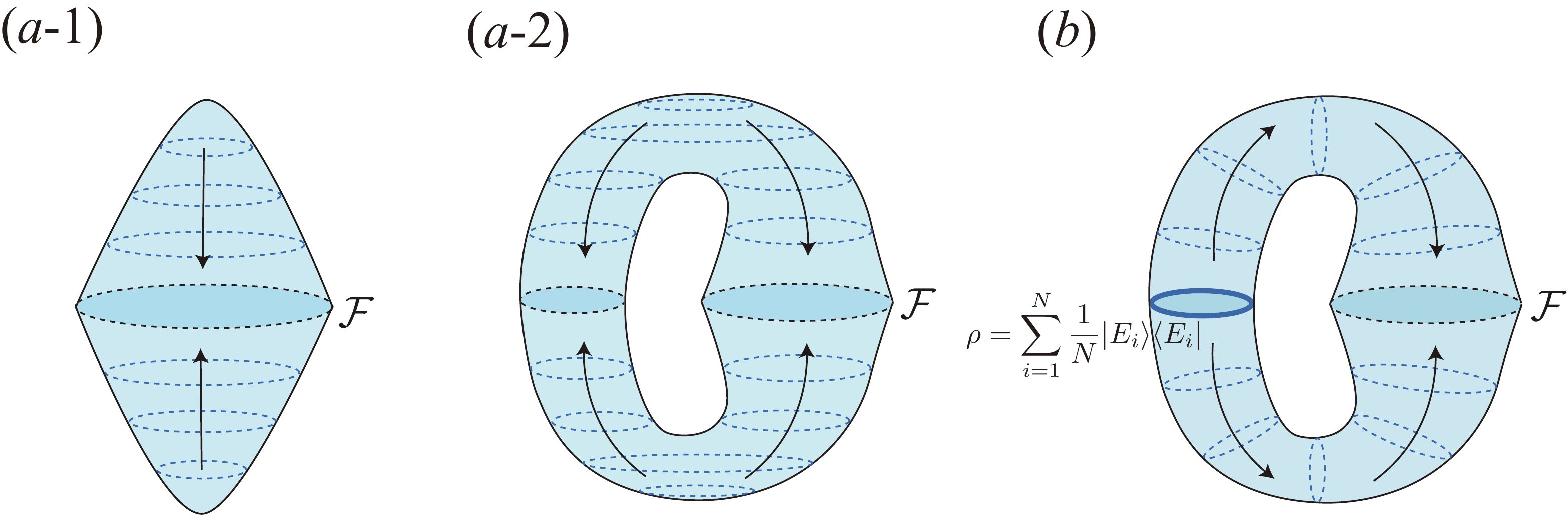}
 \end{center}
 \caption{($a$-1)($a$-2): Hartle-Hawking no boundary initial condition fixes the initial timeslice to have zero volume. When $L/\phi_b$ is sufficiently large, ($a$-2) dominates over ($a$-1), see \cite{Chen:2020tes}. ($b$): We can also view the bra-ket wormhole as taking thermal mixed state as the initial state.}
 \label{fig:explain}
\end{figure}

The second way of understanding is that the bra-ket wormhole is preparing a matter thermal mixed state as the initial state. Because the entanglement entropy obey the volume law, the entanglement entropy for large subregion is governed by the island formula. The mixed nature of the initial state cannot be seen from the matter state on $\mathcal{F}$ at the leading order, and the entropy of the state on $\mathcal{F}$ is zero. This is because a universe which is finite and has no timelike boundary cannot be entangled with exterior degrees of freedom, though it may have future spacelike boundary. Indeed, the island for the exterior degrees of freedom that are entangled with, covers the entire timeslice of the spacetime, and therefore the entanglement entropy is zero\footnote{This conclusion seems to be the same when we allow black holes to form in the spacetime, choosing the future timeslice $\mathcal{F}$ on which we do the projection onto fixed length/dilaton timeslice to include black hole interior. However, we have to be careful with the use of the projection, since the orthogonality of fixed length/dilaton states is merely approximate.}. 

This setup is what we considered in this paper. We found that we can use the island formula for spacetime initiated by a pure state, as long as it satisfies the condition (\ref{condition2}). The gravitationally prepared state is always pure regardless of the existence of the island phase. However, we cannot have mixed state as the initial state when we do not have the island phase, if we demand the state on $\mathcal{F}$ to be pure. This situation is different from the bra-ket wormhole, where having thermal mixed state as the initial state is equivalent to the emergence of the bra-ket wormhole and islands, saving the purity of the state on $\mathcal{F}$. 


\subsection{Pseudo Entanglement Wedge \label{ProjectedEW}}

In this section, we generalize the standard entanglement wedge reconstruction \cite{Almheiri:2014lwa, Dong:2016eik, Harlow:2016vwg, Cotler:2017erl, Chen:2019gbt} to the cases without a moment of time reflection symmetry. Such construction is crucial to understand the islands we considered when $\text{Re}[a_I]<a$, while leaving the case $\text{Re}[a_I]>a$ as an open question. 

Let us first explain how the moment of time reflection symmetry plays an important role in the standard entanglement wedge reconstruction even when the bulk is time-dependent. In the standard gravitational replica trick derivation of the HRT formula \cite{Dong:2016hjy, Colin-Ellerin:2020mva}, Schwinger-Keldysh formalism in the bulk spacetime and the boundary is used to prepare the boundary and bulk reduced density matrices. In that case, the HRT surface is on the bulk time slice where the bra and the ket part of the bulk geometries are glued together. This means the HRT surface sits at the moment of time reflection symmetric slice, in the geometry prepared by Schwinger-Keldysh procedure. The HRT surface defines a bulk reduced density of states thanks to this time reflection symmetry, which can be reconstructed in terms of boundary reduced density matrix through entanglement wedge reconstruction. 

This section aims to clarify the entanglement wedge reconstruction when the HRT surface or the entanglement island is located at a slice without such time reflection symmetry. We claim that in such a case, what we have in the bulk spacetime is a {\it{bulk transition matrix}} instead of a bulk state. We will see that the JLMS formula \cite{Faulkner:2013ana, Jafferis:2015del} and bulk modular Hamiltonian can also be generalized. The discussion in this section applies to the case when $a_I<a$, which does not involve reversed Euclidean time evolution, leaving the interpretation of $a_I>a$ case as open problem.

We first revisit the path integral in the gravitational replica trick. We first identify the bulk matter transition matrix, for the semiclassical geometry where we have an island in the ket part of the bulk. There are three timeslices that we need to consider, the future flat space time slice $\mathcal{T}_{\text{flat}}$ which contains $A$, the bulk time slice $\mathcal{T}_{\text{ket}}$ which contain the island $I^{\text{ket}}$ in the ket, and its bra counterpart $\mathcal{T}_{\text{bra}}$. We now define the bulk transition matrix on $A\cup I^{\text{ket}}$. We define surfaces $A^{\pm}$ and $I^{\text{ket}\pm}$ by infinitesimal Euclidean time translation of $A$ and $I$ respectively, with the sign $\pm$. We put boundary conditions on the CFT wavefunction $\psi$ at $I^{\text{ket}+}$, $I^{\text{ket}-}$, $A^{+}$ and $A^{-}$, as $\psi^{I^{\text{ket}+}}$, $\psi^{I^{\text{ket}-}}$, $\psi^{A^{+}}$ and $\psi^{A^{-}}$, see Fig.\ref{fig:EW}. With these boundary conditions, the effective matter path integral now defines the matrix element of the bulk transition matrix $\rho^{\text{Bulk}}_{\text{ket}}(A\cup I^{\text{ket}}:a_I)$,

\begin{eqnarray}\nonumber&&
\langle\mathcal{U}\psi^{I^{\text{ket}+}},\psi^{A^{-}}|\rho^{\text{Bulk}}_{\text{ket}}(A\cup I^{\text{ket}}:a_I)|\mathcal{U}\psi^{I^{\text{ket}-}},\psi^{A^{+}}\rangle
\\\nonumber&&:=\sum_{E_{\mathcal{T}_{\text{bra}}},E_{\tilde{A}},E_{\tilde{I^{\text{ket}}}}}
\langle B|e^{-[(\beta_M-\beta_I)/4-i(t_M-t_I)]H}|E_{\mathcal{T}_{\text{bra}}}\rangle
\langle E_{\mathcal{T}_{\text{bra}}}|e^{-(\beta_I/4-it_I)H}|\psi^{A^+},E_{\tilde{A}}\rangle
\\\nonumber&&
~~\times\langle\psi^{A^{-}},E_{\tilde{A}}|e^{-(\beta_I/4+it_I)H}|\psi^{I^{\text{ket}+}},E_{\tilde{I}^{\text{ket}}}\rangle
\langle\psi^{I^{\text{ket}-}},E_{\tilde{I}^{\text{ket}}}|e^{-[(\beta_M-\beta_I)/4+i(t_M-t_I)]H}|B\rangle.\\\label{MatrixE}
\end{eqnarray}\\
\begin{figure}[t]
 \begin{center}
 \includegraphics[width=4.5cm,clip]{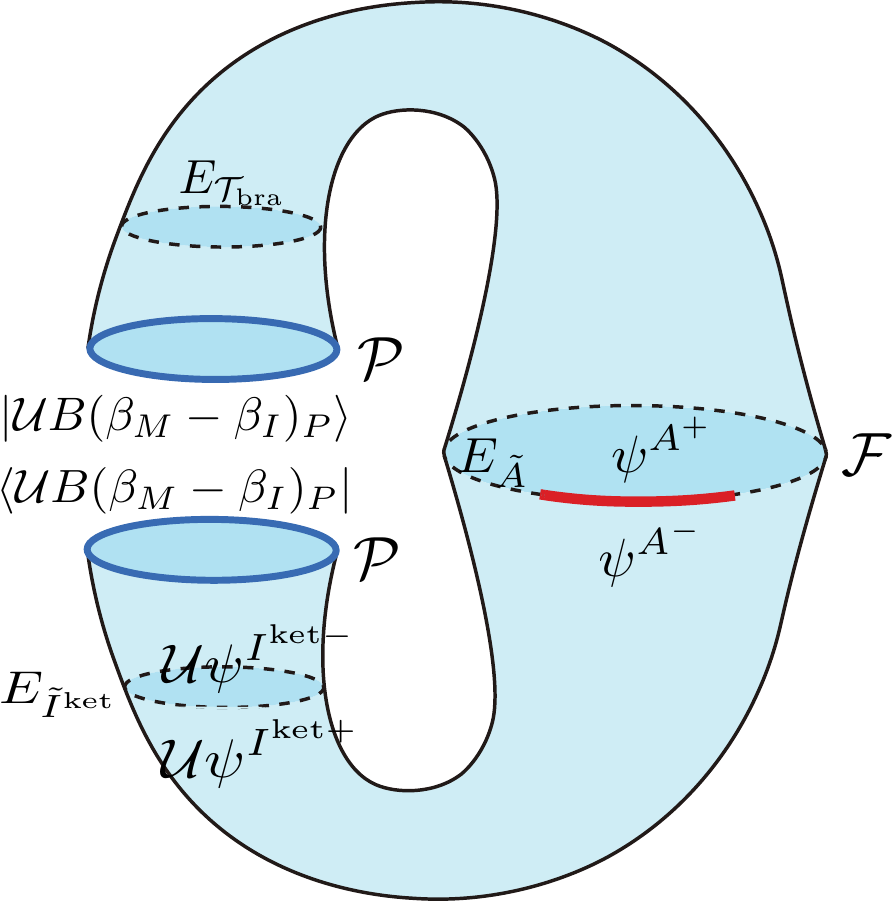}
 \end{center}
 \caption{Decomposition of the matrix element $\langle\mathcal{U}\psi^{I^{\text{ket}+}},\psi^{A^{-}}|\rho^{\text{Bulk}}_{\text{ket}}(A\cup I^{\text{ket}}:a_I)|\mathcal{U}\psi^{I^{\text{ket}-}},\psi^{A^{+}}\rangle$ in terms of basis $E_{\mathcal{T}_{\text{bra}}},E_{\tilde{A}},E_{\tilde{I}^{\text{ket}}}$ of CFT subregions. Note that the time orderings of $A^{\mp}$ and $I^{\pm}$. The initial boundary state now acts as a final state after CPT conjugation. Such CPT conjugation allows us to introduce the thermofield double state, and regard the initial state as a final state projection.}
 \label{fig:EW}
 \end{figure}
Here we defined $\beta_I:=4a_I$, and $\mathcal{U}$ is the CPT operator, which is anti-unitary and commutes with the Hamiltonian\footnote{Anti-unitary operator $\mathcal{U}$ is an anti-linear operator that satisfies $\langle \mathcal{U}a|\mathcal{U}b\rangle=\langle a|b\rangle^{*}$. Its adjoint is defined by $\langle \mathcal{U}^{\dagger}a|b\rangle=\langle a|\mathcal{U}b\rangle^{*}$.}. Here $\sum_{E_R}|E_R\rangle\langle E_R|$ is the CFT resolution of identity on a subregion $R$. Note that the definition of the transition matrix is defined in such a way that it involves {\it{reversed time evolution}} at $I$. 

The bra counterpart is similarly defined as

\begin{eqnarray}\nonumber&&
\langle\mathcal{U}\psi^{I^{\text{bra}+}},\psi^{A^{-}}|\rho^{\text{Bulk}}_{\text{bra}}(A\cup I^{\text{bra}}:a_I)|\mathcal{U}\psi^{I^{\text{bra}-}},\psi^{A^{+}}\rangle
\\\nonumber&&:=\sum_{E_{\mathcal{T}_{\text{bra}}},E_{\tilde{A}},E_{\tilde{I}^{\text{bra}}}}
\langle B|e^{-[(\beta_M-\beta_I)/4-i(t_M-t_I)]H}|\psi^{I^{\text{bra}+}},E_{\tilde{I}^{\text{bra}}}\rangle
\langle\psi^{I^{\text{bra}-}},E_{\tilde{I}^{\text{bra}}}|e^{-(\beta_I/4-it_I)H}|\psi^{A^{+}},E_{\tilde{A}}\rangle
\\&&
~~\times
\langle\psi^{A^-},E_{\tilde{A}}|e^{-(\beta_I/4+it_I)H}|
E_{\mathcal{T}_{\text{ket}}}\rangle
\langle E_{\mathcal{T}_{\text{ket}}}|e^{-[(\beta_M-\beta_I)/4+i(t_M-t_I)]H}|B\rangle.\end{eqnarray}\\
Here we defined $I^{\text{bra}\pm}$ according to the time ordering. When we take a time reflection of the ket semiclassical geometry, we have $A^{\pm}\rightarrow A^{\mp}$ and $I^{\text{ket}\pm}\rightarrow I^{\text{bra}\mp}$.
From such reflection we see that $\rho^{\text{Bulk}}_{\text{ket}}(A\cup I^{\text{ket}}:a_I)=\rho^{\text{Bulk}}_{\text{bra}}(A\cup I^{\text{bra}}:a_I)^{\dagger}$.

We will rewrite the bulk transition matrix $\rho^{\text{Bulk}}_{\text{ket}}(A\cup I^{\text{ket}}:a_I)$ in terms of thermofield double states,

\begin{equation}
|\text{TFD}(\beta,t)_{\mathcal{PF}}\rangle:=\frac{1}{\sqrt{Z_{\text{CFT}}^{\text{Thermal}}(\beta)}}\sum_Ee^{-\beta E/2-iEt}|\mathcal{U}E\rangle_{\mathcal{P}}\otimes|E\rangle_{\mathcal{F}}\in\text{CFT}_{\mathcal{P}}\otimes\text{CFT}_{\mathcal{F}},
\end{equation}
\begin{equation}
|\overline{\text{TFD}}(\beta,t)_{\mathcal{PF}}\rangle:=\frac{1}{\sqrt{Z_{\text{CFT}}^{\text{Thermal}}(\beta)}}\sum_Ee^{-\beta E/2-iEt}|E\rangle_{\mathcal{P}}\otimes|\mathcal{U}E\rangle_{\mathcal{F}}\in\text{CFT}_{\mathcal{P}}\otimes\text{CFT}_{\mathcal{F}},
\end{equation}\\
where $Z_{\text{CFT}}^{\text{Thermal}}(\beta):=\sum_ae^{-\beta E_a}$. 

Using thermofield double states and setting $\beta_I:=4a_I$, we reach the following equality,

\begin{eqnarray}\nonumber&&
\rho^{\text{Bulk}}_{\text{ket}}(A\cup I^{\text{ket}}:a_I)
=Z_{\text{CFT}}^{\text{Thermal}}(\beta_I/2)~
\text{Tr}_{\tilde{I}^{\text{ket}},\tilde{A}}
\Big[|TFD(\beta_I/2, t_I)_{\mathcal{PF}}\rangle
\langle \overline{TFD}(\beta_I/2, t_I)_{PF}|
\\&&~~\times 
e^{-[(\beta_M-\beta_I)/4-i(t_M-t_I)]H}|\mathcal{U}B\rangle
\langle \mathcal{U}B|e^{-[(\beta_M-\beta_I)/4+i(t_M-t_I)]H}\Big].\label{transitionmatrix}\end{eqnarray}\\
The derivation is explained in appendix \ref{Abulktransition}.The generalization of this equality to more generic initial states is straightforward, which gives the bulk transition matrix for generic gravitational backgrounds. Note that we have actions of $\mathcal{U}$ on the initial boundary states in (\ref{transitionmatrix}). This is because we are considering reverse time evolution for these initial-boundary states. This allows us to understand why we have entanglement island in cosmological settings. When the semiclassical geometry is analytically continued to a positive imaginary time direction, we can consider CPT conjugation at a positive Euclidean time at which the island is located, and we evolve the spacetime in reverse time direction from that time. Then, we obtain a spacetime which splits into two disconnected spacetimes, one is a {\it{closed universe}} which will be {\it{projected}} by the final boundary state, and the other spacetime will continue to the flat space where we measure the entanglement entropy. Gravity dual of such measurements using boundary states were considered in \cite{Numasawa:2016emc, Cooper:2018cmb}.

Importantly, the transition matrix $\rho^{\text{Bulk}}_{\text{ket}}(A\cup I^{\text{ket}})$ is not Hermitian, therefore not a state. However, we can still define a quantity for such transition matrix $\tilde{\rho}:=\rho/\text{Tr}\rho$, which has identical expression as standard von Neumann entropy, called {\it{pseudo entropy}} \cite{Nakata:2021ubr, Mollabashi:2020yie},

\begin{equation}
S^{\text{P}}(\tilde{\rho}):=-\text{Tr}\Big[\tilde{\rho}~\text{log}~\tilde{\rho}\Big].
\end{equation}\\
Since $\rho$ is no longer Hermitian, the definition of operator function $\text{log}$ must be specified. We use a definition of operator function $f(\rho)$ which satisfies $X^{-1}f(\rho)X=f(X^{-1}\rho X)$ and makes use of Jordan normal form. For the detailed definition, we refer to \cite{Nakata:2021ubr}. Using such definition, we have

\begin{equation}
S^{\text{P}}(\tilde{\rho})=-\sum_{\lambda(\tilde{\rho})}\lambda(\tilde{\rho})\text{log}\lambda(\tilde{\rho}).
\end{equation}\\
Here $\lambda(\tilde{\rho})$ are the eigenvalues of $\tilde{\rho}$ which can be obtained via Jordan normal form. Its Renyi entropy has an analogous expression. Since $\lambda(\tilde{\rho})$ can take any complex value, the pseudo entropy can be complex-valued. The gravity dual of this pseudo entropy is the area of the extremal surface in the bulk spacetime, anchored from boundary subregion $A$ \cite{Nakata:2021ubr}, which can be derived using gravitational replica trick \cite{Lewkowycz:2013nqa}. This equality can be generalized straightforwardly to include quantum corrections. Suppose that 
$\gamma_A$ is a bulk surface anchored from $\partial A$, and homologous and space-like separated to $A$. The $EW[A:\mathcal{M}]$ is a subregion of a timeslice bounded by $\gamma_A$ and $A$. By the path integral of bulk matter with the fixed boundary condition at $EW[A:\gamma_A]$, we obtain a bulk transition matrix $\rho^{\text{Bulk}}(EW[A:\gamma_A])$ on $EW[A:\gamma_A]$. From the gravitational replica trick, we obtain

\begin{equation}
S^P_A=\underset{\gamma_A}{\text{Min}\text{Ext}}\Big[\frac{\text{Area}[\gamma_A]}{4G_N}+S^P(\rho^{\text{Bulk}}(EW[A:\gamma_A]))\Big],\label{pSgen}
\end{equation}\\
which can be naturally called {\it{generalized pseudo entropy}}. When $\gamma_A$ realizes the extremal of (\ref{pSgen}), the region $EW[A:\gamma_A]$ is called {\it{pseudo entanglement wedge}} and $\gamma_A$ is called {\it{pseudo RT surface}}. 

In terms of pseudo entropy, the ket/bra part of our island entanglement entropy can be written as

\begin{equation}
S_A^{\text{ket}}(a_I)=2\frac{\phi_0+\phi(a_I)}{4G_N}+S^{\text{P}}(\tilde{\rho}^{\text{Bulk}}_{\text{ket}}(A\cup I^{\text{ket}}:a_I)).\end{equation}
\begin{equation}
S_A^{\text{bra}}(a_I)=2\frac{\phi_0+\phi(a_I)^{*}}{4G_N}+S^{\text{P}}(\tilde{\rho}^{\text{Bulk}}_{\text{bra}}(A\cup I^{\text{bra}}:a_I)).\end{equation}\\
This is the precise version of the island formula for the gravitationally prepared state in generic cosmological backgrounds, in terms of bulk transition matrix. Here we have $S_A^{\text{bra}}(a_I)=S_A^{\text{ket}}(a_I)^{*}$, and the entropy is given by their average

\begin{equation}
S_A(a_I)=\frac{1}{2}\Big[S_A^{\text{ket}}(a_I)+S_A^{\text{bra}}(a_I)\Big].\end{equation}\\
In the following, we will study entanglement wedge reconstruction for holographic systems as well as in the presence of island.
\subsubsection*{Pseudo Entanglement Wedge in Holography\label{pEW}}

Let us consider the case when an overlap of two CFT states has multiple dual gravity saddles $\mathcal{M}_i$ in Schwinger-Keldysh formalism. Considering a subregion $A$ on the time slice, yields a modular Hamiltonian $H_A$ and a transition matrix $\rho_A$. These saddles may not have a moment of time reflection symmetry. For the case with a moment of time reflection symmetry in holography, see \cite{Almheiri:2016blp}. 

Each bulk geometry is associated with CFT transition matrix $\rho_A^{(i)}$, which are assumed to satisfy, 

\begin{equation}
1=\text{Tr}\Big[\rho_A\Big]=\text{Tr}\Big[\sum_i\rho_A^{(i)}\Big],~~\text{Tr}\Big[\rho_A^n\Big]\approx\text{Tr}\Big[\sum_i\rho_A^{(i)}{}^n\Big]~~(n\in\mathbb{Z}_{>0}),
\end{equation}\\
In other words, the cross terms between different saddles are assumed to be suppressed. Note that this is an assumption on $\rho_A^{(i)}$, since $\rho_A^{(i)}\rho_A^{(j)}$ with $i\neq j$ are not necessarily suppressed, when we can glue $EW[A:\mathcal{M}_i]$ with $EW[A:\mathcal{M}_j]$. We can associate the corresponding modular Hamiltonians of these transition matrices $\tilde{\rho}_A^{(i)}:=\rho_A^{(i)}/p_i$ with $p_i:=\text{Tr}\rho_A^{(i)}$ as

\begin{equation}
H_A^{(i)}:=-\text{log}\tilde{\rho}_A^{(i)}.
\end{equation}\\
Then we have

\begin{equation}
S_A=-\text{Tr}\Big[\rho_A\text{log}\rho_A\Big]\approx\sum_ip_i\langle H^{(i)}_A\rangle_{\tilde{\rho}_A^{(i)}}+\sum_i -p_i\text{log}~p_i.
\end{equation}\\
Since $\langle H^{(i)}_A\rangle_{\tilde{\rho}_A^{(i)}}$ is equal to the pseudo entropy of $\tilde{\rho}_A^{(i)}$, we have

\begin{equation}
\langle H^{(i)}_A\rangle_{\tilde{\rho}_A^{(i)}}=\frac{\text{Area}[\gamma_A^i]}{4G_N}+\langle H^{\mathcal{M}_i}_{\text{bulk}:EW[A:\mathcal{M}_i]}\rangle_{\rho^{\text{Bulk}}(EW[A:\mathcal{M}_i])}.
\end{equation}\\
Here $\gamma^i_A$ is the generalized pseudo RT surface in $\mathcal{M}_i$, which is anchored from $\partial A$. The pseudo entanglement wedge $EW[A:\mathcal{M}_i]$ is a region of a time slice in $\mathcal{M}_i$, which is bounded by $A$ and $\gamma^i_A$.

With these formulae in hand, now we can compare the bulk and CFT {\it{pseudo relative entropy}}. The pseudo relative entropy between $\rho_A$ and $\sigma_A$ is defined by the usual expression

\begin{equation}
S^{\text{P}}(\rho_A|\sigma_A):=\text{Tr}\Big[\rho_A~\text{log}~\rho_A\Big]-\text{Tr}\Big[\rho_A~\text{log}~\sigma_A\Big].\end{equation}\\
Note that we again use Jordan normal form to define matrix functions. Such relative entropy is no longer positive and can be complex-valued; therefore its meaning as a distance measure for states is unclear. Let us consider a small perturbation $\rho^{(i)}_A$ as $\sigma^{(i)}_A=\rho^{(i)}_A+\delta\rho^{(i)}_A$, in such a way that the dual semiclassical geometry and the location of $\gamma^{(i)}_A$ are unchanged. In such case, we find

\begin{equation}
S^{\text{P}}(\tilde{\rho}^{(i)}_A|\tilde{\sigma}^{(i)}_A)\approx S^{\text{P}}(\rho^{\text{Bulk}}(EW[A:\mathcal{M}_i])|\sigma^{\text{Bulk}}(EW[A:\mathcal{M}_i])).\label{approxrelative}
\end{equation}\\
When $\rho_A$ and $\sigma_A$ are states instead of transition matrices, this approximate equality implies the Petz map $\mathcal{R}_{\mathcal{N},\rho_A}$ for the bulk-to-boundary isometric map $\mathcal{N}$, is indeed an approximate recovery map. In other words, it implies that the Petz map $\mathcal{R}_{\mathcal{N},\rho_A}:L(H_A)\rightarrow L(H^{\text{Bulk}}_{EW[A:\mathcal{M}]})$ for the encoding map $\mathcal{N}:L(H^{\text{Bulk}}_{EW[A:\mathcal{M}]})\rightarrow L(H_A)$ satisfies $\mathcal{R}_{\mathcal{N},\rho_A}(\mathcal{N}(\rho_{EW[A]}^{\text{Bulk}}))\approx\rho_{EW[A]}^{\text{Bulk}}$ and $\mathcal{R}_{\mathcal{N},\rho_A}(\mathcal{N}(\sigma^{\text{Bulk}}_{EW[A]}))\approx\sigma_{EW[A]}^{\text{Bulk}}$, where the error vanishes when the (\ref{approxrelative}) is an exact equality. It has not been clarified whether this same story holds for general transition matrices. We plan to explore recovery maps for general transition matrices in \cite{OPEW}. Generalization of the pseudo entanglement wedge reconstruction to entanglement pseudo island is straightforward, and is described in Appendix \ref{pEWIS}.


\subsection{Pseudo Python's Lunch\label{Python}}

While we expect that with an analog of the Petz recovery map for transition matrix, we can reconstruct bulk information, it is important to understand how difficult it is to implement such a recovery task. In an evaporating black holes after the Page time, the number of unitary gates $\mathcal{C}$ required to recover a diary thrown into the black hole, from the collected Hawking radiation is estimated to be an exponential of the black hole entropy \cite{Harlow:2013tf}

\begin{equation}
\text{log}~\mathcal{C}\approx c_0S_{\text{BH}},
\end{equation}\\
where $c_0$ is an order one constant. The connection between this exponential complexity of recovery and the bulk geometry was addressed in \cite{Brown:2019rox}. Based on the tensor network description of the bulk, it was conjectured that such exponential complexity is evaluated by an exponent of the {\it{Python's lunch}}. We briefly explain this Python's lunch conjecture. Consider an entanglement wedge EW${}_A$ of a system $A$ with a quantum Ryu-Takayanagi surface $\gamma_{\text{L}}$, and suppose that there is another quantum extremal surface $\gamma_{\text{R}}$ anchored from $A$ which is contained in EW${}_A$. The Python's lunch conjecture states that the complexity of recovery is given by

\begin{equation}
\text{log}~\mathcal{C}\approx \underset{\Sigma}{\text{Ext}}\underset{\gamma_t}{\text{Min}}\underset{t}{\text{Max}}\Big[\frac{1}{2}\Big(S_{\text{gen}}(\gamma_t)-S_{\text{gen}}(\gamma_R)\Big)\Big].\label{python}
\end{equation}\\
Here $S_{\text{gen}}(\gamma)$ is the generalized entropy of the subregion bounded by $A$ and $\gamma$, where $\Sigma$ is a timeslice which contains $A$, $\gamma_{\text{L}}$ and $\gamma_{\text{R}}$. The set $\gamma_t$ $(t\in [0,~1])$ is a continuous family of curves in $\Sigma$ anchored from $\partial A$ , whose boundary condition is given by $\gamma_{t=0}=\gamma_{\text{L}}$ and $\gamma_{t=1}=\gamma_{\text{R}}$. When the spacetime signature is Lorentzian, $\underset{\Sigma}{\text{Ext}}=\underset{\Sigma}{\text{Max}}$, while for Euclidean signature $\Sigma$ is the reflection symmetric time slice. The Python's lunch conjecture is based on tensor network description of the bulk geometry, and making use of the Grover search algorithm for the recovery task.

We will now generalize the Python's lunch conjecture to non-time-reflection symmetric case. In Euclidean case, the natural choice is to replace generalized entropy by generalized pseudo entropy,

\begin{equation}
\text{log}~\mathcal{C}\approx \underset{\Sigma}{\text{Min}}\underset{\gamma_t}{\text{Ext}}\underset{t}{\text{Max}}\Big[\frac{1}{2}\text{Re}\Big[S^P_{\text{gen}}(\gamma_t)-S^P_{\text{gen}}(\gamma_R)\Big]\Big],\label{pseudopython}
\end{equation}\\
which can be naturally called {\it{pseudo Python's lunch}} conjecture. The timeslice $\Sigma$ is now generic timeslice, and in particular not constrained to be time reflection symmetric. We leave Lorentzian generalization as a future problem, since pseudo entropy can be complex in that case so that the meaning is not clear yet. 

We now apply this {\it{pseudo Python's lunch}} conjecture to estimate the difficulty in the recovery of the bulk excitation on $\rho^{\text{Bulk}}_{\text{ket}}(A\cup I:a_I)$ at the entanglement island $I$. We assume the timeslice $\Sigma$ as the $z=-a_I$ slice for simplicity. $\gamma_{\text{R}}$ is now an empty set, and the generalized pseudo entropy $S^P_{\text{gen}}(\gamma_R)$ is equal to thermal entropy of $A$. $\gamma_{\text{L}}$ is equal to the boundary of the island, which is at $z=-a_I$ and $x_M=\pm \Delta l/2$. 

We will now estimate the size of the Python's lunch. One possible choice of the continuous family of curves $\gamma_t$ from $\gamma_{t=0}=\gamma_{\text{R}}$ to $\gamma_{t=1}=\gamma_{\text{L}}$, are the subregions $x_M\in [-t\Delta l/2,~t\Delta l/2]$ of $\Sigma$ plus $A$. In this case, the initial increase of the generalized entropy is approximately $ 2\frac{\phi_0}{4G_N}$. Therefore, we obtain an estimate for the necessary number of gates to recover the bulk excitation on the island as

\begin{equation}
\text{log}~\mathcal{C}\approx \frac{\phi_0}{4G_N}.
\end{equation}\\
We conclude that the recovery task of information on island is exponentially difficult. The argument here can be used to the Python's lunch argument for closed universes.

\section{Discussions\label{discussion}}

In this paper, we considered entanglement entropy of a gravitationally prepared state in two-dimensional gravity with finite-boundary initial conditions.  We found that when the entanglement entropy of the initial state is sufficiently large so that it could violate the entropy bound (\ref{areabound}), the entanglement island emerges near the initial boundary. Consequently, the entanglement entropy is given by the island formula and satisfies the entropy bound (\ref{areabound}) up to small corrections. We also studied possible initial conditions from extrinsic curvatures and the matter entanglement entropy. We have seen that $K\geq 0$ is necessary in order not to lose the information of the initial state in Euclidean AdS, and any small fraction of degrees of freedom should be thermally entangled with the rest in order to have consistent entanglement entropy. To summarize, we found a necessary condition on the initial state which allows us to use the island formula, for spacetime whose initial conditions are not described by the Hartle-Hawking no boundary condition.

The method in this paper can be applied to Lorentzian spacetimes in two and higher dimensions, especially de Sitter spacetimes. One obstacle for Lorentzian extension is that the rule to pick from multiple saddles is not clear from the viewpoint of saddle point analysis, especially when the imaginary part of the action is not bounded from below so that the interpretation in terms of probability breaks down. Our interpretation of this problem is based on Lefschetz thimble \cite{DiTucci:2019dji, DiTucci:2019bui}. In Lefschetz thimble, one starts with a path integral with real metric and initial boundary conditions. The oscillating path integral turns out to be approximated by {\it{complex}} saddle points and these complex saddles are determined by the initial condition. In particular, not all of the saddles are relevant for the path integral, so that the unbounded imaginary part of the action is not necessarily problematic. Closely related uses of complex saddles in the context of Lorentzian entanglement island are discussed in \cite{Colin-Ellerin:2020mva, Colin-Ellerin:2021jev}. 

Another obstacle for Lorentzian generalization is the problem of causality. Since our island lives near the initial boundary, it is natural to wonder if the formula is inconsistent with the causality. In the HRT formula, it is important that HRT surface $\gamma_A$ is space-like separated from causal wedge $\mathcal{C}(A)$ of $A$\cite{Hubeny:2012wa, Wall:2012uf, Engelhardt:2014gca, Headrick:2014cta}. If $\gamma_A$ is time-like separated from $\mathcal{C}(A)$, we have a geodesic which intersects both $\gamma_A$ and $\mathcal{C}(A)$; therefore we can consider a bulk particle on that geodesic. Though the entanglement entropy of $A$ is unchanged, the area of HRT surface $\gamma_A$ can changes in general, arriving at a contradiction. The absence of causality issue in island outside black hole horizon is discussed in \cite{Almheiri:2019yqk}.

The argument above for HRT surface does not directly apply to the gravitationally prepared state and island near the past initial boundary. This is because the spacetime we consider is complex, in particular, has Euclidean part, and the island is located at positive Euclidean time. It is interesting to understand the causality constraints systematically in complex spacetime, especially those of generalized extremal surfaces for pseudo entropy.

While we specified a necessary condition for the initial state, we have not shown that there are explicit examples. It is interesting to construct boundary state whose reduced density of state $\text{Tr}_{\text{CFT}_c}\Big[|B(\beta_M)\rangle\langle B(\beta_M)|_{\text{CFT}_c\otimes\text{CFT}_{c_p}}\Big]$ is a thermal mixed state, working explicitly in symmetric orbifold CFT for example.

The geometry we consider can be described as inserting a junction in a wormhole and taking $\mathbb{Z}_2$ orbifold. The initial matter state we consider is by construction pure, as opposed to a thermal mixed state in the bra-ket wormhole. This resembles considering non-averaged initial state by a projection. It is interesting to see how these observations are related to the eigenbranes in JT gravity \cite{Blommaert:2019wfy}, half wormholes \cite{Saad:2021rcu}, non-local interaction in spacetime branes\cite{Blommaert:2021fob}. and brane anti-brane nucleation in wormhole geometry \cite{Maldacena:2004rf, Marolf:2021kjc}. 

We also developed pseudo entanglement wedge reconstruction and pseudo JLMS formula and identified the bulk transition matrix, which should be, at least partially, recovered from a fine-grained state. Although we have found the pseudo entanglement wedge version of the JLMS formula, the recovery map for the pseudo entanglement wedge reconstruction is yet unknown in the literature. We expect there is a recovery map analogous to the Petz recovery map, which is currently under investigation in \cite{OPEW}. 


\section*{Acknowledgement}
We would like to thank R. Bousso, H. Marrochio, Y. Nomura, G. Penington and T. Takayanagi for discussions and comments. This work is supported in part by the Berkeley Center for Theoretical Physics; by the Department of Energy, Office of Science, Office of High Energy Physics under QuantISED Award DE-SC0019380 and under contract DE-AC02-05CH11231; and by the National Science Foundation under grant PHY1820912.

\appendix

\section{Other Background AdS${}_2$}
\subsection{Global AdS${}_2$}
We consider background global AdS metric

\begin{equation}
ds^2=\frac{dz^2+dx^2}{\text{sinh}^2z},~~\phi=\frac{\phi_r}{-\text{tanh}z}~~\text{for} ~~z\leq -a_c.
\end{equation}\\
We then glue the AdS spacetime at $z=-a_c$ to a flat spacetime,

\begin{equation}
ds^2=\frac{-dt_M^2+dx_M^2}{\epsilon^2}, ~~~\phi=\frac{\phi_b}{\epsilon}=\frac{\phi_r}{a_c}.
\end{equation}\\
We take the $z$ coordinate of the initial timeslice $\mathcal{P}$ as $z=-a$. On $\mathcal{P}$, the induced spatial metric and the extrinsic curvature are

\begin{equation}
\sqrt{h_{x_Mx_M}}=\frac{1}{\text{sinh}a}\left(\frac{a_c}{\epsilon}\right),~~
\widetilde{K}=\text{cosh}a.\end{equation}\\
$\sqrt{h_{x_Mx_M}}$ can take any positive value, while the extrinsic curvature is restricted to $\widetilde{K}\geq1$. We again consider Weyl transformation from the original metric to the flat space,

\begin{equation}
ds^2_g=\frac{dz^2+dx^2}{\text{sinh}^2z}\rightarrow ds^2_{g'}=\frac{dz^2+dx^2}{a_c^2}=\frac{dz_M^2+dx_M^2}{\epsilon^2}.\end{equation}\\
The Weyl factor is then

\begin{equation}
ds^2_g=\text{e}^{2\tau}ds^2_{g'},~~e^{\tau}=\frac{a_c}{-\text{sinh}z}.\end{equation}\\
For this Weyl transformation, the anomaly stress tensor is

\begin{equation}
T^{\text{Anomaly},g'\rightarrow g}_{\mu\nu}=\widetilde{T}^{\text{Anomaly},g'\rightarrow g}_{\mu\nu}-\frac{c}{24\pi}g_{\mu\nu},
\end{equation}\\
with

\begin{equation}
\widetilde{T}^{\text{Anomaly},g'\rightarrow g}_{z_Mz_M}=-\widetilde{T}^{\text{Anomaly},g'\rightarrow g}_{x_Mx_M}=-\frac{c}{24\pi}\Big(\frac{a_{\text{bra},c}}{\epsilon}\Big)^2.
\end{equation}\\
The last term in the anomaly stress tensor, $-\frac{c}{24\pi}g_{\mu\nu}$, can again be canceled by moving $\frac{c}{24\pi}\int_M \sqrt{g}$ into the matter action. The flat space after the Weyl transformation is a Euclidean strip. Its width in $(z_M, x_M)$ coordinate is $\frac{\beta_M}{2}:=a_{\text{bra}}\frac{\epsilon}{a_{\text{bra},c}}+a_{\text{ket}}\frac{\epsilon}{a_{\text{bra},c}}+\frac{\beta_b}{2}$. The stress tensor is given by
 
\begin{equation}
T^{\text{Matter},g'}_{z_Mz_M}=-T^{\text{Matter},g'}_{x_Mx_M}=-\frac{c}{6\pi}\Big(\frac{\pi}{\beta_M}\Big)^2.
\end{equation}\\
The dilaton is

\begin{equation}
\phi=\frac{\phi_b}{-\text{tanh}z}\frac{a_{\text{bra},c}}{\epsilon}+\frac{4\pi^2G_Nc}{3}\Big(\frac{1}{4\pi^2}+\frac{1}{\beta_{\text{Bulk}}^2}\Big)\Big(\frac{z}{\text{tanh}z}-1\Big).\end{equation}\\
Therefore, the overlap is

\begin{equation}
\text{log}\langle C|C\rangle= L\frac{c}{24\pi}\Big(a_{\text{bra}}\frac{a_{\text{bra},c}}{\epsilon}+a_{\text{ket}}\frac{a_{\text{ket},c}}{\epsilon}\Big)+\text{log}\langle B(\beta_M)|B(\beta_M)\rangle_{\text{Flat Cylinder}}-S_{\text{JT}_{\text{ket}}}-S_{\text{JT}_{\text{bra}}}-S_{\mathcal{P}_{\text{bra}}}-S_{\mathcal{P}_{\text{ket}}}.\label{totalactionglobal}\end{equation}\\
Here we have

\begin{equation}
S_{\text{JT}_{\text{ket}}}=-L\frac{1}{16\pi G_N}\phi_b\Big(\frac{a_{\text{ket},c}}{\epsilon}\Big)^2,~
S_{\mathcal{P}_{\text{ket}}}=L\frac{T'}{8\pi G_N}\frac{1}{\text{sinh}a}\Big(\frac{a_{\text{ket},c}}{\epsilon}\Big).\end{equation}\\
The tension $T<0$ determines $a$ uniquely, since $\text{cosh}a=-K=-T$. 

Note that $S_{\text{JT}_{\text{ket}}}$ has a wrong sign, diverges negatively when $a_c/\epsilon\rightarrow \infty$ if $\phi_b>0$. This is usually not an issue when there is no initial boundary, since demanding the global AdS geometry to have no conical deficit angle implies $L^{\mathcal{P}}=La_c/\epsilon=2\pi$. Therefore $a_c/\epsilon$ is not a variable subject to the equation of motion. On the other hand, it is not clear whether we should demand $L^{\mathcal{P}}=La_c/\epsilon=2\pi$ when there is an initial surface that hides the potential conical deficit angle. If we were to stick to positive real $a_c/\epsilon$, we should assume $\phi_b<0$ in order to have a solution with maximum partition function.

In the following, we will only consider the case with $L^{\mathcal{P}}=La_c/\epsilon=2\pi$. For simplicity, we will assume $\beta_b=0$ in the following. Then we have 

\begin{equation}
\beta_M=\frac{2aL}{\pi}.
\end{equation}\\ 
We assume $S^{\text{Thermal}}(\beta_M)/4>\frac{\pi c\beta_M}{12L}+2S_B$ and $a=a_{\text{bra}}=a_{\text{ket}}$ and $\beta_M/L\ll1$. These assumption in particular implies $\beta_{\text{Bulk}}<2\pi$ when $S_B>0$. The dilaton is then

\begin{equation}
\phi=\frac{\phi_b}{-\text{tanh}z}\frac{2\pi}{L}+\frac{4\pi^2G_Nc}{3}\Big(\frac{1}{4\pi^2}+\frac{1}{\beta_{\text{Bulk}}^2}\Big)\Big(\frac{z}{\text{tanh}z}-1\Big),\label{dilatonglobal}\end{equation}\\
and the overlap is, 

\begin{equation}
\text{log}\langle C|C\rangle=\frac{\pi cL}{12\beta_M}+\frac{\pi\phi_b}{2G_NL}-\frac{T'}{2G_N\sqrt{T^2-1}}-2S_B.\label{totalactionglobal}\end{equation}\\

We now consider whether the spacetime with EOW brane is dominant or not, compared to other geometries.

\subsubsection*{Off-diagonal Overlap of EOW Brane and No-Boundary State}

For the case when $\tilde{K}>1$, the overlap is given by

\begin{equation}
\text{log}\langle C|C\rangle_{\text{Off-diagonal}}=\Big(\frac{\phi_0}{4G_N}+\frac{\pi\phi_b}{4G_NL}\Big)+\Big(\frac{\pi\phi_b}{4G_NL}-\frac{T'}{4 G_N\sqrt{T^2-1}}\Big),\label{globaloffdiagonal}\end{equation}\\
which is suppressed compared to the diagonal overlap of EOW branes (\ref{globaloffdiagonal}) or the Hartle-Hawking no-boundary state as expected, from the condition $\frac{S_L^{\text{Thermal}}(\beta_M)}{4}=\frac{\pi Lc}{12\beta_M}>\frac{\pi c\beta_M}{12L}+2S_B$.
\subsubsection*{Spacetime Initiated by EOW Brane}
When $\tilde{K}>1$, the background geometry is given by the global AdS. The overlap is given by (\ref{totalactionglobal}). We can again derive a sufficient condition that (\ref{totalactionglobal}) is dominant compared to the bra-ket wormhole using $\frac{S_L^{\text{Thermal}}(\beta_M)}{4}=\frac{\pi Lc}{12\beta_M}>\frac{\pi c\beta_M}{12L}+2S_B$. This sufficient condition is 

\begin{equation}
\frac{T'}{4G_Nc}<-L\frac{G_Nc\pi\text{sinh}a}{64\phi_b}\label{sufconditionglobal}.
\end{equation}\\ Note in particular this condition requires $T'<0$.



\subsection{Poincare AdS${}_2$\label{Apoincare}}

We assume the background metric and the dilaton are those of Poincare AdS;

\begin{equation}
ds^2=\frac{dz^2+dx^2}{z^2},~~\phi=\frac{\phi_r}{-z}~~\text{for} ~~z\leq -a_c.
\end{equation}\\
We glue AdS spacetime at $z=-a_c$ to flat spacetime,

\begin{equation}
ds^2=\frac{-dt_M^2+dx_M^2}{\epsilon^2}, ~~\phi=\frac{\phi_b}{\ep}=\frac{\phi_r}{a_c}.
\end{equation}\\
We consider effective matter state $|C\rangle$ on Minkowski timeslice, at $t_M=0$. We again assume the $z$ coordinate of $\mathcal{P}$ as $z=-a$. On $\mathcal{P}$, the spatial metric and the extrinsic curvatures are

\begin{equation}
\sqrt{h_{x_Mx_M}}=\frac{1}{a}\left(\frac{a_c}{\epsilon}\right),~~\tilde{K}=1.\end{equation}\\
Fixing $\sqrt{h_{x_Mx_M}}$ determines the ratio $a/a_c$ only. $\sqrt{h_{x_Mx_M}}$ can take any positive value. We take $z$ coordinate of $\mathcal{P}$ as $z=-a$. We assume the topology for the ket(bra) is again the trivial one. We consider Weyl transformation from the original metric to the flat metric

\begin{equation}
ds^2_g=\frac{dz^2+dx^2}{z^2}\rightarrow ds^2_{g'}=\frac{dz^2+dx^2}{\epsilon^2}. \end{equation}\\
The Weyl factor is

\begin{equation}
ds^2_g=\text{e}^{2\tau}ds^2_{g'},~e^{\tau}=\frac{\epsilon}{-z}.\end{equation}\\
For the Weyl transformation in our case, the anomaly stress tensor is

\begin{equation}
T^{\text{Anomaly},g'\rightarrow g}_{\mu\nu}=-\frac{c}{24\pi}g_{\mu\nu},
\end{equation}\\
which can be canceled by including $\frac{c}{24\pi}\int_M \sqrt{|g|}$ in the matter action. The flat part is an infinite Euclidean strip. Its width in $(x_M,z_M)$ coordinate is $\frac{\beta_M}{2}:=a_{\text{bra}}\frac{\epsilon}{a_{\text{bra},c}}+a_{\text{ket}}\frac{\epsilon}{a_{\text{bra},c}}+\frac{\beta_b}{2}$. Then the stress tensor is given by

\begin{equation}
T^{\text{Matter},g'}_{z_Mz_M}=-T^{\text{Matter},g'}_{x_Mx_M}=-\frac{c}{6\pi}\Big(\frac{\pi}{\beta_M}\Big)^2.
\end{equation}\\
The dilaton is given by

\begin{equation}
\phi=\frac{\phi_b}{-z}\frac{a_c}{\epsilon}+\frac{4\pi^2 G_N c}{9}\Big(\frac{-z}{\beta_{\text{Bulk}}}\Big)^2.\end{equation}\\
Therefore, we have

\begin{equation}
\text{log}\langle C|C\rangle= \text{log}\langle B(\beta_M)|B(\beta_M)\rangle_{\text{Flat Cylinder}}-S_{\text{JT}_{\text{ket}}}-S_{\text{JT}_{\text{ket}}}-S_{\mathcal{P}_{\text{bra}}}-S_{\mathcal{P}_{\text{ket}}}.\end{equation}\\

\section{BCFT Pseudo Entropy\label{Abcftpseudo}}

Consider a BCFT on Euclidean two dimensional plane with coordinate $(\sigma, x)$, with two boundaries at $\sigma=\pm\beta/4$. We assume that BCFT is holographic BCFT so that the correlation functions are given by Wick contractions. Let us evaluate the pseudo entropy of an interval $A$ with boundaries at $(\sigma, x)=(\sigma,x_1),(\sigma,x_2)$, alternatively the pseudo entropy of a transition matrix $\text{Tr}_{\tilde{A}}[e^{-(\frac{\beta}{4}+\sigma)H}|B\rangle\langle B|e^{-(\frac{\beta}{4}-\sigma)H}]$. For gravity duals, see \cite{Mollabashi:2021xsd}. We will also consider its analytical continuation to real time $\sigma\rightarrow\sigma+it$, the pseudo entropy of $\text{Tr}_{\tilde{A}}[e^{-(\frac{\beta}{4}+\sigma+it)H}|B\rangle\langle B|e^{-(\frac{\beta}{4}-\sigma-it)H}]$. We define $\Delta x=x_2-x_1$. When the pseudo entropy $S_A^P$ is given by the Wick contraction between twist operators, we have

\begin{figure}[t]
 \begin{center}
 \includegraphics[width=4cm,clip]{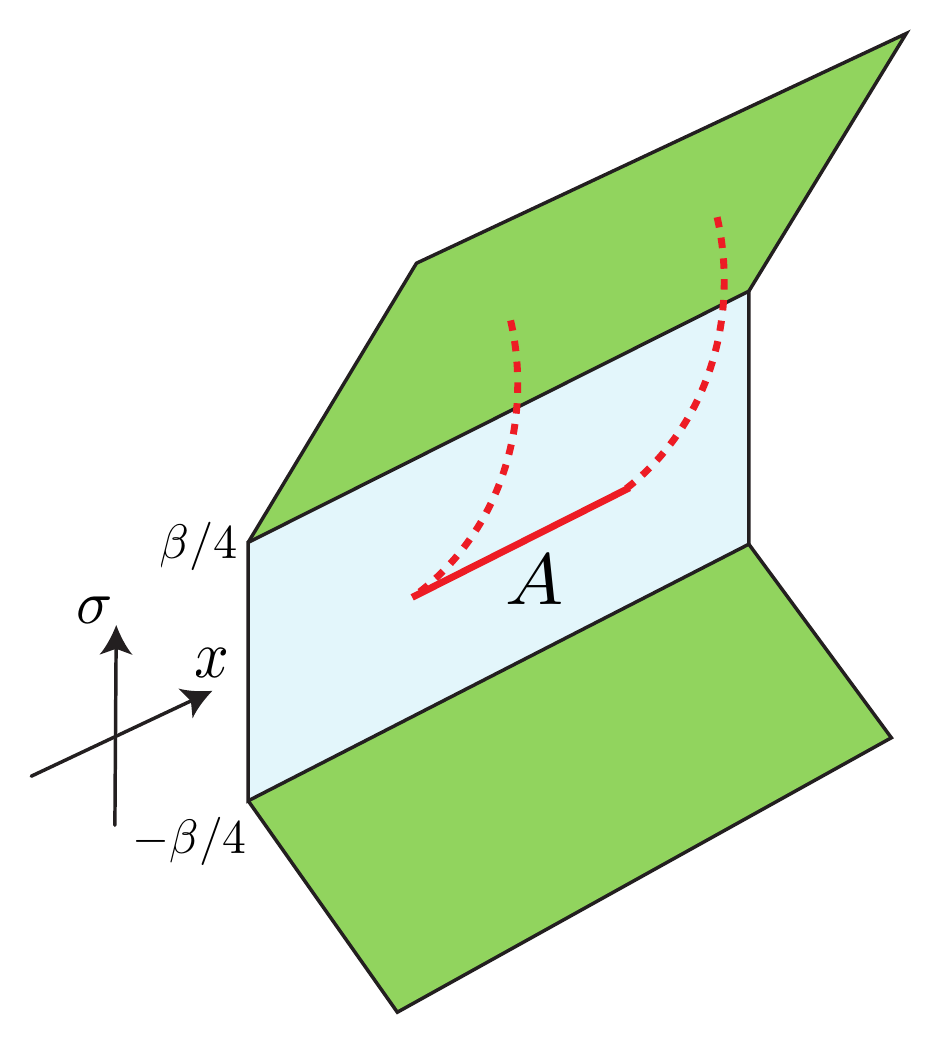}
 \end{center}
 \caption{Ryu-Takayanagi surfaces (dotted red lines) for the subregion $A$ in terms of AdS/BCFT ending at the EOW brane emanated from $\sigma=\beta/4$ boundary. This situation corresponds to the case when twist operators for $A$ contract with the boundary at $\sigma=\beta/4$. When the boundary conditions at $\sigma=\pm\beta/4$ are identical and the total system size is sufficiently large, the two EOW branes are connected, and the boundary correlation functions behave thermally. When one of two conditions is not satisfied, the EOW branes are disconnected, and RT surface can end on both EOW branes.}
 \label{fig:BCFTPE}
\end{figure}
\begin{equation}
S_A^P=\frac{c}{6}\text{log}\Big[\Big(\frac{\beta}{2\pi\epsilon}\Big)^22\text{sinh}^2\Big(\frac{\pi}{\beta}\Delta x\Big)\Big].\label{effpseudo}
\end{equation}\\
In other words we assume the dominant intermediate operator beween twist operator is the vacuum. Here $S_A^P$ is the usual thermal entropy at inverse temperature $\beta$; moreover, analytical continuation to real-time gives the same result. This implies that the state is locally at thermal equilibrium with inverse temperature $\beta$. 

Let us consider the case when two twist operators contract with the boundary at $\sigma=\pm\beta/4$, see Fig \ref{fig:BCFTPE}. The pseudo entropy is 

\begin{equation}
S_A^P(\sigma,\pm\beta/4)=\frac{c}{3}\text{log}\Big[\frac{\beta}{\pi\epsilon}\text{sin}\Big(\frac{2\pi(\mp\sigma+\beta/4)}{\beta}\Big)\Big]+2S_B,
\end{equation}\\
and its analytical continuation to real time $\sigma\rightarrow\sigma+it$ is

\begin{equation}
S_A^P(t,\sigma,~\pm\beta/4)=\frac{c}{3}\text{log}\Big[\frac{\beta}{\pi\epsilon}\text{sin}\Big(\frac{2\pi(\mp\sigma+\beta/4)\mp 2i\pi t}{\beta}\Big)\Big]+2S_B.
\end{equation}\\
When one twist operator contract with the boundary at $\sigma=\beta/4$ and another with the boundary at $\sigma=-\beta/4$, the pseudo entropy is 

\begin{equation}
S_A^P(\sigma,\beta/4,-\beta/4)=\frac{c}{6}\text{log}\Big[\Big(\frac{\beta}{\pi\epsilon}\Big)^2\text{sin}\Big(\frac{2\pi(\sigma+\beta/4)}{\beta}\Big)\text{sin}\Big(\frac{2\pi(-\sigma+\beta/4)}{\beta}\Big)\Big]+2S_B,
\end{equation}\\
and its analytical continuation to real time $\sigma\rightarrow\sigma+it$ is

\begin{equation}
S_A^P(t,\sigma,\beta/4,-\beta/4)=\frac{c}{6}\text{log}\Big[\Big(\frac{\beta}{\pi\epsilon}\Big)^2\text{sin}\Big(\frac{2\pi(\sigma+\beta/4)+2i\pi t}{\beta}\Big)\text{sin}\Big(\frac{2\pi(-\sigma+\beta/4)-2i\pi t}{\beta}\Big)\Big]+2S_B.
\end{equation}\\
When $\sigma=0$ with real time $t$, the transition matrix of interest is a state, and its entanglement entropy is given by

\begin{equation}
S_A=\frac{c}{3}\text{log}\Big[\frac{\beta}{\pi\epsilon}\text{cosh}\Big(\frac{2\pi t}{\beta}\Big)\Big]+2S_B=\frac{1}{2}\Big(S_A^P(t,0,~\beta/4)+S_A^P(t,0,-\beta/4)\Big)=S_A^P(t,0,\beta/4,-\beta/4).
\end{equation}\\
\section{Low Entropy Mixed State of Boundary States\label{Alowentropy}}

Mixed state of regularized boundary states,

\begin{equation}\text{Tr}_{\text{CFT}_c}\Big[|B(\beta_M)\rangle\langle B(\beta_M)|_{\text{CFT}_c\otimes\text{CFT}_{c_p}}\Big]=\sum_ap_a|B_a(\beta_M)\rangle\langle B_a(\beta_M)|_{\text{CFT}_{c_p}},\end{equation}\\
with $\sum_ap_a=1$, $p_a\geq0$ and $\frac{S_{c_p,L}^{\text{Thermal}}(\beta_M)}{4}=\frac{\pi c_pL}{12\beta_M}>\frac{\pi c_p\beta_M}{6L}+2S^{c_p}_{B_a}$, can be shown to violate (\ref{saturation}), assuming the number of nonzero $p_a$ is small compared to $e^{S_{c_p,L}^{\text{Thermal}}(\beta_M)}$. Since the overlap between different regularized boundary states is exponentially suppressed, the entanglement entropy of $A$ with $L/2>L_A>L/4$ is given by 

\begin{equation}
S^{\text{eff}}_{A_{c_p}}=\frac{c_p}{3}\text{log}\Big[\frac{\beta_M}{\pi\epsilon}\Big]+2\bar{S}^{c_p}_B,~S^{\text{eff}}_{\widetilde{A}_{c_p}}=\frac{c_p}{3}\text{log}\Big[\frac{\beta_M}{\pi\epsilon}\Big]+2\bar{S}^{c_p}_B.\end{equation}\\ 
Here we defined averaged boundary entropy

\begin{equation}
\bar{S}^{c_p}_B:=\sum_ap_aS^{c_p}_{B_a}-\sum_ap_a\text{log}~p_a.\end{equation}
Then we have $S(\text{CFT}_{c_p})=-\sum_ap_a\text{log}~p_a\leq \bar{S}^{c_p}_B$ and $S^{\text{eff}}_{A_{c_p}}+ S^{\text{eff}}_{\widetilde{A}_{c_p}}\approx4\bar{S}^{c_p}_B$, inconsistent with (\ref{saturation}).

\section{Details of Bulk Transition Matrix\label{Abulktransition}}

We explain details on the bulk transition matrix $\rho^{\text{Bulk}}_{\text{ket}}(A\cup I^{\text{ket}}:a_I)$. It is defined by

\begin{eqnarray}\nonumber&&
\langle\mathcal{U}\psi^{I^{\text{ket}+}},\psi^{A^{-}}|\rho^{\text{Bulk}}_{\text{ket}}(A\cup I^{\text{ket}}:a_I)|\mathcal{U}\psi^{I^{\text{ket}-}},\psi^{A^{+}}\rangle
\\\nonumber&&:=\sum_{E_{\mathcal{T}_{\text{bra}}},E_{\tilde{A}},E_{\tilde{I^{\text{ket}}}}}
\langle B|e^{-[(\beta_M-\beta_I)/4-i(t_M-t_I)]H}|E_{\mathcal{T}_{\text{bra}}}\rangle
\langle E_{\mathcal{T}_{\text{bra}}}|e^{-(\beta_I/4-it_I)H}|\psi^{A^+},E_{\tilde{A}}\rangle
\\\nonumber&&
~~\times\langle\psi^{A^{-}},E_{\tilde{A}}|e^{-(\beta_I/4+it_I)H}|\psi^{I^{\text{ket}+}},E_{\tilde{I}^{\text{ket}}}\rangle
\langle\psi^{I^{\text{ket}-}},E_{\tilde{I}^{\text{ket}}}|e^{-[(\beta_M-\beta_I)/4+i(t_M-t_I)]H}|B\rangle.\label{MatrixEap}\\
\end{eqnarray}\\
We will rewrite these matrix elements using thermofield double state. Rewriting the time evolution, we have

\begin{eqnarray}\nonumber
\langle\psi_F|e^{-(\beta/4+it)H}|\psi_P\rangle&=&
\sqrt{Z_{\text{CFT}}^{\text{Thermal}}(\beta/2)}\langle \overline{TFD}(\beta/2, -t)_{PF}|\psi_P,\mathcal{U}\psi_F\rangle\\&=&\sqrt{Z_{\text{CFT}}^{\text{Thermal}}(\beta/2)}\langle\mathcal{U}\psi_P,\psi_F|TFD(\beta/2, t)_{PF}\rangle.
\label{exTFD}
\end{eqnarray}\\
Using (\ref{exTFD}), we can express (\ref{MatrixEap}) in terms of TFD states,

\begin{eqnarray}\nonumber&&
\langle\mathcal{U}\psi^{I^{\text{ket}+}},\psi^{A^{-}}|\rho^{\text{Bulk}}_{\text{ket}}(A\cup I^{\text{ket}}:a_I)|\mathcal{U}\psi^{I^{\text{ket}-}},\psi^{A^{+}}\rangle
\\\nonumber&&
=Z_{\text{CFT}}^{\text{Thermal}}(\beta_I/2)
\\\nonumber&&
\times\sum_{E_{\mathcal{T}_{\text{bra}}},E_{\tilde{A}},E_{\tilde{I}^{\text{ket}}}}
\langle\mathcal{U}\psi^{I^{\text{ket}+}},\mathcal{U}E_{\tilde{I}^{\text{ket}}},\psi^{A^{-}},E_{\tilde{A}}|TFD(\beta_I/2, t_I)_{PF}\rangle
\langle \overline{TFD}(\beta_I/2, t_I)_{PF}|\psi^{A^+},E_{\tilde{A}},\mathcal{U}E_{\mathcal{T}_{\text{bra}}}\rangle
\\\nonumber&&~~\times
\langle \mathcal{U}E_{\mathcal{T}_{\text{bra}}}|e^{-[(\beta_M-\beta_I)/4-i(t_M-t_I)]H}|\mathcal{U}B\rangle
\langle \mathcal{U}B|e^{-[(\beta_M-\beta_I)/4+i(t_M-t_I)]H}|\mathcal{U}\psi^{I^{\text{ket}-}},\mathcal{U}E_{\tilde{I}}\rangle
\end{eqnarray}
rewriting this in terms of trace gives
\begin{eqnarray}\nonumber&&
=Z_{\text{CFT}}^{\text{Thermal}}(\beta_I/2)
\langle\mathcal{U}\psi^{I^{\text{ket}+}}
,\psi^{A^-}|\text{Tr}_{\tilde{I}^{\text{ket}},\tilde{A}}
\Big[|TFD(\beta_I/2, t_I)_{PF}\rangle
\langle \overline{TFD}(\beta_I/2, t_I)_{PF}|
\\\nonumber&&~~\times 
e^{-[(\beta_M-\beta_I)/4-i(t_M-t_I)]H}|\mathcal{U}B\rangle
\langle \mathcal{U}B|e^{-[(\beta_M-\beta_I)/4+i(t_M-t_I)]H}\Big]|\mathcal{U}\psi^{I^{\text{ket}-}},\psi^{A^+}\rangle.\\\end{eqnarray}
From this equality, we immediately have (\ref{transitionmatrix}).


\section{Pseudo Entanglement Wedge Reconstruction of the Initial State\label{pEWIS}}

The pseudo entanglement wedge in holography can be generalized straightforwardly to the entanglement island. Let us assume the full density matrix can be decomposed as

\begin{equation}
\text{Tr}_{H^{\text{CFT}}_A}\Big[\Big(\tilde{\rho}_A^{\text{CFT}}\Big)^n\Big]\approx \text{Tr}_{H^{\text{CFT}}_A}\Big[\Big(\tilde{\rho}_A^{\text{CFT:ket}}/2\Big)^n+\Big(\tilde{\rho}_A^{\text{CFT:bra}}/2\Big)^n\Big]~~(n\in\mathbb{Z}_{>0}),
\end{equation}\\
corresponding to two semiclassical saddles. In other words, the cross terms are assumed to be suppressed. Note that $1=\text{Tr}_{H^{\text{CFT}}_A}\Big[\tilde{\rho}_A^{\text{CFT:ket}}\Big]=\text{Tr}_{H^{\text{CFT}}_A}\Big[\tilde{\rho}_A^{\text{CFT:bra}}\Big]$. In the bulk, the ket/bra bulk transition matrices $\tilde{\rho}^{\text{Bulk}}_{\text{ket}}(A\cup I^{\text{ket}}:a_I)$ and $\tilde{\rho}^{\text{Bulk}}_{\text{bra}}(A\cup I^{\text{bra}}:a_I)$ are defined, therefore we can also define bulk modular Hamiltonians as

\begin{equation}
H^{\text{ket}}_{\text{bulk}:A\cup I^{\text{ket}}}:=-\text{log}~\tilde{\rho}^{\text{Bulk}}_{\text{ket}}(A\cup I^{\text{ket}}:a_I),~~
H^{\text{bra}}_{\text{bulk}:A\cup I^{\text{bra}}}:=-\text{log}~\tilde{\rho}^{\text{Bulk}}_{\text{bra}}(A\cup I^{\text{bra}}:a_I).
\end{equation}\\
Using modular Hamiltonians $H_A^{\text{CFT:ket}}:=-\text{log}\tilde{\rho}_A^{\text{CFT:ket}}$ and $H_A^{\text{CFT:bra}}:=-\text{log}\tilde{\rho}_A^{\text{CFT:bra}}$, we have

\begin{eqnarray}\nonumber&&
S_A^{\text{ket}}(a_I)=\langle H_A^{\text{CFT:ket}}\rangle_{\tilde{\rho}^{\text{CFT:ket}}_A}=2\frac{\phi_0+\phi(a_{I^{\text{ket}}})}{4G_N}+\langle H^{\text{ket}}_{\text{bulk}:A\cup I^{\text{ket}}}\rangle_{\tilde{\rho}^{\text{Bulk}}_{\text{ket}}},\\&&
S_A^{\text{bra}}(a_I)=\langle H_A^{\text{CFT:bra}}\rangle_{\tilde{\rho}^{\text{CFT:bra}}_A}=2\frac{\phi_0+\phi(a_{I^{\text{bra}}})}{4G_N}+\langle H^{\text{bra}}_{\text{bulk}:A\cup I^{\text{bra}}}\rangle_{\tilde{\rho}^{\text{Bulk}}_{\text{bra}}}.
\end{eqnarray}\\
Therefore the equalities between bulk and boundary pseudo relative entropy are similarly given by

\begin{eqnarray}\nonumber
S^{\text{P}}(e^{-H^{\text{CFT:ket}}_{A}(\rho)}|e^{-H^{\text{CFT:ket}}_{A}(\sigma)})&\approx& S^{\text{P}}(e^{-H^{\text{ket}}_{\text{bulk}:A\cup I}(\rho)}|e^{-H^{\text{ket}}_{\text{bulk}:A\cup I}(\sigma)}),
\\
S^{\text{P}}(e^{-H^{\text{CFT:bra}}_{A}(\rho)}|e^{-H^{\text{CFT:bra}}_{A}(\sigma)})&\approx& S^{\text{P}}(e^{-H^{\text{bra}}_{\text{bulk}:A\cup I}(\rho)}|e^{-H^{\text{bra}}_{\text{bulk}:A\cup I}(\sigma)}).\end{eqnarray}
\bibliographystyle{JHEP}
\bibliography{dS}

\providecommand{\href}[2]{#2}\begingroup\raggedright\begin{thebibliography}{100}

\bibitem{Penington:2019npb}
G.~Penington, \emph{{Entanglement Wedge Reconstruction and the Information
  Paradox}}, \href{https://doi.org/10.1007/JHEP09(2020)002}{\emph{JHEP}
  {\bfseries 09} (2020) 002}
  [\href{https://arxiv.org/abs/1905.08255}{{\ttfamily 1905.08255}}].

\bibitem{Almheiri:2019psf}
A.~Almheiri, N.~Engelhardt, D.~Marolf and H.~Maxfield, \emph{{The entropy of
  bulk quantum fields and the entanglement wedge of an evaporating black
  hole}}, \href{https://doi.org/10.1007/JHEP12(2019)063}{\emph{JHEP} {\bfseries
  12} (2019) 063} [\href{https://arxiv.org/abs/1905.08762}{{\ttfamily
  1905.08762}}].

\bibitem{Almheiri:2019hni}
A.~Almheiri, R.~Mahajan, J.~Maldacena and Y.~Zhao, \emph{{The Page curve of
  Hawking radiation from semiclassical geometry}},
  \href{https://doi.org/10.1007/JHEP03(2020)149}{\emph{JHEP} {\bfseries 03}
  (2020) 149} [\href{https://arxiv.org/abs/1908.10996}{{\ttfamily
  1908.10996}}].

\bibitem{Penington:2019kki}
G.~Penington, S.~H. Shenker, D.~Stanford and Z.~Yang, \emph{{Replica wormholes
  and the black hole interior}},
  \href{https://arxiv.org/abs/1911.11977}{{\ttfamily 1911.11977}}.

\bibitem{Almheiri:2019qdq}
A.~Almheiri, T.~Hartman, J.~Maldacena, E.~Shaghoulian and A.~Tajdini,
  \emph{{Replica Wormholes and the Entropy of Hawking Radiation}},
  \href{https://doi.org/10.1007/JHEP05(2020)013}{\emph{JHEP} {\bfseries 05}
  (2020) 013} [\href{https://arxiv.org/abs/1911.12333}{{\ttfamily
  1911.12333}}].

\bibitem{Ryu:2006bv}
S.~Ryu and T.~Takayanagi, \emph{{Holographic derivation of entanglement entropy
  from AdS/CFT}},
  \href{https://doi.org/10.1103/PhysRevLett.96.181602}{\emph{Phys. Rev. Lett.}
  {\bfseries 96} (2006) 181602}
  [\href{https://arxiv.org/abs/hep-th/0603001}{{\ttfamily hep-th/0603001}}].

\bibitem{Ryu:2006ef}
S.~Ryu and T.~Takayanagi, \emph{{Aspects of Holographic Entanglement Entropy}},
  \href{https://doi.org/10.1088/1126-6708/2006/08/045}{\emph{JHEP} {\bfseries
  08} (2006) 045} [\href{https://arxiv.org/abs/hep-th/0605073}{{\ttfamily
  hep-th/0605073}}].

\bibitem{Hubeny:2007xt}
V.~E. Hubeny, M.~Rangamani and T.~Takayanagi, \emph{{A Covariant holographic
  entanglement entropy proposal}},
  \href{https://doi.org/10.1088/1126-6708/2007/07/062}{\emph{JHEP} {\bfseries
  07} (2007) 062} [\href{https://arxiv.org/abs/0705.0016}{{\ttfamily
  0705.0016}}].

\bibitem{Takayanagi:2011zk}
T.~Takayanagi, \emph{{Holographic Dual of BCFT}},
  \href{https://doi.org/10.1103/PhysRevLett.107.101602}{\emph{Phys. Rev. Lett.}
  {\bfseries 107} (2011) 101602}
  [\href{https://arxiv.org/abs/1105.5165}{{\ttfamily 1105.5165}}].

\bibitem{Fujita:2011fp}
M.~Fujita, T.~Takayanagi and E.~Tonni, \emph{{Aspects of AdS/BCFT}},
  \href{https://doi.org/10.1007/JHEP11(2011)043}{\emph{JHEP} {\bfseries 11}
  (2011) 043} [\href{https://arxiv.org/abs/1108.5152}{{\ttfamily 1108.5152}}].

\bibitem{Lewkowycz:2013nqa}
A.~Lewkowycz and J.~Maldacena, \emph{{Generalized gravitational entropy}},
  \href{https://doi.org/10.1007/JHEP08(2013)090}{\emph{JHEP} {\bfseries 08}
  (2013) 090} [\href{https://arxiv.org/abs/1304.4926}{{\ttfamily 1304.4926}}].

\bibitem{Faulkner:2013ana}
T.~Faulkner, A.~Lewkowycz and J.~Maldacena, \emph{{Quantum corrections to
  holographic entanglement entropy}},
  \href{https://doi.org/10.1007/JHEP11(2013)074}{\emph{JHEP} {\bfseries 11}
  (2013) 074} [\href{https://arxiv.org/abs/1307.2892}{{\ttfamily 1307.2892}}].

\bibitem{Engelhardt:2014gca}
N.~Engelhardt and A.~C. Wall, \emph{{Quantum Extremal Surfaces: Holographic
  Entanglement Entropy beyond the Classical Regime}},
  \href{https://doi.org/10.1007/JHEP01(2015)073}{\emph{JHEP} {\bfseries 01}
  (2015) 073} [\href{https://arxiv.org/abs/1408.3203}{{\ttfamily 1408.3203}}].

\bibitem{Jafferis:2015del}
D.~L. Jafferis, A.~Lewkowycz, J.~Maldacena and S.~J. Suh, \emph{{Relative
  entropy equals bulk relative entropy}},
  \href{https://doi.org/10.1007/JHEP06(2016)004}{\emph{JHEP} {\bfseries 06}
  (2016) 004} [\href{https://arxiv.org/abs/1512.06431}{{\ttfamily
  1512.06431}}].

\bibitem{Randall:1999vf}
L.~Randall and R.~Sundrum, \emph{{An Alternative to compactification}},
  \href{https://doi.org/10.1103/PhysRevLett.83.4690}{\emph{Phys. Rev. Lett.}
  {\bfseries 83} (1999) 4690}
  [\href{https://arxiv.org/abs/hep-th/9906064}{{\ttfamily hep-th/9906064}}].

\bibitem{Karch:2000ct}
A.~Karch and L.~Randall, \emph{{Locally localized gravity}},
  \href{https://doi.org/10.1088/1126-6708/2001/05/008}{\emph{JHEP} {\bfseries
  05} (2001) 008} [\href{https://arxiv.org/abs/hep-th/0011156}{{\ttfamily
  hep-th/0011156}}].

\bibitem{Almheiri:2014lwa}
A.~Almheiri, X.~Dong and D.~Harlow, \emph{{Bulk Locality and Quantum Error
  Correction in AdS/CFT}},
  \href{https://doi.org/10.1007/JHEP04(2015)163}{\emph{JHEP} {\bfseries 04}
  (2015) 163} [\href{https://arxiv.org/abs/1411.7041}{{\ttfamily 1411.7041}}].

\bibitem{Pastawski:2015qua}
F.~Pastawski, B.~Yoshida, D.~Harlow and J.~Preskill, \emph{{Holographic quantum
  error-correcting codes: Toy models for the bulk/boundary correspondence}},
  \href{https://doi.org/10.1007/JHEP06(2015)149}{\emph{JHEP} {\bfseries 06}
  (2015) 149} [\href{https://arxiv.org/abs/1503.06237}{{\ttfamily
  1503.06237}}].

\bibitem{Dong:2016eik}
X.~Dong, D.~Harlow and A.~C. Wall, \emph{{Reconstruction of Bulk Operators
  within the Entanglement Wedge in Gauge-Gravity Duality}},
  \href{https://doi.org/10.1103/PhysRevLett.117.021601}{\emph{Phys. Rev. Lett.}
  {\bfseries 117} (2016) 021601}
  [\href{https://arxiv.org/abs/1601.05416}{{\ttfamily 1601.05416}}].

\bibitem{Harlow:2016vwg}
D.~Harlow, \emph{{The Ryu\textendash{}Takayanagi Formula from Quantum Error
  Correction}}, \href{https://doi.org/10.1007/s00220-017-2904-z}{\emph{Commun.
  Math. Phys.} {\bfseries 354} (2017) 865}
  [\href{https://arxiv.org/abs/1607.03901}{{\ttfamily 1607.03901}}].

\bibitem{Cotler:2017erl}
J.~Cotler, P.~Hayden, G.~Penington, G.~Salton, B.~Swingle and M.~Walter,
  \emph{{Entanglement Wedge Reconstruction via Universal Recovery Channels}},
  \href{https://doi.org/10.1103/PhysRevX.9.031011}{\emph{Phys. Rev. X}
  {\bfseries 9} (2019) 031011}
  [\href{https://arxiv.org/abs/1704.05839}{{\ttfamily 1704.05839}}].

\bibitem{Hayden:2018khn}
P.~Hayden and G.~Penington, \emph{{Learning the Alpha-bits of Black Holes}},
  \href{https://doi.org/10.1007/JHEP12(2019)007}{\emph{JHEP} {\bfseries 12}
  (2019) 007} [\href{https://arxiv.org/abs/1807.06041}{{\ttfamily
  1807.06041}}].

\bibitem{Chen:2019gbt}
C.-F. Chen, G.~Penington and G.~Salton, \emph{{Entanglement Wedge
  Reconstruction using the Petz Map}},
  \href{https://doi.org/10.1007/JHEP01(2020)168}{\emph{JHEP} {\bfseries 01}
  (2020) 168} [\href{https://arxiv.org/abs/1902.02844}{{\ttfamily
  1902.02844}}].

\bibitem{Hayden:2007cs}
P.~Hayden and J.~Preskill, \emph{{Black holes as mirrors: Quantum information
  in random subsystems}},
  \href{https://doi.org/10.1088/1126-6708/2007/09/120}{\emph{JHEP} {\bfseries
  09} (2007) 120} [\href{https://arxiv.org/abs/0708.4025}{{\ttfamily
  0708.4025}}].

\bibitem{Almheiri:2019yqk}
A.~Almheiri, R.~Mahajan and J.~Maldacena, \emph{{Islands outside the horizon}},
   \href{https://arxiv.org/abs/1910.11077}{{\ttfamily 1910.11077}}.

\bibitem{Rozali:2019day}
M.~Rozali, J.~Sully, M.~Van~Raamsdonk, C.~Waddell and D.~Wakeham,
  \emph{{Information radiation in BCFT models of black holes}},
  \href{https://doi.org/10.1007/JHEP05(2020)004}{\emph{JHEP} {\bfseries 05}
  (2020) 004} [\href{https://arxiv.org/abs/1910.12836}{{\ttfamily
  1910.12836}}].

\bibitem{Almheiri:2019psy}
A.~Almheiri, R.~Mahajan and J.~E. Santos, \emph{{Entanglement islands in higher
  dimensions}},
  \href{https://doi.org/10.21468/SciPostPhys.9.1.001}{\emph{SciPost Phys.}
  {\bfseries 9} (2020) 001} [\href{https://arxiv.org/abs/1911.09666}{{\ttfamily
  1911.09666}}].

\bibitem{Bousso:2019ykv}
R.~Bousso and M.~Toma\v{s}evi\'c, \emph{{Unitarity From a Smooth Horizon?}},
  \href{https://doi.org/10.1103/PhysRevD.102.106019}{\emph{Phys. Rev. D}
  {\bfseries 102} (2020) 106019}
  [\href{https://arxiv.org/abs/1911.06305}{{\ttfamily 1911.06305}}].

\bibitem{Marolf:2020xie}
D.~Marolf and H.~Maxfield, \emph{{Transcending the ensemble: baby universes,
  spacetime wormholes, and the order and disorder of black hole information}},
  \href{https://doi.org/10.1007/JHEP08(2020)044}{\emph{JHEP} {\bfseries 08}
  (2020) 044} [\href{https://arxiv.org/abs/2002.08950}{{\ttfamily
  2002.08950}}].

\bibitem{Balasubramanian:2020hfs}
V.~Balasubramanian, A.~Kar, O.~Parrikar, G.~S\'arosi and T.~Ugajin,
  \emph{{Geometric secret sharing in a model of Hawking radiation}},
  \href{https://doi.org/10.1007/JHEP01(2021)177}{\emph{JHEP} {\bfseries 01}
  (2021) 177} [\href{https://arxiv.org/abs/2003.05448}{{\ttfamily
  2003.05448}}].

\bibitem{Gautason:2020tmk}
F.~F. Gautason, L.~Schneiderbauer, W.~Sybesma and L.~Thorlacius, \emph{{Page
  Curve for an Evaporating Black Hole}},
  \href{https://doi.org/10.1007/JHEP05(2020)091}{\emph{JHEP} {\bfseries 05}
  (2020) 091} [\href{https://arxiv.org/abs/2004.00598}{{\ttfamily
  2004.00598}}].

\bibitem{Anegawa:2020ezn}
T.~Anegawa and N.~Iizuka, \emph{{Notes on islands in asymptotically flat 2d
  dilaton black holes}},
  \href{https://doi.org/10.1007/JHEP07(2020)036}{\emph{JHEP} {\bfseries 07}
  (2020) 036} [\href{https://arxiv.org/abs/2004.01601}{{\ttfamily
  2004.01601}}].

\bibitem{Hashimoto:2020cas}
K.~Hashimoto, N.~Iizuka and Y.~Matsuo, \emph{{Islands in Schwarzschild black
  holes}}, \href{https://doi.org/10.1007/JHEP06(2020)085}{\emph{JHEP}
  {\bfseries 06} (2020) 085}
  [\href{https://arxiv.org/abs/2004.05863}{{\ttfamily 2004.05863}}].

\bibitem{Hartman:2020swn}
T.~Hartman, E.~Shaghoulian and A.~Strominger, \emph{{Islands in Asymptotically
  Flat 2D Gravity}}, \href{https://doi.org/10.1007/JHEP07(2020)022}{\emph{JHEP}
  {\bfseries 07} (2020) 022}
  [\href{https://arxiv.org/abs/2004.13857}{{\ttfamily 2004.13857}}].

\bibitem{Bousso:2020kmy}
R.~Bousso and E.~Wildenhain, \emph{{Gravity/ensemble duality}},
  \href{https://doi.org/10.1103/PhysRevD.102.066005}{\emph{Phys. Rev. D}
  {\bfseries 102} (2020) 066005}
  [\href{https://arxiv.org/abs/2006.16289}{{\ttfamily 2006.16289}}].

\bibitem{Chen:2020uac}
H.~Z. Chen, R.~C. Myers, D.~Neuenfeld, I.~A. Reyes and J.~Sandor,
  \emph{{Quantum Extremal Islands Made Easy, Part I: Entanglement on the
  Brane}}, \href{https://doi.org/10.1007/JHEP10(2020)166}{\emph{JHEP}
  {\bfseries 10} (2020) 166}
  [\href{https://arxiv.org/abs/2006.04851}{{\ttfamily 2006.04851}}].

\bibitem{Chandrasekaran:2020qtn}
V.~Chandrasekaran, M.~Miyaji and P.~Rath, \emph{{Including contributions from
  entanglement islands to the reflected entropy}},
  \href{https://doi.org/10.1103/PhysRevD.102.086009}{\emph{Phys. Rev. D}
  {\bfseries 102} (2020) 086009}
  [\href{https://arxiv.org/abs/2006.10754}{{\ttfamily 2006.10754}}].

\bibitem{Li:2020ceg}
T.~Li, J.~Chu and Y.~Zhou, \emph{{Reflected Entropy for an Evaporating Black
  Hole}}, \href{https://doi.org/10.1007/JHEP11(2020)155}{\emph{JHEP} {\bfseries
  11} (2020) 155} [\href{https://arxiv.org/abs/2006.10846}{{\ttfamily
  2006.10846}}].

\bibitem{Bak:2020enw}
D.~Bak, C.~Kim, S.-H. Yi and J.~Yoon, \emph{{Unitarity of entanglement and
  islands in two-sided Janus black holes}},
  \href{https://doi.org/10.1007/JHEP01(2021)155}{\emph{JHEP} {\bfseries 01}
  (2021) 155} [\href{https://arxiv.org/abs/2006.11717}{{\ttfamily
  2006.11717}}].

\bibitem{Chen:2020tes}
Y.~Chen, V.~Gorbenko and J.~Maldacena, \emph{{Bra-ket wormholes in
  gravitationally prepared states}},
  \href{https://doi.org/10.1007/JHEP02(2021)009}{\emph{JHEP} {\bfseries 02}
  (2021) 009} [\href{https://arxiv.org/abs/2007.16091}{{\ttfamily
  2007.16091}}].

\bibitem{Dong:2020uxp}
X.~Dong, X.-L. Qi, Z.~Shangnan and Z.~Yang, \emph{{Effective entropy of quantum
  fields coupled with gravity}},
  \href{https://doi.org/10.1007/JHEP10(2020)052}{\emph{JHEP} {\bfseries 10}
  (2020) 052} [\href{https://arxiv.org/abs/2007.02987}{{\ttfamily
  2007.02987}}].

\bibitem{Chen:2020hmv}
H.~Z. Chen, R.~C. Myers, D.~Neuenfeld, I.~A. Reyes and J.~Sandor,
  \emph{{Quantum Extremal Islands Made Easy, Part II: Black Holes on the
  Brane}}, \href{https://doi.org/10.1007/JHEP12(2020)025}{\emph{JHEP}
  {\bfseries 12} (2020) 025}
  [\href{https://arxiv.org/abs/2010.00018}{{\ttfamily 2010.00018}}].

\bibitem{Hernandez:2020nem}
J.~Hernandez, R.~C. Myers and S.-M. Ruan, \emph{{Quantum extremal islands made
  easy. Part III. Complexity on the brane}},
  \href{https://doi.org/10.1007/JHEP02(2021)173}{\emph{JHEP} {\bfseries 02}
  (2021) 173} [\href{https://arxiv.org/abs/2010.16398}{{\ttfamily
  2010.16398}}].

\bibitem{Marolf:2020rpm}
D.~Marolf and H.~Maxfield, \emph{{Observations of Hawking radiation: the Page
  curve and baby universes}},
  \href{https://doi.org/10.1007/JHEP04(2021)272}{\emph{JHEP} {\bfseries 04}
  (2021) 272} [\href{https://arxiv.org/abs/2010.06602}{{\ttfamily
  2010.06602}}].

\bibitem{Matsuo:2020ypv}
Y.~Matsuo, \emph{{Islands and stretched horizon}},
  \href{https://arxiv.org/abs/2011.08814}{{\ttfamily 2011.08814}}.

\bibitem{Numasawa:2020sty}
T.~Numasawa, \emph{{Four coupled SYK models and Nearly AdS$_2$ gravities: Phase
  Transitions in Traversable wormholes and in Bra-ket wormholes}},
  \href{https://arxiv.org/abs/2011.12962}{{\ttfamily 2011.12962}}.

\bibitem{Colin-Ellerin:2020mva}
S.~Colin-Ellerin, X.~Dong, D.~Marolf, M.~Rangamani and Z.~Wang,
  \emph{{Real-time gravitational replicas: Formalism and a variational
  principle}},  \href{https://arxiv.org/abs/2012.00828}{{\ttfamily
  2012.00828}}.

\bibitem{Geng:2020fxl}
H.~Geng, A.~Karch, C.~Perez-Pardavila, S.~Raju, L.~Randall, M.~Riojas et~al.,
  \emph{{Information Transfer with a Gravitating Bath}},
  \href{https://arxiv.org/abs/2012.04671}{{\ttfamily 2012.04671}}.

\bibitem{Caceres:2020jcn}
E.~Caceres, A.~Kundu, A.~K. Patra and S.~Shashi, \emph{{Warped Information and
  Entanglement Islands in AdS/WCFT}},
  \href{https://doi.org/10.1007/JHEP07(2021)004}{\emph{JHEP} {\bfseries 07}
  (2021) 004} [\href{https://arxiv.org/abs/2012.05425}{{\ttfamily
  2012.05425}}].

\bibitem{Akal:2020twv}
I.~Akal, Y.~Kusuki, N.~Shiba, T.~Takayanagi and Z.~Wei, \emph{{Entanglement
  Entropy in a Holographic Moving Mirror and the Page Curve}},
  \href{https://doi.org/10.1103/PhysRevLett.126.061604}{\emph{Phys. Rev. Lett.}
  {\bfseries 126} (2021) 061604}
  [\href{https://arxiv.org/abs/2011.12005}{{\ttfamily 2011.12005}}].

\bibitem{Bousso:2021sji}
R.~Bousso and A.~Shahbazi-Moghaddam, \emph{{Island Finder and Entropy Bound}},
  \href{https://arxiv.org/abs/2101.11648}{{\ttfamily 2101.11648}}.

\bibitem{Kawabata:2021hac}
K.~Kawabata, T.~Nishioka, Y.~Okuyama and K.~Watanabe, \emph{{Probing Hawking
  radiation through capacity of entanglement}},
  \href{https://arxiv.org/abs/2102.02425}{{\ttfamily 2102.02425}}.

\bibitem{Miyata:2021ncm}
A.~Miyata and T.~Ugajin, \emph{{Evaporation of black holes in flat space
  entangled with an auxiliary universe}},
  \href{https://arxiv.org/abs/2104.00183}{{\ttfamily 2104.00183}}.

\bibitem{Kawabata:2021vyo}
K.~Kawabata, T.~Nishioka, Y.~Okuyama and K.~Watanabe, \emph{{Replica wormholes
  and capacity of entanglement}},
  \href{https://arxiv.org/abs/2105.08396}{{\ttfamily 2105.08396}}.

\bibitem{Harlow:2020bee}
D.~Harlow and E.~Shaghoulian, \emph{{Global symmetry, Euclidean gravity, and
  the black hole information problem}},
  \href{https://doi.org/10.1007/JHEP04(2021)175}{\emph{JHEP} {\bfseries 04}
  (2021) 175} [\href{https://arxiv.org/abs/2010.10539}{{\ttfamily
  2010.10539}}].

\bibitem{Chen:2020ojn}
Y.~Chen and H.~W. Lin, \emph{{Signatures of global symmetry violation in
  relative entropies and replica wormholes}},
  \href{https://arxiv.org/abs/2011.06005}{{\ttfamily 2011.06005}}.

\bibitem{Hsin:2020mfa}
P.-S. Hsin, L.~V. Iliesiu and Z.~Yang, \emph{{A violation of global symmetries
  from replica wormholes and the fate of black hole remnants}},
  \href{https://arxiv.org/abs/2011.09444}{{\ttfamily 2011.09444}}.

\bibitem{Yonekura:2020ino}
K.~Yonekura, \emph{{Topological violation of global symmetries in quantum
  gravity}},  \href{https://arxiv.org/abs/2011.11868}{{\ttfamily 2011.11868}}.

\bibitem{Saad:2018bqo}
P.~Saad, S.~H. Shenker and D.~Stanford, \emph{{A semiclassical ramp in SYK and
  in gravity}},  \href{https://arxiv.org/abs/1806.06840}{{\ttfamily
  1806.06840}}.

\bibitem{Saad:2019pqd}
P.~Saad, \emph{{Late Time Correlation Functions, Baby Universes, and ETH in JT
  Gravity}},  \href{https://arxiv.org/abs/1910.10311}{{\ttfamily 1910.10311}}.

\bibitem{Cotler:2020lxj}
J.~Cotler and K.~Jensen, \emph{{Gravitational Constrained Instantons}},
  \href{https://arxiv.org/abs/2010.02241}{{\ttfamily 2010.02241}}.

\bibitem{Cotler:2021cqa}
J.~Cotler and K.~Jensen, \emph{{Wormholes and black hole microstates in
  AdS/CFT}},  \href{https://arxiv.org/abs/2104.00601}{{\ttfamily 2104.00601}}.

\bibitem{Nomura:2019dlz}
Y.~Nomura, \emph{{Interior of a unitarily evaporating black hole}},
  \href{https://doi.org/10.1103/PhysRevD.102.026001}{\emph{Phys. Rev. D}
  {\bfseries 102} (2020) 026001}
  [\href{https://arxiv.org/abs/1911.13120}{{\ttfamily 1911.13120}}].

\bibitem{Langhoff:2020jqa}
K.~Langhoff and Y.~Nomura, \emph{{Ensemble from Coarse Graining: Reconstructing
  the Interior of an Evaporating Black Hole}},
  \href{https://doi.org/10.1103/PhysRevD.102.086021}{\emph{Phys. Rev. D}
  {\bfseries 102} (2020) 086021}
  [\href{https://arxiv.org/abs/2008.04202}{{\ttfamily 2008.04202}}].

\bibitem{Nomura:2020ewg}
Y.~Nomura, \emph{{From the Black Hole Conundrum to the Structure of Quantum
  Gravity}}, \href{https://doi.org/10.1142/S021773232130007X}{\emph{Mod. Phys.
  Lett. A} {\bfseries 36} (2021) 2130007}
  [\href{https://arxiv.org/abs/2011.08707}{{\ttfamily 2011.08707}}].

\bibitem{Hartman:2020khs}
T.~Hartman, Y.~Jiang and E.~Shaghoulian, \emph{{Islands in cosmology}},
  \href{https://doi.org/10.1007/JHEP11(2020)111}{\emph{JHEP} {\bfseries 11}
  (2020) 111} [\href{https://arxiv.org/abs/2008.01022}{{\ttfamily
  2008.01022}}].

\bibitem{Balasubramanian:2020coy}
V.~Balasubramanian, A.~Kar and T.~Ugajin, \emph{{Entanglement between two
  disjoint universes}},
  \href{https://doi.org/10.1007/JHEP02(2021)136}{\emph{JHEP} {\bfseries 02}
  (2021) 136} [\href{https://arxiv.org/abs/2008.05274}{{\ttfamily
  2008.05274}}].

\bibitem{Balasubramanian:2020xqf}
V.~Balasubramanian, A.~Kar and T.~Ugajin, \emph{{Islands in de Sitter space}},
  \href{https://doi.org/10.1007/JHEP02(2021)072}{\emph{JHEP} {\bfseries 02}
  (2021) 072} [\href{https://arxiv.org/abs/2008.05275}{{\ttfamily
  2008.05275}}].

\bibitem{Geng:2021wcq}
H.~Geng, Y.~Nomura and H.-Y. Sun, \emph{{An Information Paradox and Its
  Resolution in de Sitter Holography}},
  \href{https://arxiv.org/abs/2103.07477}{{\ttfamily 2103.07477}}.

\bibitem{Aalsma:2021bit}
L.~Aalsma and W.~Sybesma, \emph{{The Price of Curiosity: Information Recovery
  in de Sitter Space}},  \href{https://arxiv.org/abs/2104.00006}{{\ttfamily
  2104.00006}}.

\bibitem{Balasubramanian:2021wgd}
V.~Balasubramanian, A.~Kar and T.~Ugajin, \emph{{Entanglement between two
  gravitating universes}},  \href{https://arxiv.org/abs/2104.13383}{{\ttfamily
  2104.13383}}.

\bibitem{Kames-King:2021etp}
J.~Kames-King, E.~Verheijden and E.~Verlinde, \emph{{No Page Curves for the de
  Sitter Horizon}},  \href{https://arxiv.org/abs/2108.09318}{{\ttfamily
  2108.09318}}.

\bibitem{Langhoff:2021uct}
K.~Langhoff, C.~Murdia and Y.~Nomura, \emph{{The Multiverse in an Inverted
  Island}},  \href{https://arxiv.org/abs/2106.05271}{{\ttfamily 2106.05271}}.

\bibitem{Bekenstein:1980jp}
J.~D. Bekenstein, \emph{{A Universal Upper Bound on the Entropy to Energy Ratio
  for Bounded Systems}},
  \href{https://doi.org/10.1103/PhysRevD.23.287}{\emph{Phys. Rev. D} {\bfseries
  23} (1981) 287}.

\bibitem{Fischler:1998st}
W.~Fischler and L.~Susskind, \emph{{Holography and cosmology}},
  \href{https://arxiv.org/abs/hep-th/9806039}{{\ttfamily hep-th/9806039}}.

\bibitem{Bousso:1999xy}
R.~Bousso, \emph{{A Covariant entropy conjecture}},
  \href{https://doi.org/10.1088/1126-6708/1999/07/004}{\emph{JHEP} {\bfseries
  07} (1999) 004} [\href{https://arxiv.org/abs/hep-th/9905177}{{\ttfamily
  hep-th/9905177}}].

\bibitem{Bousso:1999cb}
R.~Bousso, \emph{{Holography in general space-times}},
  \href{https://doi.org/10.1088/1126-6708/1999/06/028}{\emph{JHEP} {\bfseries
  06} (1999) 028} [\href{https://arxiv.org/abs/hep-th/9906022}{{\ttfamily
  hep-th/9906022}}].

\bibitem{Bousso:2000nf}
R.~Bousso, \emph{{Positive vacuum energy and the N bound}},
  \href{https://doi.org/10.1088/1126-6708/2000/11/038}{\emph{JHEP} {\bfseries
  11} (2000) 038} [\href{https://arxiv.org/abs/hep-th/0010252}{{\ttfamily
  hep-th/0010252}}].

\bibitem{Bousso:2000md}
R.~Bousso, \emph{{Bekenstein bounds in de Sitter and flat space}},
  \href{https://doi.org/10.1088/1126-6708/2001/04/035}{\emph{JHEP} {\bfseries
  04} (2001) 035} [\href{https://arxiv.org/abs/hep-th/0012052}{{\ttfamily
  hep-th/0012052}}].

\bibitem{Bousso:2015mna}
R.~Bousso, Z.~Fisher, S.~Leichenauer and A.~C. Wall, \emph{{Quantum focusing
  conjecture}}, \href{https://doi.org/10.1103/PhysRevD.93.064044}{\emph{Phys.
  Rev. D} {\bfseries 93} (2016) 064044}
  [\href{https://arxiv.org/abs/1506.02669}{{\ttfamily 1506.02669}}].

\bibitem{Gibbons:1977mu}
G.~W. Gibbons and S.~W. Hawking, \emph{{Cosmological Event Horizons,
  Thermodynamics, and Particle Creation}},
  \href{https://doi.org/10.1103/PhysRevD.15.2738}{\emph{Phys. Rev. D}
  {\bfseries 15} (1977) 2738}.

\bibitem{Shiromizu:1993mt}
T.~Shiromizu, K.~Nakao, H.~Kodama and K.-I. Maeda, \emph{{Can large black holes
  collide in de Sitter space-time? An inflationary scenario of an inhomogeneous
  universe}}, \href{https://doi.org/10.1103/PhysRevD.47.R3099}{\emph{Phys. Rev.
  D} {\bfseries 47} (1993) R3099}.

\bibitem{Hayward:1993tt}
S.~A. Hayward, T.~Shiromizu and K.-i. Nakao, \emph{{A Cosmological constant
  limits the size of black holes}},
  \href{https://doi.org/10.1103/PhysRevD.49.5080}{\emph{Phys. Rev. D}
  {\bfseries 49} (1994) 5080}
  [\href{https://arxiv.org/abs/gr-qc/9309004}{{\ttfamily gr-qc/9309004}}].

\bibitem{Maeda:1997fh}
K.~Maeda, T.~Koike, M.~Narita and A.~Ishibashi, \emph{{Upper bound for entropy
  in asymptotically de Sitter space-time}},
  \href{https://doi.org/10.1103/PhysRevD.57.3503}{\emph{Phys. Rev. D}
  {\bfseries 57} (1998) 3503}
  [\href{https://arxiv.org/abs/gr-qc/9712029}{{\ttfamily gr-qc/9712029}}].

\bibitem{Anous:2020lka}
T.~Anous, J.~Kruthoff and R.~Mahajan, \emph{{Density matrices in quantum
  gravity}}, \href{https://doi.org/10.21468/SciPostPhys.9.4.045}{\emph{SciPost
  Phys.} {\bfseries 9} (2020) 045}
  [\href{https://arxiv.org/abs/2006.17000}{{\ttfamily 2006.17000}}].

\bibitem{Akal:2020wfl}
I.~Akal, Y.~Kusuki, T.~Takayanagi and Z.~Wei, \emph{{Codimension two holography
  for wedges}}, \href{https://doi.org/10.1103/PhysRevD.102.126007}{\emph{Phys.
  Rev. D} {\bfseries 102} (2020) 126007}
  [\href{https://arxiv.org/abs/2007.06800}{{\ttfamily 2007.06800}}].

\bibitem{Caputa:2017urj}
P.~Caputa, N.~Kundu, M.~Miyaji, T.~Takayanagi and K.~Watanabe, \emph{{Anti-de
  Sitter Space from Optimization of Path Integrals in Conformal Field
  Theories}}, \href{https://doi.org/10.1103/PhysRevLett.119.071602}{\emph{Phys.
  Rev. Lett.} {\bfseries 119} (2017) 071602}
  [\href{https://arxiv.org/abs/1703.00456}{{\ttfamily 1703.00456}}].

\bibitem{Caputa:2017yrh}
P.~Caputa, N.~Kundu, M.~Miyaji, T.~Takayanagi and K.~Watanabe, \emph{{Liouville
  Action as Path-Integral Complexity: From Continuous Tensor Networks to
  AdS/CFT}}, \href{https://doi.org/10.1007/JHEP11(2017)097}{\emph{JHEP}
  {\bfseries 11} (2017) 097}
  [\href{https://arxiv.org/abs/1706.07056}{{\ttfamily 1706.07056}}].

\bibitem{Boruch:2020wax}
J.~Boruch, P.~Caputa and T.~Takayanagi, \emph{{Path-Integral Optimization from
  Hartle-Hawking Wave Function}},
  \href{https://doi.org/10.1103/PhysRevD.103.046017}{\emph{Phys. Rev. D}
  {\bfseries 103} (2021) 046017}
  [\href{https://arxiv.org/abs/2011.08188}{{\ttfamily 2011.08188}}].

\bibitem{Boruch:2021hqs}
J.~Boruch, P.~Caputa, D.~Ge and T.~Takayanagi, \emph{{Holographic path-integral
  optimization}}, \href{https://doi.org/10.1007/JHEP07(2021)016}{\emph{JHEP}
  {\bfseries 07} (2021) 016}
  [\href{https://arxiv.org/abs/2104.00010}{{\ttfamily 2104.00010}}].

\bibitem{Bousso:1998na}
R.~Bousso and S.~W. Hawking, \emph{{Lorentzian condition in quantum gravity}},
  \href{https://doi.org/10.1103/PhysRevD.59.103501}{\emph{Phys. Rev. D}
  {\bfseries 59} (1999) 103501}
  [\href{https://arxiv.org/abs/hep-th/9807148}{{\ttfamily hep-th/9807148}}].

\bibitem{DiTucci:2019dji}
A.~Di~Tucci and J.-L. Lehners, \emph{{No-Boundary Proposal as a Path Integral
  with Robin Boundary Conditions}},
  \href{https://doi.org/10.1103/PhysRevLett.122.201302}{\emph{Phys. Rev. Lett.}
  {\bfseries 122} (2019) 201302}
  [\href{https://arxiv.org/abs/1903.06757}{{\ttfamily 1903.06757}}].

\bibitem{DiTucci:2019xcr}
A.~Di~Tucci, J.~Feldbrugge, J.-L. Lehners and N.~Turok, \emph{{Quantum
  Incompleteness of Inflation}},
  \href{https://doi.org/10.1103/PhysRevD.100.063517}{\emph{Phys. Rev. D}
  {\bfseries 100} (2019) 063517}
  [\href{https://arxiv.org/abs/1906.09007}{{\ttfamily 1906.09007}}].

\bibitem{DiTucci:2019bui}
A.~Di~Tucci, J.-L. Lehners and L.~Sberna, \emph{{No-boundary prescriptions in
  Lorentzian quantum cosmology}},
  \href{https://doi.org/10.1103/PhysRevD.100.123543}{\emph{Phys. Rev. D}
  {\bfseries 100} (2019) 123543}
  [\href{https://arxiv.org/abs/1911.06701}{{\ttfamily 1911.06701}}].

\bibitem{Calabrese:2005in}
P.~Calabrese and J.~L. Cardy, \emph{{Evolution of entanglement entropy in
  one-dimensional systems}},
  \href{https://doi.org/10.1088/1742-5468/2005/04/P04010}{\emph{J. Stat. Mech.}
  {\bfseries 0504} (2005) P04010}
  [\href{https://arxiv.org/abs/cond-mat/0503393}{{\ttfamily
  cond-mat/0503393}}].

\bibitem{Nakata:2021ubr}
Y.~Nakata, T.~Takayanagi, Y.~Taki, K.~Tamaoka and Z.~Wei, \emph{{New
  holographic generalization of entanglement entropy}},
  \href{https://doi.org/10.1103/PhysRevD.103.026005}{\emph{Phys. Rev. D}
  {\bfseries 103} (2021) 026005}
  [\href{https://arxiv.org/abs/2005.13801}{{\ttfamily 2005.13801}}].

\bibitem{Mollabashi:2020yie}
A.~Mollabashi, N.~Shiba, T.~Takayanagi, K.~Tamaoka and Z.~Wei, \emph{{Pseudo
  Entropy in Free Quantum Field Theories}},
  \href{https://doi.org/10.1103/PhysRevLett.126.081601}{\emph{Phys. Rev. Lett.}
  {\bfseries 126} (2021) 081601}
  [\href{https://arxiv.org/abs/2011.09648}{{\ttfamily 2011.09648}}].

\bibitem{Mollabashi:2021xsd}
A.~Mollabashi, N.~Shiba, T.~Takayanagi, K.~Tamaoka and Z.~Wei, \emph{{Aspects
  of Pseudo Entropy in Field Theories}},
  \href{https://arxiv.org/abs/2106.03118}{{\ttfamily 2106.03118}}.

\bibitem{Nishioka:2021cxe}
T.~Nishioka, T.~Takayanagi and Y.~Taki, \emph{{Topological pseudo entropy}},
  \href{https://arxiv.org/abs/2107.01797}{{\ttfamily 2107.01797}}.

\bibitem{OPEW}
M.~Miyaji and G.~Penington, \emph{{Work in progress}}, .

\bibitem{Brown:2019rox}
A.~R. Brown, H.~Gharibyan, G.~Penington and L.~Susskind, \emph{{The
  Python\textquoteright{}s Lunch: geometric obstructions to decoding Hawking
  radiation}}, \href{https://doi.org/10.1007/JHEP08(2020)121}{\emph{JHEP}
  {\bfseries 08} (2020) 121}
  [\href{https://arxiv.org/abs/1912.00228}{{\ttfamily 1912.00228}}].

\bibitem{Maldacena:2016upp}
J.~Maldacena, D.~Stanford and Z.~Yang, \emph{{Conformal symmetry and its
  breaking in two dimensional Nearly Anti-de-Sitter space}},
  \href{https://doi.org/10.1093/ptep/ptw124}{\emph{PTEP} {\bfseries 2016}
  (2016) 12C104} [\href{https://arxiv.org/abs/1606.01857}{{\ttfamily
  1606.01857}}].

\bibitem{Goel:2020yxl}
A.~Goel, L.~V. Iliesiu, J.~Kruthoff and Z.~Yang, \emph{{Classifying boundary
  conditions in JT gravity: from energy-branes to $\alpha$-branes}},
  \href{https://doi.org/10.1007/JHEP04(2021)069}{\emph{JHEP} {\bfseries 04}
  (2021) 069} [\href{https://arxiv.org/abs/2010.12592}{{\ttfamily
  2010.12592}}].

\bibitem{Antonini:2021xar}
S.~Antonini and B.~Swingle, \emph{{Holographic boundary states and
  dimensionally-reduced braneworld spacetimes}},
  \href{https://arxiv.org/abs/2105.02912}{{\ttfamily 2105.02912}}.

\bibitem{Cooper:2018cmb}
S.~Cooper, M.~Rozali, B.~Swingle, M.~Van~Raamsdonk, C.~Waddell and D.~Wakeham,
  \emph{{Black hole microstate cosmology}},
  \href{https://doi.org/10.1007/JHEP07(2019)065}{\emph{JHEP} {\bfseries 07}
  (2019) 065} [\href{https://arxiv.org/abs/1810.10601}{{\ttfamily
  1810.10601}}].

\bibitem{Antonini:2019qkt}
S.~Antonini and B.~Swingle, \emph{{Cosmology at the end of the world}},
  \href{https://doi.org/10.1038/s41567-020-0909-6}{\emph{Nature Phys.}
  {\bfseries 16} (2020) 881}
  [\href{https://arxiv.org/abs/1907.06667}{{\ttfamily 1907.06667}}].

\bibitem{Herzog:2015ioa}
C.~P. Herzog, K.-W. Huang and K.~Jensen, \emph{{Universal Entanglement and
  Boundary Geometry in Conformal Field Theory}},
  \href{https://doi.org/10.1007/JHEP01(2016)162}{\emph{JHEP} {\bfseries 01}
  (2016) 162} [\href{https://arxiv.org/abs/1510.00021}{{\ttfamily
  1510.00021}}].

\bibitem{Miyaji:2021ktr}
M.~Miyaji, T.~Takayanagi and T.~Ugajin, \emph{{Spectrum of End of the World
  Branes in Holographic BCFTs}},
  \href{https://arxiv.org/abs/2103.06893}{{\ttfamily 2103.06893}}.

\bibitem{Israel:1966rt}
W.~Israel, \emph{{Singular hypersurfaces and thin shells in general
  relativity}}, \href{https://doi.org/10.1007/BF02710419}{\emph{Nuovo Cim. B}
  {\bfseries 44S10} (1966) 1}.

\bibitem{Oshikawa:1996dj}
M.~Oshikawa and I.~Affleck, \emph{{Boundary conformal field theory approach to
  the critical two-dimensional Ising model with a defect line}},
  \href{https://doi.org/10.1016/S0550-3213(97)00219-8}{\emph{Nucl. Phys. B}
  {\bfseries 495} (1997) 533}
  [\href{https://arxiv.org/abs/cond-mat/9612187}{{\ttfamily
  cond-mat/9612187}}].

\bibitem{Takayanagi:2019tvn}
T.~Takayanagi and K.~Tamaoka, \emph{{Gravity Edges Modes and Hayward Term}},
  \href{https://doi.org/10.1007/JHEP02(2020)167}{\emph{JHEP} {\bfseries 02}
  (2020) 167} [\href{https://arxiv.org/abs/1912.01636}{{\ttfamily
  1912.01636}}].

\bibitem{Hayward:1993my}
G.~Hayward, \emph{{Gravitational action for space-times with nonsmooth
  boundaries}}, \href{https://doi.org/10.1103/PhysRevD.47.3275}{\emph{Phys.
  Rev. D} {\bfseries 47} (1993) 3275}.

\bibitem{Saad:2021rcu}
P.~Saad, S.~H. Shenker, D.~Stanford and S.~Yao, \emph{{Wormholes without
  averaging}},  \href{https://arxiv.org/abs/2103.16754}{{\ttfamily
  2103.16754}}.

\bibitem{Blommaert:2019wfy}
A.~Blommaert, T.~G. Mertens and H.~Verschelde, \emph{{Eigenbranes in
  Jackiw-Teitelboim gravity}},
  \href{https://doi.org/10.1007/JHEP02(2021)168}{\emph{JHEP} {\bfseries 02}
  (2021) 168} [\href{https://arxiv.org/abs/1911.11603}{{\ttfamily
  1911.11603}}].

\bibitem{Blommaert:2021fob}
A.~Blommaert, L.~V. Iliesiu and J.~Kruthoff, \emph{{Gravity factorized}},
  \href{https://arxiv.org/abs/2111.07863}{{\ttfamily 2111.07863}}.

\bibitem{Dong:2016hjy}
X.~Dong, A.~Lewkowycz and M.~Rangamani, \emph{{Deriving covariant holographic
  entanglement}}, \href{https://doi.org/10.1007/JHEP11(2016)028}{\emph{JHEP}
  {\bfseries 11} (2016) 028}
  [\href{https://arxiv.org/abs/1607.07506}{{\ttfamily 1607.07506}}].

\bibitem{Goto:2020wnk}
K.~Goto, T.~Hartman and A.~Tajdini, \emph{{Replica wormholes for an evaporating
  2D black hole}},  \href{https://arxiv.org/abs/2011.09043}{{\ttfamily
  2011.09043}}.

\bibitem{Colin-Ellerin:2021jev}
S.~Colin-Ellerin, X.~Dong, D.~Marolf, M.~Rangamani and Z.~Wang,
  \emph{{Real-time gravitational replicas: Low dimensional examples}},
  \href{https://arxiv.org/abs/2105.07002}{{\ttfamily 2105.07002}}.

\bibitem{Calabrese:2004eu}
P.~Calabrese and J.~L. Cardy, \emph{{Entanglement entropy and quantum field
  theory}}, \href{https://doi.org/10.1088/1742-5468/2004/06/P06002}{\emph{J.
  Stat. Mech.} {\bfseries 0406} (2004) P06002}
  [\href{https://arxiv.org/abs/hep-th/0405152}{{\ttfamily hep-th/0405152}}].

\bibitem{Miyaji:2014mca}
M.~Miyaji, S.~Ryu, T.~Takayanagi and X.~Wen, \emph{{Boundary States as
  Holographic Duals of Trivial Spacetimes}},
  \href{https://doi.org/10.1007/JHEP05(2015)152}{\emph{JHEP} {\bfseries 05}
  (2015) 152} [\href{https://arxiv.org/abs/1412.6226}{{\ttfamily 1412.6226}}].

\bibitem{Affleck:1991tk}
I.~Affleck and A.~W.~W. Ludwig, \emph{{Universal noninteger 'ground state
  degeneracy' in critical quantum systems}},
  \href{https://doi.org/10.1103/PhysRevLett.67.161}{\emph{Phys. Rev. Lett.}
  {\bfseries 67} (1991) 161}.

\bibitem{Numasawa:2016emc}
T.~Numasawa, N.~Shiba, T.~Takayanagi and K.~Watanabe, \emph{{EPR Pairs, Local
  Projections and Quantum Teleportation in Holography}},
  \href{https://doi.org/10.1007/JHEP08(2016)077}{\emph{JHEP} {\bfseries 08}
  (2016) 077} [\href{https://arxiv.org/abs/1604.01772}{{\ttfamily
  1604.01772}}].

\bibitem{Almheiri:2016blp}
A.~Almheiri, X.~Dong and B.~Swingle, \emph{{Linearity of Holographic
  Entanglement Entropy}},
  \href{https://doi.org/10.1007/JHEP02(2017)074}{\emph{JHEP} {\bfseries 02}
  (2017) 074} [\href{https://arxiv.org/abs/1606.04537}{{\ttfamily
  1606.04537}}].

\bibitem{Harlow:2013tf}
D.~Harlow and P.~Hayden, \emph{{Quantum Computation vs. Firewalls}},
  \href{https://doi.org/10.1007/JHEP06(2013)085}{\emph{JHEP} {\bfseries 06}
  (2013) 085} [\href{https://arxiv.org/abs/1301.4504}{{\ttfamily 1301.4504}}].

\bibitem{Hubeny:2012wa}
V.~E. Hubeny and M.~Rangamani, \emph{{Causal Holographic Information}},
  \href{https://doi.org/10.1007/JHEP06(2012)114}{\emph{JHEP} {\bfseries 06}
  (2012) 114} [\href{https://arxiv.org/abs/1204.1698}{{\ttfamily 1204.1698}}].

\bibitem{Wall:2012uf}
A.~C. Wall, \emph{{Maximin Surfaces, and the Strong Subadditivity of the
  Covariant Holographic Entanglement Entropy}},
  \href{https://doi.org/10.1088/0264-9381/31/22/225007}{\emph{Class. Quant.
  Grav.} {\bfseries 31} (2014) 225007}
  [\href{https://arxiv.org/abs/1211.3494}{{\ttfamily 1211.3494}}].

\bibitem{Headrick:2014cta}
M.~Headrick, V.~E. Hubeny, A.~Lawrence and M.~Rangamani, \emph{{Causality
  \textbackslash{}\& holographic entanglement entropy}},
  \href{https://doi.org/10.1007/JHEP12(2014)162}{\emph{JHEP} {\bfseries 12}
  (2014) 162} [\href{https://arxiv.org/abs/1408.6300}{{\ttfamily 1408.6300}}].

\bibitem{Maldacena:2004rf}
J.~M. Maldacena and L.~Maoz, \emph{{Wormholes in AdS}},
  \href{https://doi.org/10.1088/1126-6708/2004/02/053}{\emph{JHEP} {\bfseries
  02} (2004) 053} [\href{https://arxiv.org/abs/hep-th/0401024}{{\ttfamily
  hep-th/0401024}}].

\bibitem{Marolf:2021kjc}
D.~Marolf and J.~E. Santos, \emph{{AdS Euclidean wormholes}},
  \href{https://arxiv.org/abs/2101.08875}{{\ttfamily 2101.08875}}.

\end{thebibliography}\endgroup

\end{document}